\documentclass[10pt,a4paper,ptm]{article}
\usepackage[utf8]{inputenc}
\usepackage[english]{babel}
\usepackage{amsmath}
\usepackage{amsfonts}
\usepackage{amssymb}
\usepackage{graphicx}
\usepackage{xcolor,tikz,datetime,multirow,lscape,multicol,float,soul}
\usepackage[outline]{contour}
\contourlength{.2pt}
\usepackage[round]{natbib}
\usepackage[affil-it]{authblk}
\usepackage[breaklinks=true,bookmarks=true,colorlinks=true,linkcolor=blue,citecolor=blue]{hyperref}
\newsavebox{\astrutbox}
\sbox{\astrutbox}{\rule[-5pt]{0pt}{20pt}}

\title{\bf Direct numerical simulation of open-channel flow over a fully-rough wall at moderate relative submergence}

\author[1]{{\bf MARCO MAZZUOLI}}
\author[2]{{\bf MARKUS UHLMANN}}
\affil[1]{{\small Department of Civil, Chemical and Environmental Engineering (DICCA), University of Genoa, 
Via Montallegro 1, 16145 Genova, Italy}}
\affil[2]{{\small Institute for Hydromechanics, Karlsruhe Institute of Technology,
76131 Karlsruhe, Germany}}

\usepackage{soul,xcolor,multirow,ifthen,float,morefloats}
\newcommand{\figdir}{./figure/}

\newcommand{\refdir}{./}

\newcommand{\pbt}[1]{#1}

\newcommand{\mut}[1]{#1}

\usepackage{geometry}
\newlength\halflineskip
\newlength\affilskip
\usepackage{fancyhdr}
\usepackage[font=small,labelfont=sc]{caption}
\usepackage{titlesec}
\titleformat{\section}
  {\bf\large}
  {\thesection}{1em}{\vspace*{-.2cm}}
\usepackage{etoolbox}
\AtBeginEnvironment{tabular}{\small}

\usepackage{array}
\newif\ifbib
\bibfalse
\newif\ifall
\alltrue
\newif\ifone
\newif\iftwo
\newif\ifthree
\newif\iffour
\newif\iffive
\newif\ifsix
\newif\ifseven
\newif\ifeight
\newif\ifnine
\newif\iften
\newif\ifeleven
\newif\iftwelve
\newif\ifthirteen
\newif\iffourteen
\newif\iffifteen
\newif\ifsixteen
\newif\ifseventeen
\newif\ifeighteen
\newif\ifnineteen
\newif\iftwenty
\newif\iftwentyone
\newif\iftwentytwo
\onefalse
\twofalse
\threefalse
\fourfalse
\fivefalse
\sixfalse
\sevenfalse
\eightfalse
\ninefalse
\tenfalse
\elevenfalse
\twelvefalse
\thirteenfalse
\fourteenfalse
\fifteenfalse
\sixteenfalse
\seventeenfalse
\eighteenfalse
\nineteenfalse
\twentyfalse
\twentyonefalse
\twentytwofalse
\ifall
\onetrue
\twotrue
\threetrue
\fourtrue
\fivetrue
\sixtrue
\seventrue
\eighttrue
\ninetrue
\tentrue
\eleventrue
\twelvetrue
\thirteentrue
\fourteentrue
\fifteentrue
\sixteentrue
\seventeentrue
\eighteentrue
\nineteentrue
\twentytrue
\twentyonetrue
\twentytwotrue
\fi
\fourteenfalse
\nineteentrue
\sixteenfalse
\seventeenfalse
\nineteenfalse
\twentyfalse
\twentyonefalse
\twentytwofalse

\date{May, 2017}

\begin{document}
%
\maketitle
\begin{abstract}
Direct numerical simulation of open-channel flow over a bed of spheres arranged 
in a regular pattern has been carried out at \mut{bulk Reynolds number and}
roughness Reynolds number \mut{(based on sphere diameter)} of approximately \mut{$6900$ and}
$120$\mut{, respectively,} for which the flow regime is fully-rough.
\mut{The open-channel height was approximately $5.5$ times the diameter of the spheres.}
Extending the results obtained by 
Chan-Braun {\it et al.\ }({\it J.\ Fluid Mech.}, vol.\ 684, 2011, 441) 
for an open-channel flow in the transitionally-rough regime, the present 
purpose is to show how the flow structure changes as the fully-rough regime is attained and, 
for the first time, to enable a direct comparison with experimental observations. 
\mut{
  Different statistical tools were used to investigate the flow field in the roughness sublayer 
  and in the logarithmic region.
}
The results indicate that\mut{, in the vicinity of the roughness elements,} 
the average flow field is affected 
both by Reynolds number effects and by the geometrical features of 
the roughness, 
\mut{
  while at larger wall-distances this is not the case, and roughness
  concepts can be applied.         
  Thus, the roughness function is computed which in the present set-up 
  can be expected to depend on the relative submergence.    
}
\mut{
  The flow-roughness interaction occurs mostly in the region above the
  virtual origin of the velocity profile, 
  and the effect of form-induced velocity fluctuations is maximum
  at the level of sphere crests.
  In particular,  
}
the root mean square of fluctuations about the \mut{streamwise component of the} average velocity field 
\mut{
  reflects the geometry of the spheres in the roughness sublayer and attains a maximum value just 
  above the roughness elements. 
  The latter is significantly weakened and shifted towards larger
  wall-distances as compared to the transitionally-rough regime or the case of a smooth wall. %
}%
\pbt{%
  The spanwise length scale of turbulent velocity fluctuations in the
  vicinity of the sphere crests shows the same dependence on the
  distance from the wall as that observed over a smooth wall, and both
  vary with Reynolds number in a similar fashion. %
}%
Moreover, the hydrodynamic force and torque experienced by the roughness elements are 
investigated and the footprint left by vortex structures on the stress acting on the 
sphere surface is observed.
Finally, the possibility either to adopt an analogy between the hydrodynamic forces 
associated with the interaction of turbulent structures with a flat smooth wall or 
with the surface of the spheres is also discussed, distinguishing the skin-friction 
from the form-drag contributions both in the transitionally-rough and in the fully-rough regimes.
\end{abstract}
%
%
%
\section{Introduction}
The present investigation is motivated by the problem of erosion and deposition 
of sediment at the bottom of fluvial and estuarine environments as well as along hill slopes, 
which result from the action of the surface water flow. 
The ultimate goal is to understand the morphological evolution of the sediment bed. 
Other applications of channel flows can be found in chemistry, biological fluid 
dynamics and industrial engineering.

Providing reliable predictions of the river-bed evolution requires a clear picture 
of the interaction between the flow and the bottom roughness \mut{due to sediments and bedforms.}
Typically, in the regimes of practical interest, the flow is turbulent and the bottom 
is not smooth, but it is characterised by the presence of natural or artificial protrusions 
that affect the structure of the flow over a region few times thicker than the size of the 
protrusions, namely the \textit{roughness sublayer}.
Hence, a detailed description of turbulence \mut{in the vicinity of individual roughness 
elements} is necessary to comprehend the dynamics of solid-fluid interaction and 
possibly formulate consistent relationships with quantities observable at \mut{larger} scales. 
This knowledge becomes crucial in the case of shallow open-channels which are characterised 
by small values of the \mut{\textit{relative submergence} (also termed \textit{inundation ratio})} 
$H/k$, where $H$ \mut{denotes} the open-channel height and $k$ the roughness size, 
since velocity fluctuations originating at the bottom may significantly affect the entire flow field.
However, even in the simple case in which monosized regular roughness elements are considered, 
obtaining accurate measurements of the velocity and pressure fluctuations in the crevices 
between the roughness elements is extremely difficult \citep[e.g.][]{hong2011,amir2014}, 
and discrepant pictures of the origin of the turbulent vortices and of their influence on the flow structure have been provided \citep[e.g.][for a review]{marusic2010}.

It is well established that a\mut{n open-channel flow (or similarly a boundary layer)} 
over a smooth wall at sufficiently high Reynolds numbers develops a \textit{viscous sublayer}, 
dominated by viscous effects, and \textit{\mut{a logarithmic} region}, where the fluid 
viscosity plays a negligible role, matched together through a \textit{buffer layer} 
characterised by strong normal \mut{Reynolds stresses}.
Let us consider roughness elements located on a flat smooth wall 
\mut{ with a certain arrangement, 
}
defining the geometrical 
roughness size $k$ as the average distance 
from the wall to the crest of the roughness elements. 
\mut{
  Let, for the moment, large values of $H/k$ be considered, 
}
for which effects associated with the free-slip boundary condition at the free 
surface of the open channel can be neglected in the vicinity of the bottom wall. 
Therefore, for increasing values of the roughness Reynolds number $k^+=k u_\tau/\nu$, 
where $u_\tau$ and $\nu$ denote the friction velocity and the kinematic viscosity of the 
fluid, three flow regimes \mut{can be} identified: the \textit{hydraulically-smooth} regime, 
the \textit{transitionally-rough} regime and the \textit{fully-rough} regime \citep{jimenez2004}.
\mut{
  Let henceforth $y$ and $y_0$ denote the wall-normal coordinate and the position of the 
  virtual origin; the latter one is defined as the plane where a smooth wall should be placed 
  (in absence of the roughness elements) to observe the logarithmic region originating 
  at the same distance from $y_0$ as in the rough-wall case.
}
In the hydraulically-smooth regime, 
\mut{
  the roughness elements are entirely contained within the viscous sublayer and the velocity 
  profile, 
  as a function of the distance from the virtual wall,  
}
practically collapses upon the profile that could be obtained over a smooth wall at 
the same bulk Reynolds number 
\mut{
  $\mut{Re_{bh}}=hU_{bh}/\nu$,
} where $h$ equals $H-y_0$ and 
\mut{
  $U_{bh}$
}
denotes the bulk velocity defined as $
\mut{
  U_{bh}
}
=1/h\int^H_{y_0}\left\langle\overline{u}\right\rangle\,dy$, the operator 
$\left\langle\overline{\cdot}\right\rangle$ indicating the statistical average defined more 
precisely below. 
Then, by gradually increasing the Reynolds number \mut{$k^+$} until the transitionally-rough 
regime is attained, the viscous sublayer is significantly thinned with respect to the 
hydraulically-smooth regime, while the mean velocity, \mut{normalised by $u_\tau$}, is 
reduced (shifted) in the \mut{logarithmic} region as an effect of the increasing momentum 
transfer to the roughness elements. 
Finally, for $k^+\gtrsim 55-90$, the buffer layer disappears and the fully-rough flow 
regime is reached \citep{ligrani1986}.

\mut{
  The shift of the velocity profile in the logarithmic region with respect to that observed 
  in absence of the roughness elements (i.e. over a smooth-wall), namely the 
  \emph{roughness function} $\Delta\left\langle\overline{u}\right\rangle^+$, 
  can be used as a universal parameter to classify the flow
  regime in wall bounded turbulent flows characterised by large values of the relative 
  submergence so that the flow structure becomes independent of $H/k$ 
  \citep[$H/k>40$,][]{jimenez2004}. 
  Instead, for small values of the relative submergence, some mechanisms of the 
  wall-turbulence are possibly affected while the effects of the geometrical 
  features of the roughness elements can be recognised in the entire domain.
  These cases can be suitably studied as flows over obstacles, since roughness 
  concept cannot be applied to any region of the flow field.
  However, for \emph{moderate relative submergence}, the effects 
  associated with individual roughness elements tend to vanish far from the bottom 
  and a logarithmic region can be clearly distinguished over the
  roughness sublayer \citep[e.g.][]{bayazit1976}. 
  In the latter case, which is the object of the present investigation,
  the hydraulically-smooth, transitionally- and fully-rough regimes can be identified 
  on the basis of the roughness function and of the relative submergence $H/k$ which 
  becomes a parameter of the problem.

  The roughness characteristics of a natural bed (for instance a river channel) are not homogeneous
  (due to the non-regular shape and arrangement of roughness elements and to the presence of 
  multiple scales defining the roughness geometry) and can vary with time
  (e.g. due to sediment transport and bedform evolution). 
  We presently consider the particular case of fixed and identical roughness elements
  arranged with a regular pattern, which limits the geometrical characterization of 
  the roughness to that of a minimal roughness unit (i.e. a single roughness elements 
  and its closest neighbours) and allows us to consider the roughness
  characteristics 
  as spatially  
  homogeneous and constant with time.
}
Although this approach limits the scope of application of the present results to specific 
problems, it allows us to \mut{identify} clearly the mechanism of flow-roughness interaction.

Let us focus our attention on the flow structure in the
\mut{
  logarithmic region, 
  where
} 
the velocity profile behaves like a logarithmic similarity law (i.e. Prandtl's celebrated 
``law of the wall'') which can be expressed in terms of wall units as follows:
\begin{equation}
  \left\langle\overline{u}\right\rangle^+(y^+) 
  = 
  \dfrac{1}{\kappa}\ln (y^+-y^+_0) 
  + 
  C^+_{I}(k^+)
\label{eq1}
\end{equation}
or equivalently in terms of $k$:
\begin{equation}
  \left\langle\overline{u}\right\rangle^+(y) 
  = 
  \dfrac{1}{\kappa}\ln \dfrac{y-y_0}{k} 
  + 
  C^+_{II}
  \:\: .
\label{eq1a}
\end{equation}
\mut{It turns out that } 
$\frac{d\left\langle\overline{u}\right\rangle^+}{dy}$ only depends on $y$, $k$ and 
$u_{\tau}$, 
while 
$(y^+-y^+_0)\frac{\partial\left\langle\overline{u}\right\rangle^+}{\partial y^+}$ 
is constant, namely the inverse of the Von \mut{K\'arm\'an} constant $\kappa$\mut{, $y_0$ denoting 
the aforementioned virtual origin of the wall-normal coordinate}.
The integration constant $C^+_{I}$ was experimentally found to tend to
$5.1$ for $k^+\rightarrow 0$ (hydraulically-smooth regime) and to
$C^+_{II}-\frac{1}{\kappa}\ln k^+$, with $C^+_{II}$ constant, for
$k^+\rightarrow \infty$ (fully-rough regime)
\citep{nikuradse1933,schlichting1968,pimenta1975,ligrani1986}. 
The value of the constant $C^+_{II}$, in the fully-rough regime
\mut{
  at large relative submergence, 
}
depends on the shape and the arrangement of roughness elements and tends 
to $8.5$ for the case of a boundary layer over a plane, well-packed, 
sandy bottom \citep{ligrani1986}.
The fact that $\left\langle\overline{u}\right\rangle^+$ is independent of $k^+$ 
in the fully-rough regime follows the 
disappearance of the buffer layer and the attenuation of viscous effects above the roughness 
\citep{tani1987}. 
This feature is in line with Townsend's hypothesis, according
to which in a boundary layer at high values of $k^+$ the
turbulence structure above the roughness sublayer is practically
unaffected by the roughness shape and arrangement. 
From equation \eqref{eq1a} it can be observed that, once the fully-rough regime is attained, 
\mut{
  if the relative submergence is sufficiently large that a logarithmic region can be identified, 
}
the velocity profile experiences a shift which increases proportionally to the logarithm of 
$k$ \citep[e.g.][]{nikuradse1933,ligrani1986}.
This suggests that the flow structure may remain basically unaffected by further increases 
of $k^+$, and that increases of $k^+$ result in the progressive truncation of the velocity 
profile at distance \mut{approximately $k^+$} from the wall.
\mut{
  Deviations of such a distance from $k^+$ are associated with the position of $y_0$ that in turn
  reflects the geometrical features of the roughness.
}
%

\mut{
After the pioneering systematic work of \citet{nikuradse1933} on rough
pipe flows, \citet{schlichting1936} was the first who experimentally
investigated the effects on open-channel flow of combining different
(regular, homogeneous) arrangements of spheres mounted on a smooth
wall and varying their size. 
Since then, large efforts have been devoted to investigate the relevance of the roughness geometrical features on the turbulence structure.
\pbt{
  Providing an exhaustive roughness characterization is still one of the major 
  challenges in the field. 
}
A number of recent experimental studies showed that the size, shape and arrangement of roughness elements can significantly affect the bottom drag and the flow structure in the near bottom region 
}
\pbt{
  \citep[][to cite a few examples]{
  amir2011,cooper2013,florens2013,willingham2014,placidi2015,bossuyt2017}
}
In the attempt of synthesizing the complexity of the roughness
geometry within a single parameter, the \textit{sand-grain roughness}
(or \textit{effective roughness}), $k_s$, is commonly used 
which is either found to be proportional to the roughness size $k$
or proportional to the channel height (boundary layer thickness),
depending on the precise geometrical features of the rough surface. 
\mut{
  In the fully rough regime, for large values of $H/k$, $k_s$ is defined as the roughness height which
  produces the same roughness function as that measured by
  \citet{nikuradse1933}, cf.\ \cite{flack2010}. 
  Concerning the value of $k_s$, \citet{schlichting1936} observed
  that, for mono-sized,   
  spherical roughness elements in a hexagonal arrangement, $k/k_s$
  ranged from $0.26$ to $4.41$ only by varying the distance between the grains.
  Using a single parameter to characterise rough surfaces is undoubtedly convenient as long as 
  the relative submergence and the roughness Reynolds number are sufficiently high ($H/k>40$ 
  and $k_s^+>50$, \citet{jimenez2004}).
  However, 
  a single parameter does not suffice, at moderate relative submergence, 
  to characterize the roughness function, which ultimately depends also on $H/k$ %
  \pbt{%
    and on the other length scales characterising the roughness geometry. 
  }
  At moderate relative submergence, the value of the roughness function at which the fully-rough
  regime is attained can be different from that measured by \citet{nikuradse1933} and, consequently,
  the parameter $k_s^+$ can be no longer used unambiguously to determine the flow regime.
  For instance, \citet{amir2014} have recently carried out experiments of a moderately 
  shallow open-channel flow in the fully-rough regime.
  They could observe the mean velocity profile to follow a logarithmic curve in the core 
  of the flow field and, in one of their tests, they measured the roughness function equal 
  to $6.5$, which corresponds to a value of $k_s^+$ barely larger than $50$, 
  although the flow regime was fully-rough.
}
Indeed, the quantity $k_s$ is defined heuristically, leaving us free to interpret it 
as the hydrodynamic response of the flow to the disturbance induced by the roughness.
For example, in this line, \citet{orlandi2003} and \citet{flores2006} investigated 
the effect of superimposing a disturbance of the velocity field in the vicinity of a 
smooth-wall over an otherwise undisturbed channel-flow and observed the development 
of turbulent fluctuations associated with the disturbance that were similar to those 
induced by a physical roughness.
\mut{
  In fact, the velocity close to the wall could be locally and instantaneously nullified 
  by the disturbance.
}
Although the bottom was not rough, \citet{orlandi2003} and \citet{flores2006} simulated the fully-rough regime and, in principle, they could have estimated the value of 
\mut{
  the roughness function and of $k_s$.
}

In the present
\mut{
  direct numerical
}
simulations, 
\mut{
  the effect of the relative submergence cannot be neglected, and 
}
the grain size $k$ \mut{will be used} as the length scale instead of $k_s$, 
\mut{
  since the results cannot be generalised to other rough-wall flows
  characterised by the same roughness function.
}

\mut{Nonetheless, }\citet{chan2011}, who performed two direct numerical simulations of 
\mut{moderately} shallow open-channel flow in the hydraulically-smooth and the 
transitionally-rough regimes, \mut{noted} that the \mut{values of the roughness function 
were in the range of} those obtained for rough boundary layers at the same 
\mut{roughness} Reynolds numbers \mut{$k_s^+$}.
\citet{chan2011,chan2013} could observe the presence of a buffer layer just above the crest 
of the roughness elements in which the velocity field is affected by %
  \pbt{%
  viscosity, the friction velocity, the channel height and bulk velocity as well as the
  roughness size. %
  }%
  \mut{%
  It is widely recognised that the features of the turbulence structure related to the 
  geometrical characteristics of the roughness are lost at a certain distance from the wall 
  in well developed boundary layers at sufficiently high Reynolds numbers. 
  Experimental results confirm a fair agreement with the ``law of the wall'' 
  \citep[e.g][]{ligrani1986,bandyopadhyay1987,SchultzFlack2007}.
  However, some of the effects associated with the roughness or the presence of a pressure 
  gradient were not found completely to disappear in experiments
  carried out in plane channel flow 
  \citep[e.g.][]{grass1991,hong2011,hong2012} and open-channel flow  
  \citep[e.g.][]{balachandar1999,tachie2000,amir2014}.
}
%
%
In particular, \citet{hong2011} found that Townsend's hypothesis for the Reynolds stress statistics was supported \mut{above} the roughness sublayer ($y\gtrsim 2 k$) but the presence of roughness-related small-scale turbulence affected the dissipation rate in the entire flow field, while \citet{george2007} showed that the effect of the mean pressure gradient on turbulence statistics in pressure-gradient-driven channel (or pipe) flows can be assumed negligible only in a certain region in the vicinity of the bottom and at moderate Reynolds numbers. %
\pbt{%
  Such evidences has questioned the universality of Townsend's similarity hypothesis, 
  thereby challenging researchers to define its limitations more clearly. 
  Even though this lies outside the purpose of the present contribution, 
  it is worthwhile to mention that %
}%
\mut{
  the influence of the roughness on the turbulence structure could be
  presumably amplified if the size  
  of the roughness elements is of the same order of magnitude as the
  open-channel height, %
}%
\pbt{%
  without necessarily degenerating into the flow around a sequence of obstacles. %
}%
\pbt{%
  For instance, \citet{amir2011} observed that, in a boundary layer over
  genuinely three-dimensional roughness, 
  inner and outer scales were distinguishable and separated 
  as long as the size of the roughness did not exceed $15\%$ of the boundary layer thickness,
  which is well above the value ($2.5\%$) indicated by \citet{jimenez2004}. %
}%

From these considerations it follows that an investigation of the
flow-roughness interaction at the scale of the roughness elements
is needed in order to push significantly further our understanding of the
dynamics of moderately shallow open-channel flow. 
Indeed, an exhaustive description of the flow-roughness interactions, in particular between 
the transitionally-rough and the fully-rough regimes, is still missing and should in principle 
require the systematic exploration of different \mut{wall} configurations 
\mut{%
  and relative submergence,} 
as well as the possibility to make accurate measurements of velocity and pressure fields 
in the vicinity of the roughness elements.

An effect of roughness, in the fully-rough regime, is to introduce and sustain turbulent 
fluctuations characterised by length and time scales of the same order of magnitude as 
$k$ and $k/\mut{U_{bh}}$, respectively.
In fact, in a series of experiments of open-channel flow in the fully-rough regime in 
which a ``random'' distribution of %
closely-packed spheres resting %
on the wall was used, 
\citet{grass1991} found that the fluctuations of the streamwise velocity component in the 
vicinity of sphere crests were spanwise-correlated over a specific wavelength, similarly 
to what happens over a smooth wall. 
\mut{
  However, in that case, the wavelength was proportional to the size of the roughness elements.
}
In the experiment of \citet{grass1991} the arrangement of the spheres on the wall was not regular, 
thus the measurements of velocity fluctuations in the vicinity of sphere crests could be affected 
also by the local geometrical configuration of the rough wall. 
A more general result could be obtained by systematically computing the fluctuations about the 
time-averaged flow field 
\mut{
  at many points distributed in the vicinity of the roughness elements. 
}
This task could be more easily addressed with a numerical approach, 
as \citet{grass1991} themselves suggested.

\mut{
  \citet{amir2014} 
  investigated the hydrodynamic force exerted on spheres mounted in hexagonal
  pattern on the wall, by instrumenting several spheres with four-point pressure probes. 
}
They computed the statistics of pressure fluctuations for several cases differing in the bed 
slope and flow depth and were able to estimate the integral time scale of drag and lift 
fluctuations on the basis of measurements made from spheres equispaced in a row.
However, by adopting the four-point pressure technique, \citet{amir2014} did not consider the  
distribution of the stress on the surface of individual spheres, and they could not account for 
the distribution 
\mut{
  of the shear stress and how it is affected by Reynolds number effects.
}
Also this task can be more easily achieved numerically.

In the present work the study of \citet{chan2011,chan2013} has been extended to higher 
Reynolds numbers such that the fully-rough regime is attained.
We have kept the geometrical particle arrangement identical as in the earlier study, 
i.e. a square pattern with a \mut{relative submergence} of $H/k\simeq 5.5$.
By increasing the bulk Reynolds number, the diameter of the spherical roughness elements 
in the present simulation well exceeds the value of $100$ wall units.

The organization of the paper is as follows. 
After describing the chosen setup and the numerical approach, an analysis of the flow 
structure below and over the crest of the spheres is provided in section \ref{sec1}.
A comparison between length- and time-scales of the turbulent flow over the roughness elements 
and that obtained over a flat smooth wall for the same value of the bulk Reynolds number was 
also made.
Moreover, the distribution of the stress on the surface of the roughness elements is analysed 
in section \ref{sec2}.

Finally, although the Reynolds numbers investigated by \citet{amir2014} are larger than those 
reproducible nowadays by DNS, a comparison between their experimental results and present 
numerical results in the fully-rough regime was possible and is shown in section \ref{sec3}.
The paper closes with the description of a conceptual model of the interaction between 
turbulent structures and the roughness elements, and, in section \ref{sec4}, with the 
conclusions of the present results.
%
\section{Flow configuration and numerical approach}\label{sec0}
%
\begin{figure}
\ifone
\setlength{\unitlength}{0.353mm}
\begin{picture}(0,100)(0,0)
\put(2,2){\includegraphics[trim=0cm 0cm 0cm 0cm, clip, width=1.\textwidth]{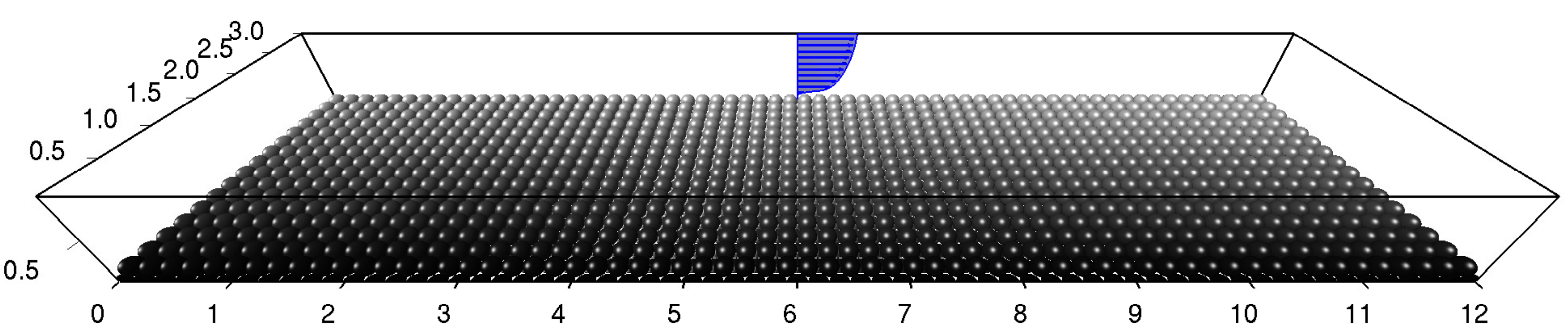}}
\put(199,77){$\left\langle\overline{u}\right\rangle$}
\put(376,5){$x/H$}
\put(-12,33){$y/H$}
\put(67,82){$z/H$}
\put(364,13.5){\vector(1,0){20}}
\put(76.5,75){\vector(3,2){14}}
\put(11,34){\vector(-3,4){12}}
\end{picture}
\fi
\caption{Sketch of the computational domain.}
\label{sketch}
\end{figure}
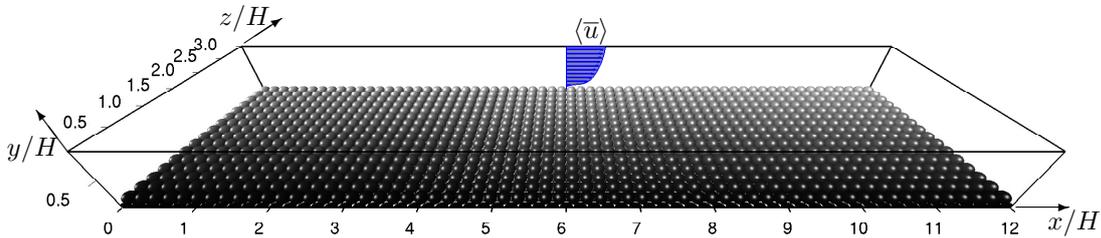
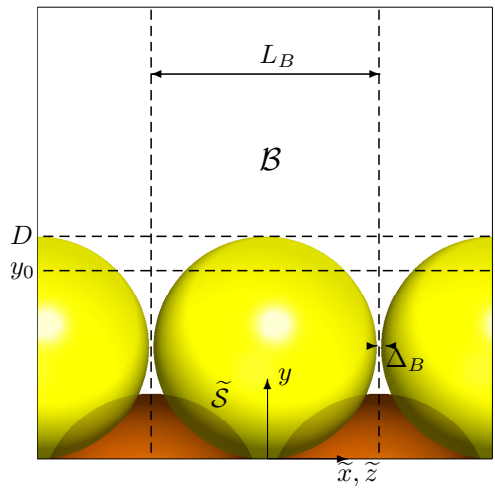
\begin{figure}
\setlength{\unitlength}{0.353mm}
\begin{picture}(0,175)(0,0)
\put(55,0){
\put(0,5){
\put(50,0){\includegraphics[trim=0cm 0cm 0cm 0cm, clip, width=.45\textwidth]{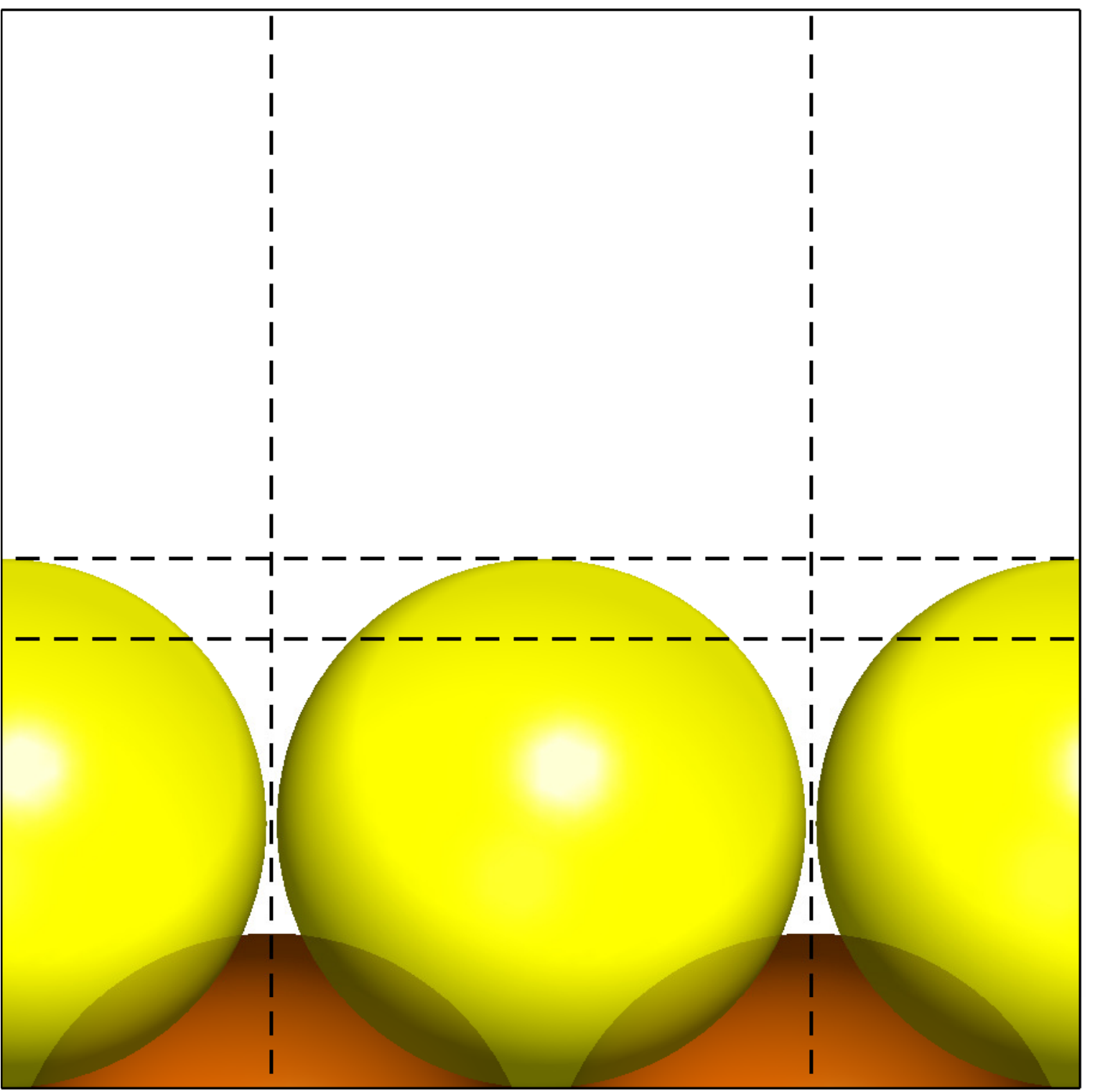}}
\put(136,0){
\put(4,30){$y$}
\put(26,-8){$\widetilde{x},\widetilde{z}$}
\put(0,.5){\vector(1,0){30}}
\put(0,.5){\vector(0,1){30}}
}
}
\put(40,75){$y_0$}
\put(40,86){$D$}
\put(92.5,150){\vector(1,0){85.5}}
\put(178,150){\vector(-1,0){85.5}}
\put(133,154){$L_B$}
\put(133,114){\large$\mathcal{B}$}
\put(180,40){
\put(1,8){\vector(-1,0){1}}
\put(-4,8){\vector(1,0){1}}
\put(0,0){$\Delta_B$}
}
\put(115,25){$\widetilde{\mathcal{S}}$}
}
\end{picture}
\caption{Side view of a detail of the bottom roughness.}
\label{sketch2}
\end{figure}

Direct numerical simulations of the incompressible Navier-Stokes equation were performed over a rectangular domain of dimensions $ L_x $, $ L_y $ and $ L_z $ in the streamwise, wall-normal and spanwise directions, respectively.
Let us indicate hereafter the simulations of the transitionally- and fully-rough open-channel 
flow with D50 and D120, respectively. 
The flow configuration is sketched in figure~\ref{sketch}, details of the roughness geometry 
are shown in figure~\ref{sketch2}.
The arrangement of roughness elements is the same as that adopted by \citet{chan2011} 
for their simulations, \mut{hence} it consists in one layer of $1024$ monosized 
rigid spheres crystallised on the wall at the vertices of a square grid of side 
$L_B=D+\Delta_B $ aligned with the $x$- and $z$-directions (yellow spheres 
in figure~\ref{sketch2}), and a second layer of spherical caps %
\pbt{%
  (highlighted in red in figure~\ref{sketch2}) %
}%
with the same 
arrangement as the first one shifted by $ (L_B/2,y_2,L_B/2) $, where 
$y_2=D/2-\sqrt{2}(D/2+\Delta_x)$ 
\mut{
  and $\Delta_x$ denotes the computational grid spacing.
} 
Figure~\ref{sketch2} also shows that, by exploiting the symmetry
properties of the roughness arrangement, a cuboidal sub-domain 
$\mathcal{B}=[-L_B/2,L_B/2\mut{[}\times[0,H]\times[-L_B/2,L_B/2\mut{[}$ can be
defined \mut{in the local coordinates system $(\widetilde{x},y,\widetilde{z})$ with 
the same orientation as $(x,y,z)$ and} origin in the projection on the wall of the 
top-layer sphere center\mut{. 
The sub-domain $\mathcal{B}$} is geometrically periodic 
in the streamwise and spanwise directions and invariant to the exchange between 
$\widetilde{x}$- and $\widetilde{z}$-axis, i.e. a square arrangement.
There are no gaps between the spheres and the wall, while the \mut{minimal} 
distance $ \Delta_B $ is required between the spheres 
\pbt{%
  by the immersed boundary method proposed by \citet{Uhlmann2005}. %
}%
Since roughness elements are spherical, hereinafter the roughness size and the 
relative Reynolds number will be referred to as $D$ and $D^+$, instead of $k$ and $k^+$, 
respectively.
The numerical approach used by \citet{chan2011} was also used for the present simulations. It 
consists in a second-order accurate fractional-step method.
In particular, a semi-implicit scheme is employed for the viscous terms along with a three-step (low storage) Runge-Kutta method for the nonlinear terms.
Standard centred second-order finite-difference approximations of the spatial derivatives 
are used over a staggered uniform Cartesian grid of spacing $\Delta_x=\Delta_y=\Delta_z$.
The spherical roughness elements are represented by means of an
immersed boundary technique, and further details on the 
numerical method can be found in \citet{Uhlmann2005} and \citet{uhlmann:06b}.
This numerical procedure underwent an extensive validation and has been recently used to 
perform the direct numerical simulation of flows in which particles were either fixed or free to move 
\citep{uhlmann:08a,chan2011,garcia2012,uhlmann:14a,Aman2014a,Aman2014b,chouippe:15a,mazzuoli2016,uhlmann:16a}. %
\pbt{%
  The spacing $\Delta_B$ was set equal to $2\Delta_x$. We do not expect 
  the results to be significantly affected by the value of $\Delta_B$, 
  as along as it remains a small fraction of the particle diameter $D$ 
  and, consequently, it does not entail any significant change in the 
  width (and spanwise separation) of the inter-particle grooves. %
}%

The size of the computational domain, the small-scale resolution and further details are 
shown in table \ref{tab1}.
The present parameter point corresponds to $D^+=119$ and $\mut{Re_{bH}}=6865$, and 
it will henceforth be denoted as case D120. The table also shows the parameters of 
case D50 of \cite{chan2011} which will be used frequently in the discussion below. 
The continuity and momentum equations were solved numerically, obtaining the velocity 
components and pressure, denoted by $(u, v, w)$ and $p$ respectively, throughout the whole 
computational domain including the space occupied by the spheres\mut{.} 
\mut{P}eriodicity conditions were applied at the boundaries of the domain in the streamwise 
and spanwise directions, and the no-slip condition was forced at the fluid-spheres interface 
by means of the immersed boundary method proposed by \citet{Uhlmann2005}, i.e. through force 
terms directly added to the momentum equations (direct forcing method). %
\pbt{%
  In the immersed boundary method, the velocity field at the end
  of a full time step slightly deviates from the desired value (zero)
  at the surface of the spheres due to the effect of the projection step. 
  The magnitude of this error amounts to $6.6\cdot10^{-3}\,u_\tau$ on average, 
  with the local instantaneous maximum measuring $5.8\cdot10^{-1}\,u_\tau$.
  We believe that this level of error is not significant with respect to
  the statistical analysis performed in the present manuscript. %
}%
Finally, no-slip and free-slip boundary conditions were imposed to the flow field at the wall 
$(y=0)$ and at the open-surface $(y=H)$ of the computational domain, respectively.

The mass flow is maintained steady throughout the simulations by a uniform pressure gradient 
which drives the flow and is updated at each time step. 
Thus, while the bulk velocity defined as $U_{bH}=1/H\int^H_{0}\left\langle\overline{u}\right\rangle\,dy$ 
and consequently the bulk Reynolds number $\mut{Re_{bH}}=U_{bH} H/\nu$ were constant, 
the Reynolds numbers $Re_\tau$ and $D^+$ fluctuated about the average value indicated 
in table \ref{tab1}. 
\mut{
  The definition of the Reynolds number $R_{bh}=U_{bh}h/\nu$ based on the effective 
  flow depth $h=H-y_0$ and on the bulk velocity 
  $U_{bh}=1/h\int_0^h\left\langle\overline{u}\right\rangle\,dy$ is recalled here for 
  the sake of clarity. 
  The ratio between $U_{bh}$ and $U_{bH}$ was found equal to $1.16$ and $1.17$ for 
  the simulations D50 and D120, respectively. 
}
\setlength{\tabcolsep}{0.35em} 
{\renewcommand{\arraystretch}{1.5}%
\begin{table}
  \begin{center}
    \begin{tabular}{l c c c c c c c c c r r r r c c}
    \hline
    $\mathrm{run}$ &   
    $ \dfrac{U_{bh}}{u_{\tau}}$ &
    $\dfrac{H}{D}$ &
    $ \mut{Re_{bH}} $ &
    $ Re_\tau $ &
    $ D^+ $ &
    $ \Delta_x^+ $ &
    $ \dfrac{L_x}{H} $ &
    $ \dfrac{L_y}{H} $ &
    $ \dfrac{L_z}{H} $ &
    $ \dfrac{D}{\Delta_x} $ &
    $ N_x $ &
    $ N_y $ &
    $ N_z $ &
    $ \dfrac{T_{obs}}{T_{bH}}$ &
    source \vspace*{.05cm}
    \\
    \hline
    D50  & $12.2$ & $5.6$ & $2872$ & $234$ & $49$  & $1.1$ & $12$ &
    $1$ & $3$ & $56$  & $3072$ & $256$ & $768$  & 120 & 
    CB2011
    \\
    D120 & $12.4$ & $5.4$ & $6865$ & $544$ & $119$ & $1.1$ & $12$ &
    $1$ & $3$ & $106$ & $6912$ & $576$ & $1728$ & 
    60 & present 
    \\
    \hline
    \end{tabular}
  \end{center}
  \caption{Parameters of the simulations. $N_x$, $N_y$ and $N_z$
    denote the number of grid points in the stramwise, wall-normal and
    spanwise directions, respectively. $T_{obs}$ denotes the
    simulation time 
    (excluding the initial transient), 
    normalised by the bulk time-unit
    $T_{bH}=H/U_{bH}$. 
    The data of case D50 is from \citet{chan2011}. 
  }
  \label{tab1}
\end{table}
}
The grid spacing was sufficiently small to resolve the vortices associated with the 
dissipative turbulent scales ($\Delta^+_x=1.1$) while the size of the computational box 
was found large enough to include the large vortex structures of the flow. 
In fact, the pre-multiplied spectra of turbulent velocity fluctuations (figure omitted) 
show that only weak energy is associated with structures larger than the domain presently 
considered which, anyway, should not influence the spectra in the vicinity of the bottom
\citep{alamo2004,chan2012}. 

Moreover, additional DNS of open-channel flow over a smooth wall have
been performed for the same box-size and bulk Reynolds number as case
D120 using a pseudo-spectral method \citep{KMM1987}. These smooth-wall
data, as well as those reported for $\mut{Re_{bH}}=2870$ by
\citet{chan2011}, will be used for the purpose of comparison below.
\section{Results and discussion}
%
The main concern of the present investigation is how an increase of the bulk Reynolds number 
changes the flow structure such that the flow regime becomes fully-rough.
Prior to \mut{defining the roughness sublayer for the present simulation and} exploring the 
numerical results, some considerations are formulated on the basis of the geometrical 
configuration of the solid boundary, in order to enable appropriate statistical tools to 
investigate the fluid-roughness interaction.

Turbulence in the vicinity of a spherical roughness element\mut{, namely in the roughness
sublayer,} is clearly neither homogeneous nor isotropic and in principle turbulence statistics should
be described as functions of three space dimensions. %
For the present square particle arrangement in a doubly-periodic
computational domain statistics are invariant with respect to spanwise
or streamwise shifts by integer multiples of the distance between two
spheres. 
Furthermore, statistics are symmetric with respect to the
$(x,y)$-plane through the center of any sphere. 
These features are exploited by the averaging operators used here,
which are formally defined in appendix~\ref{apxA}. 

What will be hereafter referred to as \textit{average flow field} and indicated with
$\left\langle\overline{\boldsymbol{u}}\right\rangle_B$, can be defined
in the \textit{sphere-boxes}, namely the cuboidal periodic subdomain $\mathcal{B}$ introduced 
in section \ref{sec0} (figure~\ref{sketch2}), and estimated by combining the time-average 
operator and the sphere-box-average operator.
Note that $\left\langle\overline{\boldsymbol{u}}\right\rangle_B$ is a three-dimensional quantity.  
Indeed, $\mathcal{O}(100)$ flow fields were obtained systematically collecting one snapshot 
every $0.6$ bulk time-units ($T_b=H/U_{bH}$). 
Thus, the average flow field $\left\langle\overline{\boldsymbol{u}}\right\rangle_B$ was obtained 
in post-processing phase over $\mathcal{O}(10^5)$ samples.
Additionally, the plane-averaged velocity and pressure fields, 
\mut{
  as well as the variance of their fluctuations,
}
were computed and collected during run time with a frequency $100$ times larger than the 
sampling of the snapshots.
Then, plane-averaged samples were also averaged over time in post-processing phase.
The simulations were preliminarily run until turbulence was fully developed before starting 
the sampling procedure.
Turbulent fluctuations around the time-averaged
($\overline{\boldsymbol{u}}$), 
sphere-box/time-averaged 
($\left\langle\overline{\boldsymbol{u}}\right\rangle_B$) 
and plane/time-averaged 
($\left\langle\overline{\boldsymbol{u}}\right\rangle$) 
velocity fields can be defined as follows
\begin{eqnarray}
\boldsymbol{u}'   &=&\boldsymbol{u}-\overline{\boldsymbol{u}}\label{eq3a} \\
\boldsymbol{u}''  &=&\boldsymbol{u}-\left\langle\overline{\boldsymbol{u}}\right\rangle_B\label{eq3b}\\
\boldsymbol{u}''' &=&\boldsymbol{u}-\left\langle\overline{\boldsymbol{u}}\right\rangle\label{eq3c}\:\: ,
\end{eqnarray}
respectively, each helping to interpret physical phenomena at different scales.
In a similar way, the fluctuations of pressure and other scalar or vectorial quantities can 
be also defined.
The definition (\ref{eq3a}) of \textit{time-fluctuations} can be useful to investigate 
the evolution of turbulence structures of size much larger than $D$ (i.e. comparable with 
that of the computational domain).
The definition (\ref{eq3b}) seems more appropriate to study the interaction of individual 
roughness elements with the flow, selecting the turbulence scales larger than the size of 
$\mathcal{B}$.
In the general case of an arbitrary arrangement of monosized spheres, fluctuations (\ref{eq3b}) would not contain the information associated with the geometrical properties of individual roughness elements (\textit{individual form contribution}) because the spheres are identical, but that related to their arrangement (\textit{collective form contribution}).
Since the present arrangement is regular, the collective form contribution is also not present in fluctuations (\ref{eq3b}), while it definitely affects the average flow field  \mut{$\left\langle\overline{\boldsymbol{u}}\right\rangle_B$}.
\mut{
  In fact, the quantity 
  $\left\langle\overline{\boldsymbol{u}}\right\rangle_B
  -
  \left\langle\overline{\boldsymbol{u}}\right\rangle$, 
  which is equal to 
  $\boldsymbol{u}'' 
  - 
  \boldsymbol{u}'''$, 
  is also called \textit{spatial disturbance} (or \textit{form disturbance}) 
  and depends on the size of the sub-domain 
  $\mathcal{B}$ \citep{nikora2001}.
  It can be verified that the stress related to the spatial disturbance of the generic velocity components $\phi$ and $\psi$ is equal to the \textit{dispersive stress} (or \textit{form-induced stress}) reduced by the product of the respective plane/time-averaged velocity, which reads 
  \begin{equation}
  \left\langle\overline{
  \left(\phi'' 
  - 
  \phi'''\right)
  \left(\psi'' 
  - 
  \psi'''\right)
  }\right\rangle 
  =
  \left\langle\overline{
  \left\langle\overline{\phi}\right\rangle_B
  \left\langle\overline{\psi}\right\rangle_B
  }\right\rangle 
  - 
  \left\langle\overline{\phi}\right\rangle
  \left\langle\overline{\psi}\right\rangle \:\:,
  \label{eq11}
  \end{equation}
  where the dispersive stress is the first term on the right side of \eqref{eq11}.
  The same definition of the spatial disturbance can be extended also to the 
  fluctuations of pressure or any scalar quantity. 
}
\mut{
  Hence, 
}
turbulent fluctuations defined by (\ref{eq3c}) contain both the time-fluctuations and those associated with the flow pattern around the roughness elements.
%

  At this point let us also define the stress tensor for future
  reference, viz.
  \begin{equation}\label{equ-define-total-stress-tensor}
    \boldsymbol{\tau}
    =
    \boldsymbol{\tau}_\nu
    -
    p_{tot}\mathbf{I}
    \,,
  \end{equation}
  where
  $\boldsymbol{\tau}_\nu=\varrho\nu\left(\nabla\mathbf{u}+\left(\nabla\mathbf{u}\right)^T\right)$
  is the viscous stress (with $\varrho$ the fluid density), and
  $\mathbf{I}$ is the identity tensor.
  The pressure field in fully-developed plane-channel flow can be
  written as follows 
  \begin{equation}\label{equ-define-total-pressure}
    p_{tot}(\mathbf{x},t)
    =
    p_l(x,t)
    +
    p(\mathbf{x},t)
    \,,
  \end{equation}
  where $p_l$ corresponds to the linear variation in the streamwise direction due to the imposed 
  driving pressure gradient (with $\langle p_l\rangle_x=0$) and $p$ is the instantaneous fluctuation 
  whose box-average is zero. 

\subsection{The velocity and vorticity fields}\label{sec1}
\begin{figure}
\iftwo
\setlength{\unitlength}{0.353mm}
\begin{picture}(0,195)(0,0)
\put(55,0){\includegraphics[width=.62\textwidth]{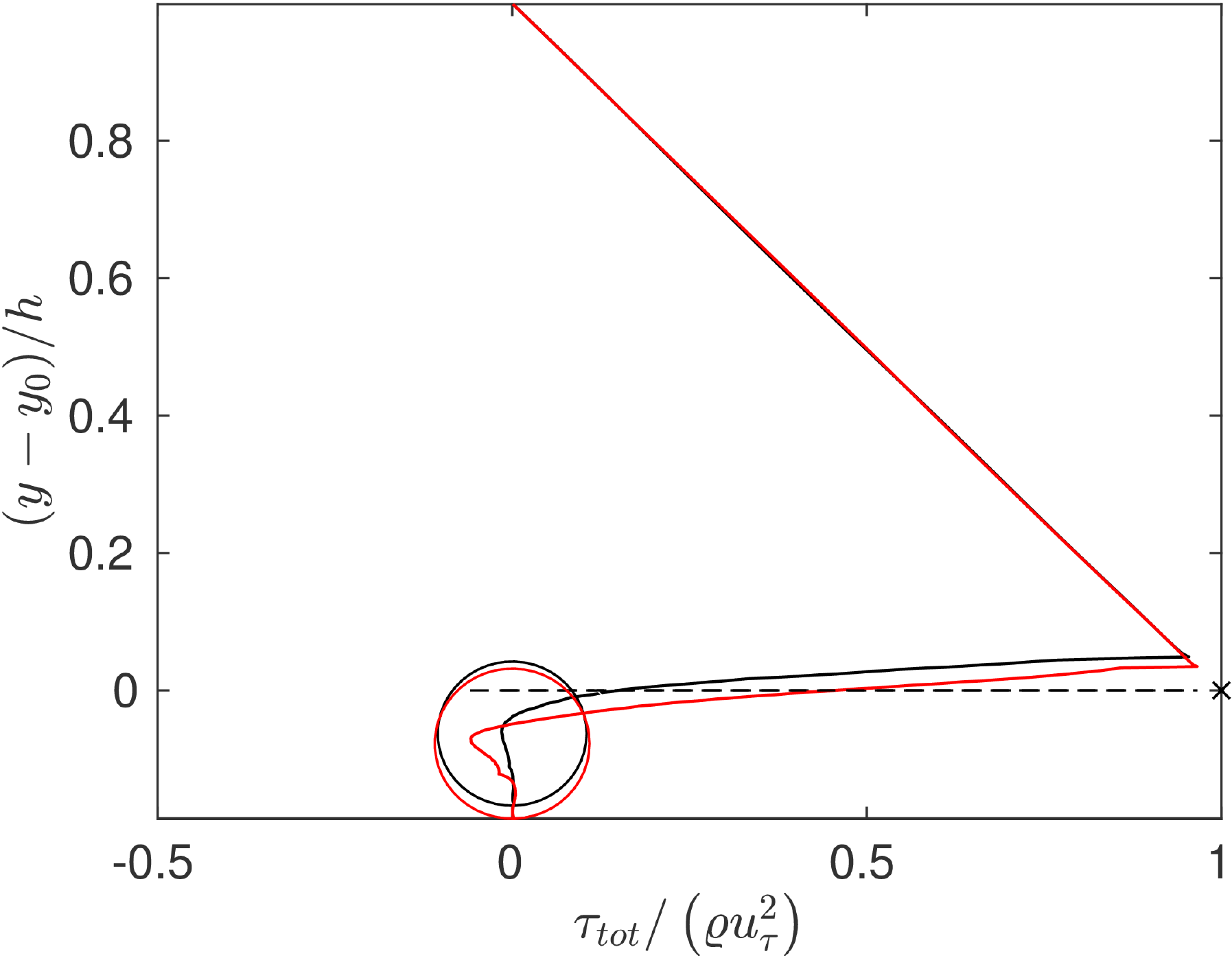}}
\end{picture}
\fi
\caption{Wall-normal shear stress as function of the distance from the
  virtual wall (broken line). Black line: run D50; red line: D120. The
  symbol $\times$ indicates the value of the bottom shear stress
  extrapolated  
  down to a wall-normal distance $y=y_0$ using the slope of the
  linear shear-stress profile far from the wall. 
}
\label{fig2}
\end{figure}
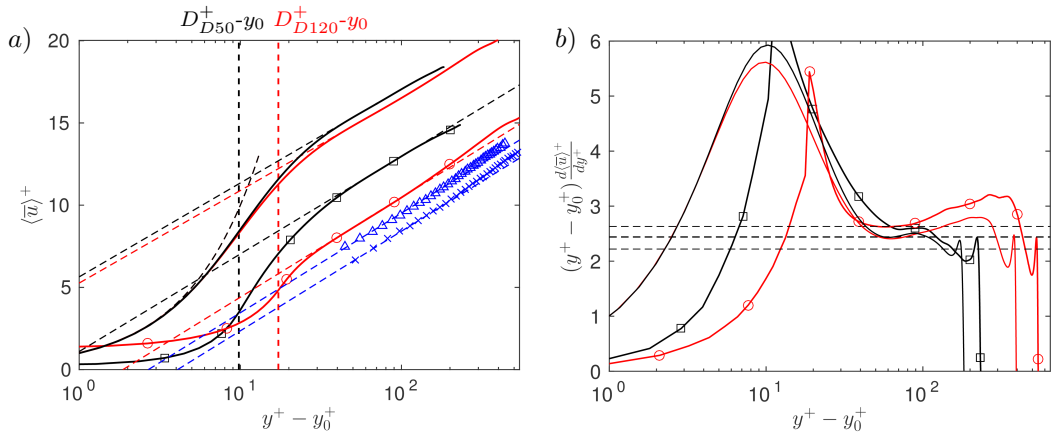
\begin{figure}
\ifthree
\setlength{\unitlength}{0.353mm}
\begin{picture}(0,160)(0,0)
\put(-2,0){\includegraphics[trim=0cm 0cm 0cm 0cm, clip, width=.49\textwidth]{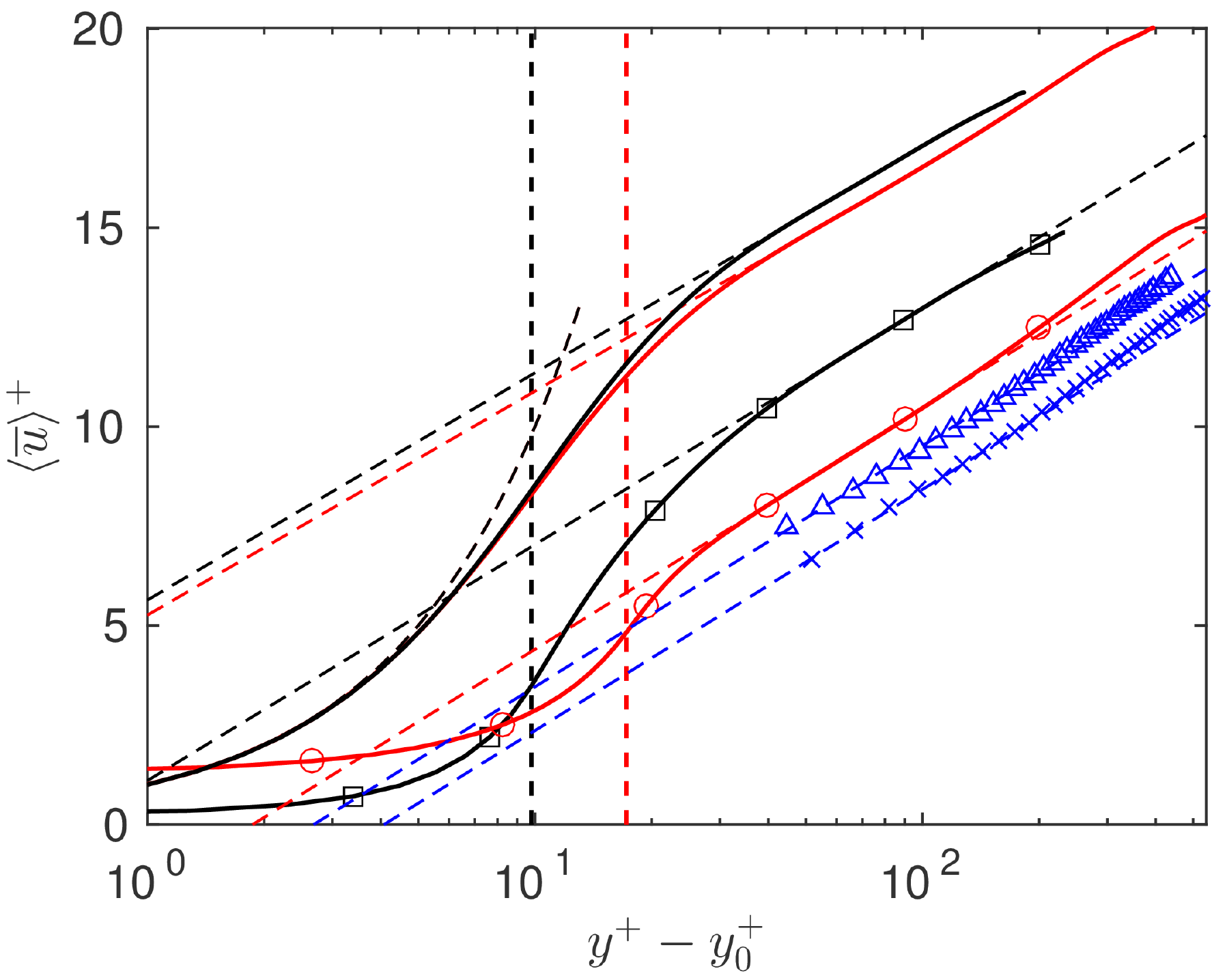}}
\put(197,0){\includegraphics[trim=0cm 0cm 0cm 0cm, clip, width=.485\textwidth]{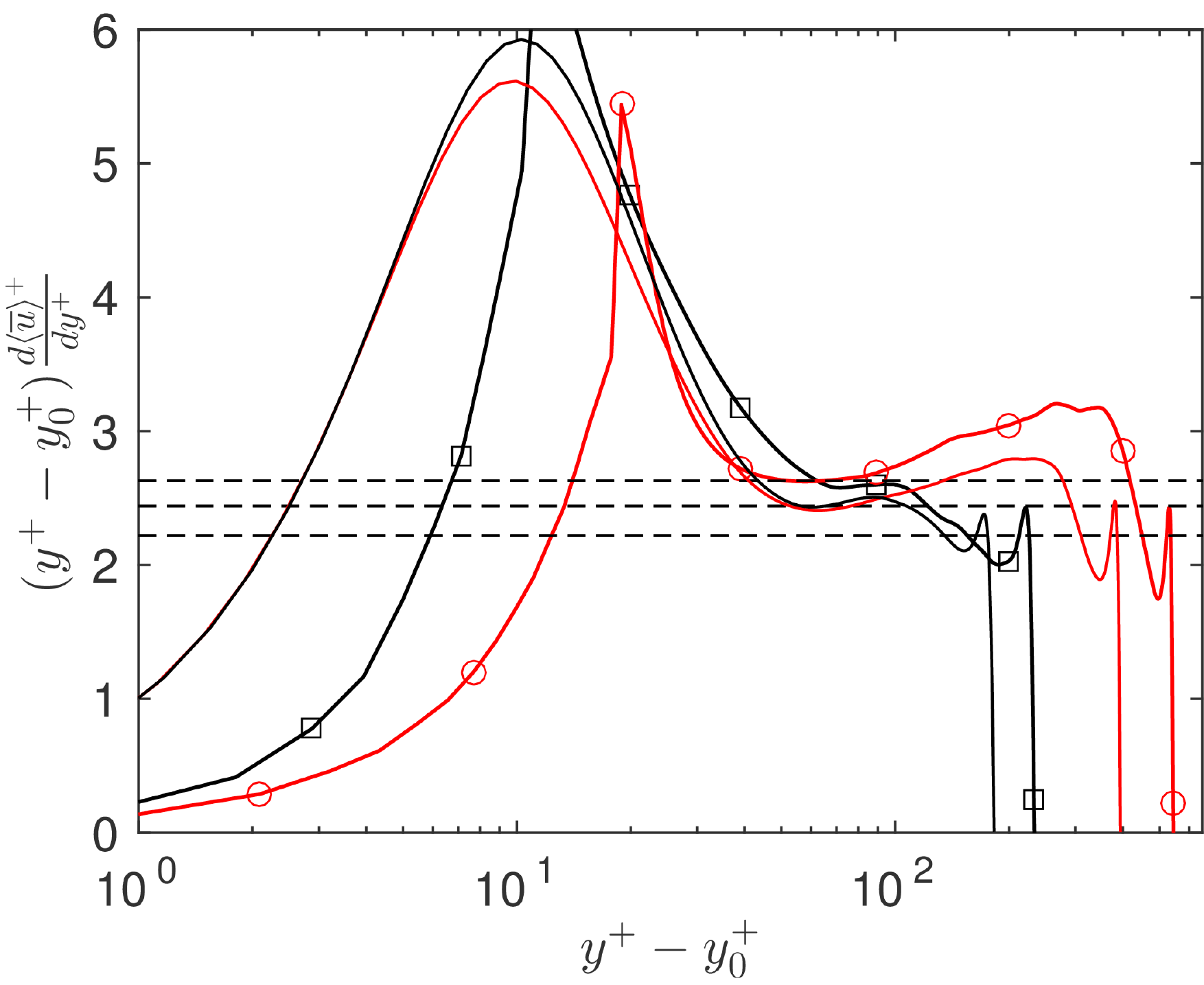}}
\put(-6,145){$a)$}
\put(198,145){$b)$}
\put(93,153){\footnotesize \color{red!}$D^+_{D120}$-$y_0$}
\put(58,153){\footnotesize $D^+_{D50}$-$y_0$}
\end{picture}
\fi
\caption{Profiles of $a)$ $\left\langle \overline{u}\right\rangle^+$
  and of $b)$ $(y^+-y^+_0)\frac{d\left\langle
      \overline{u}\right\rangle^+}{dy^+}$ as function of the distance
  from the virtual wall. Lines ---$\square$--- and
  \mut{---$\bigcirc$---} indicate the case D50 and D120, respectively,
  while solid lines indicate the simulations performed over a smooth
  wall at $Re_{b}=2900$ and $Re_{b}=6864$, respectively. %
  \mut{%
    The broken lines in panel $b)$ indicates the value of the logarithmic
    constant $1/\kappa=2.44$ (central line) $\pm 0.12$ (upper and lower lines). %
  }%
  \pbt{%
    Symbols {\color{blue!}$\triangle$} and {\color{blue!}$\times$} indicate the values
    measured by \citet{amir2014} in their experiments no.~$1$ and no.~$4$, respectively. %
  }%
} 
\label{fig1}
\end{figure}

Since in the transitionally-rough regime viscous effects are still relevant over the roughness, 
and are presumably relevant also in the fully-rough regime at least along the crevices between 
the roughness elements, viscous scales will be used as reference scales.
Hence, the friction velocity $u_\tau=\sqrt{\mut{\tau_w}/\varrho}$ was estimated, where the 
value of the \mut{wall} shear stress $\mut{\tau_w}$ was extrapolated down to the wall-normal 
distance $y=y_0$ using the linear profile of the wall-normal \mut{total} shear stress
\mut{
  $\tau_{tot}
  =
  \rho\nu
  \frac{\langle\overline{u}\rangle}{dy}
  -
  \rho
  \langle\overline{u^{\prime\prime\prime} v^{\prime\prime\prime}}\rangle$
} 
far from the rough wall, as shown in figure~\ref{fig2} 
\mut{
  \citep{chan2011}
}. 
Figure~\ref{fig1}a shows the velocity profiles which were computed for the runs D50 and D120 and for the respective simulations performed at the same bulk Reynolds numbers in absence of the roughness elements.
The value of $y_0$ for the simulation D50 was set equal to $0.8 D$ according to the indications of \citet{chan2011}.
Indeed, it was found that the profiles of 
$(y^+
-
y^+_0)
\frac{d\left\langle\overline{u}\right\rangle^+}
{d y^+}$ 
(which is equal to an inverse von K\'arm\'an ``constant'') for the run D50 approached, 
in the \mut{logarithmic} region, the curve obtained from the respective 
simulation over a smooth wall at $\mut{Re_{bH}}\approx2900$ (see figure~\ref{fig1}b).
A similar agreement between the rough and smooth wall profiles 
was obtained for the run D120 by positioning the virtual wall $y_0$ at the distance $0.85 D$. 
\mut{
   This choice allows us to compare velocity profiles obtained for the smooth- and 
   rough-wall cases in the logarithmic region and to estimate the roughness function. 
   It is worthwhile to mention that other methodologies adopted to estimate the position 
   of $y_0$, like the location of the drag force centroid \citep{jackson1981}, 
   provided approximately to the same value \citep[see also Appendix A1 of][]{chan2011}. 
}
Since, in the \mut{logarithmic} region, the curves of figure~\ref{fig1}b 
should be independent of $y^+$, it was also possible to estimate the value for the 
von K\'arm\'an constant, $\kappa$, as the inverse of the value attained at the local 
minimum of the curves, as suggested (among others) by \citet{balachandar1999}, 
\citet{tachie2000} and \citet{george2007}. %
\pbt{%
  \citet{hoyas2006} showed that the fact not to observe a wide region 
  where $1/\kappa$ is constant is an effect of the higher-order terms 
  usually included in the wake-component of the profile.
  Nonetheless, these authors observed the existence of a substantial logarithmic region. %
}%
Therefore, the von K\'arm\'an constant was estimated equal to $0.388$ and $0.381$ for the simulations D50 and D120, respectively.
Slight deviations of the value of $\kappa$ from the value $0.41$, estimated by \citet{coles1968} 
for boundary layers, were often reported in the literature \citep[e.g.][]{osterlund2000,marusic2013} 
and, in particular for the present channel-flow configuration, can be related to pressure 
gradient effects \citep{george2007}. 
\mut{
  However, it should be noted that the present values of $\kappa$ are in the range estimated by
  \citet{marusic2013} ($\kappa=0.39\pm 0.02$), while a local minimum of 
  $(y^+-y^+_0)\frac{d\left\langle\overline{u}\right\rangle^+}{d y^+}$ is attained approximately at 
  $y^+-y^+_0 = 50$ consistently with other numerical \citep{hoyas2006} and experimental
  \citep[e.g.][]{mckeon2004,tachie2000} results.
}
Some considerations can be formulated for the simulation D50 on the basis of the velocity profiles of figure~\ref{fig1}a.
\citet{chan2011} found that the flow regime for the simulation D50 was transitionally-rough.
This also appears from the velocity profile (black-square line in figure~\ref{fig1}a) which 
shows the presence of the buffer layer above the crest of the spheres.
The thickness of the viscous sub-layer $\delta^+_{sub}$ was estimated from the intersection of the 
logarithmic law and the linear law of the wall (cf.\ figure~\ref{fig1}a), which yields
$\delta^+_{sub}=11.5$ in the smooth-wall case at $\mut{Re_{bH}}=2900$. 
In case D50 we formally apply the same procedure, although a linear law is not observed; 
this yields $\delta^+_{sub}=5.0$ in this case. 
Then, the constant $C^+_I(D^+\rightarrow 0)$ shown in equation (\ref{eq1}) can be determined by 
extrapolating the logarithmic profile \mut{down} to $y^+-y^+_0=1$ and it was equal to $5.5$ for the 
smooth-wall case at $\mut{Re_{bH}}\approx2900$. 
\mut{
  According to \citet{ligrani1986}, in the transitionally-rough regime, the
  roughness function, $\Delta\left\langle\overline{u}\right\rangle^+$, 
  can be estimated as follows: 
}
\begin{equation}
    \Delta \left\langle\overline{u}\right\rangle^+
    = 
    C^+_I(D^+\rightarrow 0) 
    - 
    \delta^+_{sub} 
    + 
    \dfrac{1}{\kappa}
    \ln \delta_{sub}^+
\label{eq2}
\end{equation}
which was equal to $4.4$, where $C^+_I(D^+\rightarrow 0)$ indicates the value of $C^+_I$ 
for the smooth-wall case and the value of $\kappa$ is that associated with the smooth-wall 
simulation ($\kappa\approx0.41$).
%
%
The value of $C^+_{II}$ can be also estimated on the basis of the expression (\ref{eq1a}) 
evaluated at $y=D+y_0$: 
\begin{equation}
C^+_{II} = \left\langle\overline{u}\right\rangle^+\left(D+y_0\right)\:\: .
\label{eq3}
\end{equation}
\mut{
  If the dependence on the relative submergence is neglected for a moment, 
  then $C^+_{II}$ can be interpreted as equal to $B^+-\frac{1}{\kappa}\ln\frac{k_s}{D}$.
  The value of $B^+$ can be obtained for instance from the diagram of figure 1 of 
  \citet{ligrani1986}, leaving $k^+_s$ approximately equal to $30$ for the run D50. 
}
%
However, $C^+_{II}$ is affected \mut{not only} by effects associated with the arrangement of the 
roughness elements\mut{, but also by} those related to the \mut{relative submergence} $H/D$. 
%
%
\mut{
  Indeed, on the basis of the range indicated in figure~8 of
  \citet{chan2011} for the case D50,  
}
the actual value of $k^+_s$ could be \mut{presumably} larger than the value presently estimated. 
An expression similar to \eqref{eq2} can be written also for the fully-rough regime:
\begin{equation}
  \Delta \left\langle\overline{u}\right\rangle^+= C^+_I(D^+\rightarrow 0) - C^+_{II} + \dfrac{1}{\kappa}\ln (D^+)
\label{eq4}
\end{equation}
where the values of $\kappa$ and $y_0$ are those attained over a smooth-wall, 
i.e. $0.41$ and $0$, respectively.
Although the values of $\Delta \left\langle\overline{u}\right\rangle^+$ in expressions \eqref{eq2} and \eqref{eq4} are evaluated at different distances from $y_0$ (i.e. $\delta_{sub}$ and $D$, respectively), their 
dependence on the distance in the range $\vert D-\delta_{sub}\vert$ was 
small (i.e. $\sim 0.8$ for $D^+\sim 100$). 
Note that this dependence of $\Delta \left\langle\overline{u}\right\rangle^+$ on the distance from the wall is due to 
deviations of $\kappa$ which are small.
%
%
Similarly to boundary layers for which $k_s$ can be proportional to $k$ ($k$-roughness) or to the boundary layer thickness ($d$-roughness), for shallow open-channels $k_s$ possibly depends on both $H$ and $k$ (as well as on the roughness geometrical features) and, therefore, the present estimate of $C^+_{II}$ should be associated only with the particular configuration of the present simulations.
However, by noting that, in the fully rough regime, $\Delta \left\langle\overline{u}\right\rangle^+$ monotonically increases with $D^+$ for a given configuration of the roughness \citep[e.g. see][]{jimenez2004}, combining equations \eqref{eq2} and \eqref{eq4}, the following inequality can be found which poses an upper limit to the value of $C_{II}$:
\begin{equation}
C^+_{II} \leq \delta^+_{sub} - \dfrac{1}{\kappa}\ln \delta_{sub}^+ + \dfrac{1}{\kappa}\ln (D^+)\:\: .
\label{eq9}
\end{equation}
Since the sum of the first two terms on the right hand side of \eqref{eq9} is minimum for $\delta^+_{sub}=1/\kappa$ (which is an admissible value of $\delta^+_{sub}$ in the transitionally-rough regime) and is equal to $0.26$ and to $0.08$ for $\kappa$ equal to $0.41$ and to $0.38$, respectively, 
it is possible to approximate the inequality \eqref{eq9} with the following one:
\begin{equation}
C^+_{II} \lesssim \dfrac{1}{\kappa}\ln (D^+)
\label{eq10}
\end{equation}
where $\kappa=0.41$.
The inequality \eqref{eq10}, is 
\mut{
  a bound for the value of $C^+_{II}$ 
}
independent of the particular flow configuration.
\mut{
  This does not imply the independence for $C^+_{II}$, which instead is a function of both 
  the roughness geometrical features and the relative submergence. 
}
In other words, for a given value of the roughness Reynolds number $k^+$, the configuration (arrangement and shape of roughness elements, \mut{relative submergence},~$\ldots$) which minimises the flow resistance is the configuration for which $C^+_{II}$ tends to $\frac{1}{\kappa}\ln (k^+)$.
\setlength{\tabcolsep}{0.35em} 
{\renewcommand{\arraystretch}{1.5}%
\begin{table}
  \begin{center}
    \begin{tabular}{l c c c c c c c c}
    \hline
    $\mathrm{run}$ &   
    $\dfrac{H}{D}$ &
    $ \mut{Re_{bH}} $ &
    $ D^+ $ &
    $ \frac{1}{\kappa}\ln(D^+) $ &
    $ C_{II}^+ $ &
    $ \Delta \left\langle\overline{u}\right\rangle^+ $ &
    arrangement & 
    source \vspace*{.05cm}
    \\
    \hline
    D50   & $5.6$ & $2872$  &  $49$ &  $9.5$ & $11.0$ & $4.0$  & square    & CB2011 \\
    D120  & $5.4$ & $6865$  & $119$ & $11.6$ & $10.5$ & $6.3$  & square    & present \\
    no. 1 & $2.5$ & $4915$  & $170$ & $12.5$ & $11.1$ & $6.5$  & hexagonal & AM2014 \\
    no. 4 & $5.0$ & $15508$ & $243$ & $13.4$ & $11.1$ & $7.4$  & hexagonal & AM2014 \\
    no. 2 & $2.5$ & $10678$ & $376$ & $14.5$ & $10.9$ & $8.7$  & hexagonal & AM2014 \\
    no. 5 & $5.0$ & $32373$ & $526$ & $15.3$ & $10.6$ & $9.8$  & hexagonal & AM2014 \\
    no. 8 & $7.5$ & $63305$ & $658$ & $15.8$ & $10.2$ & $10.7$ & hexagonal & AM2014 \\
    no. 6 & $5.0$ & $45763$ & $724$ & $16.1$ & $10.4$ & $10.8$ & hexagonal & AM2014 \\
    \hline
    \end{tabular}
  \end{center}
  \caption{
  Values of the parameters, of the integration constant $C_{II}^+$ and of the shift of the velocity profile $\Delta \left\langle\overline{u}\right\rangle^+$ (computed at $y^+=y_0^+ + D^+$) for the present simulation D120, for the simulation D50 \citep{chan2011}, and for six of the experiments carried out by \citet{amir2014} (original numbering). 
  Note that \citet{amir2014} chose $y_0=D$ as opposed to the present choice $y_0=0.85 D$.
  }
  \label{tab2}
\end{table}
}
For the simulation D120, $\Delta \left\langle\overline{u}\right\rangle^+=6.3$
\mut{
  (provided that the $C^+_{I}(D^+\rightarrow 0)$ is equal to $5.2$ for the smooth-wall case at 
  $\mut{Re_{bH}}\approx6800$) 
}
and the left and right sides of the inequality \eqref{eq10} are equal to $10.5$ and $11.6$, respectively. The fact that $C^+_{II}$ approaches its upper limit indicates that the present flow configuration is fairly conductive,
\mut{
  i.e. that the quantity $\varrho \mut{U_{bh}}^2/\mut{\tau_w}$ is relatively large (see table \ref{tab1}).
}
%
Table \ref{tab2} shows the values of $C_{II}^+$ and $\frac{1}{\kappa}\ln(D^+)$ which were also computed for five experiments of \citet{amir2014}. These authors investigated open-channel flow over a layer of spheres at rest on a smooth wall in a hexagonal pattern and assumed that $y_0=D$.
Although the range of values of $D^+$ and their choice of $y_0$ are somewhat different, the experimental results of \citet{amir2014} suggest that the effect of increasing $D^+$ is to decrease the conductivity.
In particular, at their smallest value $D^+=170$, which is not too far off the value of our present case D120, the deviation of $C_{II}^+$ from its upper limit is similar to that computed for the present run D120. 
\mut{
  Indeed, table \ref{tab2} shows that the roughness function is also very similar. 
}
The same can be said of their case number $4$, where $D^+=243$, but where the \mut{relative submergence} $H/D=5$ roughly matches the present value. %
\pbt{%
  The values of the mean velocity measured in the experiments numbers $1$ and $4$ of 
  \citet{amir2014} are also shown in figure~\ref{fig1}a as a function of $y^+-y^+_0$, 
  where $y^+_0$ is assumed equal to $0.85~D^+$.
  A region where the mean velocity followed a logarithmic profile can be  
  observed also in the latter two experiments. %
}%
In case D50, the inequality \eqref{eq10} is no longer valid mainly because the viscous sublayer is significantly thick in the transitionally-rough regime. 
However, it can be verified that the equality \eqref{eq9} is approximately satisfied 
if also the dependency of $\Delta \left\langle\overline{u}\right\rangle^+$ on the distance 
from the wall is taken into account. 
This indicates that the present square arrangement of the spheres almost maximises the flow conductivity in the transitionally-rough regime.

\mut{%
Ignoring 
for a moment the dependency on $H/D$, as already proposed for the 
transitionally-rough regime above, it would be found that
$\ln \frac{k_s}{D}=(8.5-C_{II}^+)\kappa$ and the value of $k^+_s$ can be estimated to measure 
approximately $50$.
As mentioned above, this value is not a reliable indicator of the flow regime which can be 
significantly underestimated.
In fact, the experiment number~1 of \citet{amir2014} was characterized by a value of the roughness 
function relatively small compared with that of \citet{nikuradse1933}, although the flow regime 
was fully-rough.
}
%
%
Also using a $k$-criterion based on the value of the roughness parameter 
\mut{
   $k^+_0=D^+/30$ (also termed \emph{roughness length}) 
}
to classify the flow regime, as suggested by \citet{jayatilleke1966} and \citet{reynolds1974}, the flow of simulation D120 falls well into the fully-rough range \citep[see the diagram in figure 3.3a of][for a comparison]{pimenta1975}.
In fact, it is shown in the following that the flow structure related to the simulation D120 exhibits the typical characteristics of the fully-rough regime.
%
\begin{figure}
\setlength{\unitlength}{0.353mm}
\begin{picture}(0,180)(0,0)
\put(-1,-2){\includegraphics[trim=0cm 1.1cm 0cm 0cm, clip, width=.495\textwidth]{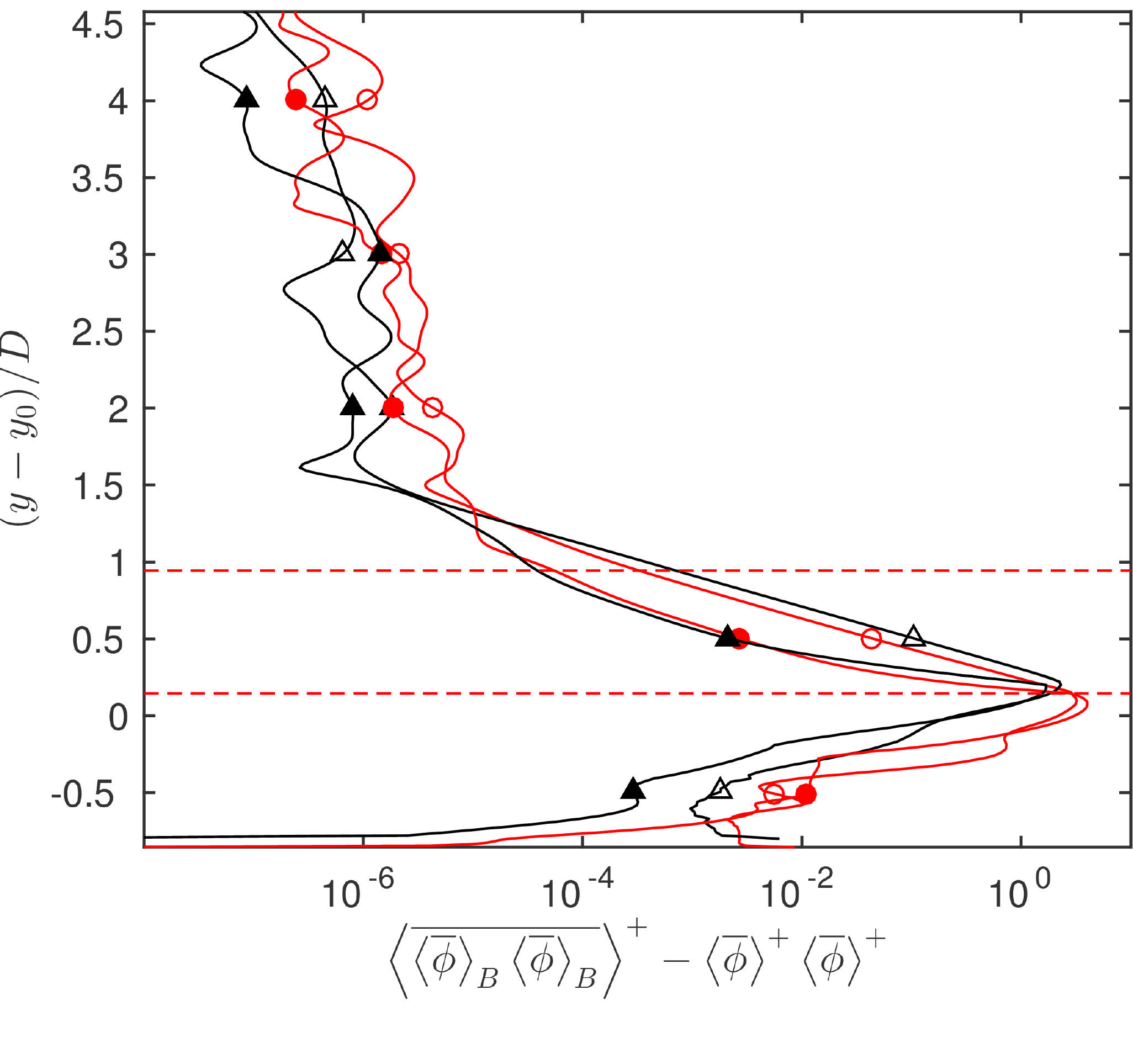}}
\put(197,0){\includegraphics[trim=0cm 0cm 0cm 0cm, clip, width=.49\textwidth]{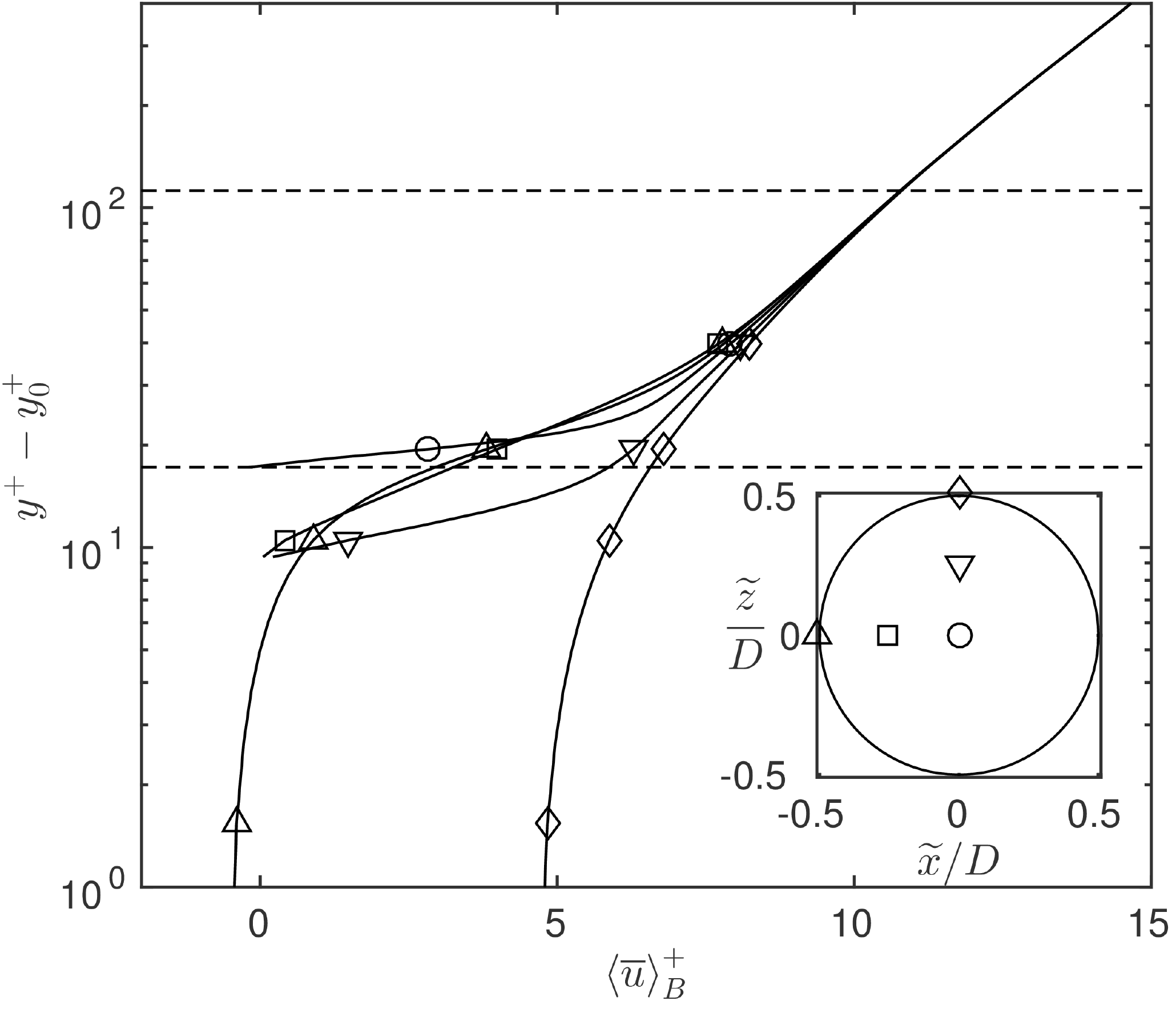}}
\put(-2,160){$a)$}
\put(198,160){$b)$}
\put(25,74){{\color{red!}\scriptsize\textit{roughness}}}
\put(25,64){{\color{red!}\scriptsize\textit{sublayer\ bound}}}
\put(25,52){{\color{red!}\scriptsize$y=D$}}
\put(223,137){\scriptsize\textit{roughness}}
\put(223,127){\scriptsize\textit{sublayer\ bound}}
\put(345,92.5){\scriptsize$D^+-y_0^+$}
\end{picture}
\caption{
  \mut{
    Panel $(a)$ shows the profiles of the variance of the quantity 
    $\left\langle \overline{\phi}\right\rangle_B$ reduced by the square product of 
    $\left\langle \overline{\phi}\right\rangle$, which equals the variance of 
    $\phi''-\phi'''$. Full symbols indicate $\phi\equiv u$ (normalized by 
    $u_\tau$) while empty symbols indicate $\phi\equiv p$ (normalized by 
    $\varrho u_\tau^2$). 
    $\vartriangle$, $\blacktriangle$: run D50; $\square$, $\blacksquare$: run D120. 
    Panel $(b)$ shows the profiles of the average velocity field computed for the run D120 
    at the following $(\widetilde{x},\widetilde{z})$ coordinates that are also indicated in 
    the topview of the small inset: $\bigcirc$: $(0,0)$, $\square$: $(-L_B/4,0)$, 
    $\vartriangle$: $(-L_B/2,0)$, $\triangledown$: $(0,L_B/4)$, $\diamond$: $(0,L_B/2)$.
  }
}
\label{fig3rev1}
\end{figure}
\begin{figure}
\setlength{\unitlength}{0.353mm}
\begin{picture}(0,150)(0,0)
\put(0,0){\includegraphics[trim=0cm 1.5cm 0cm 0cm, clip, width=.47\textwidth]{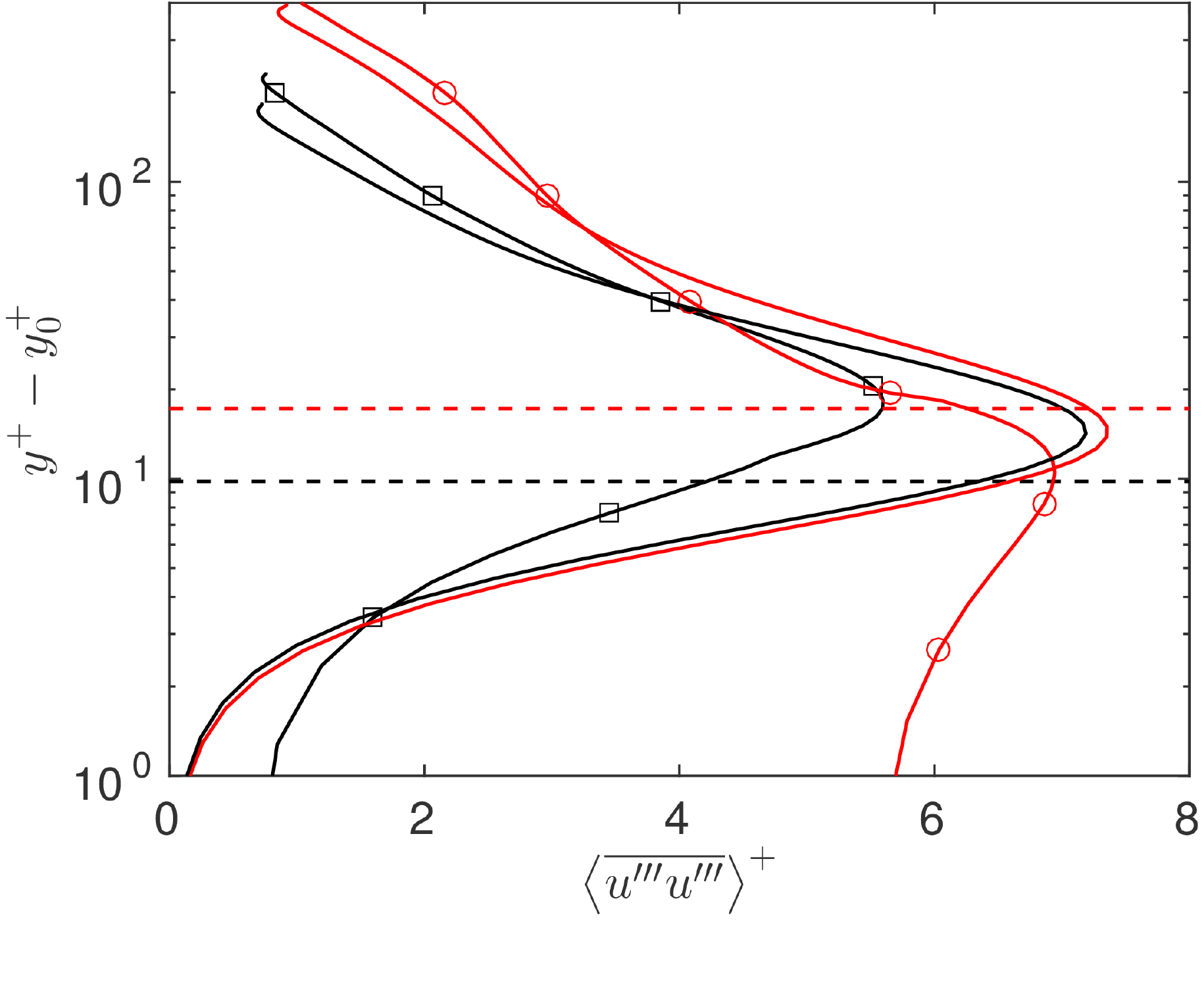}}
\put(200,0){\includegraphics[trim=0cm 1.5cm 0cm 0cm, clip, width=.47\textwidth]{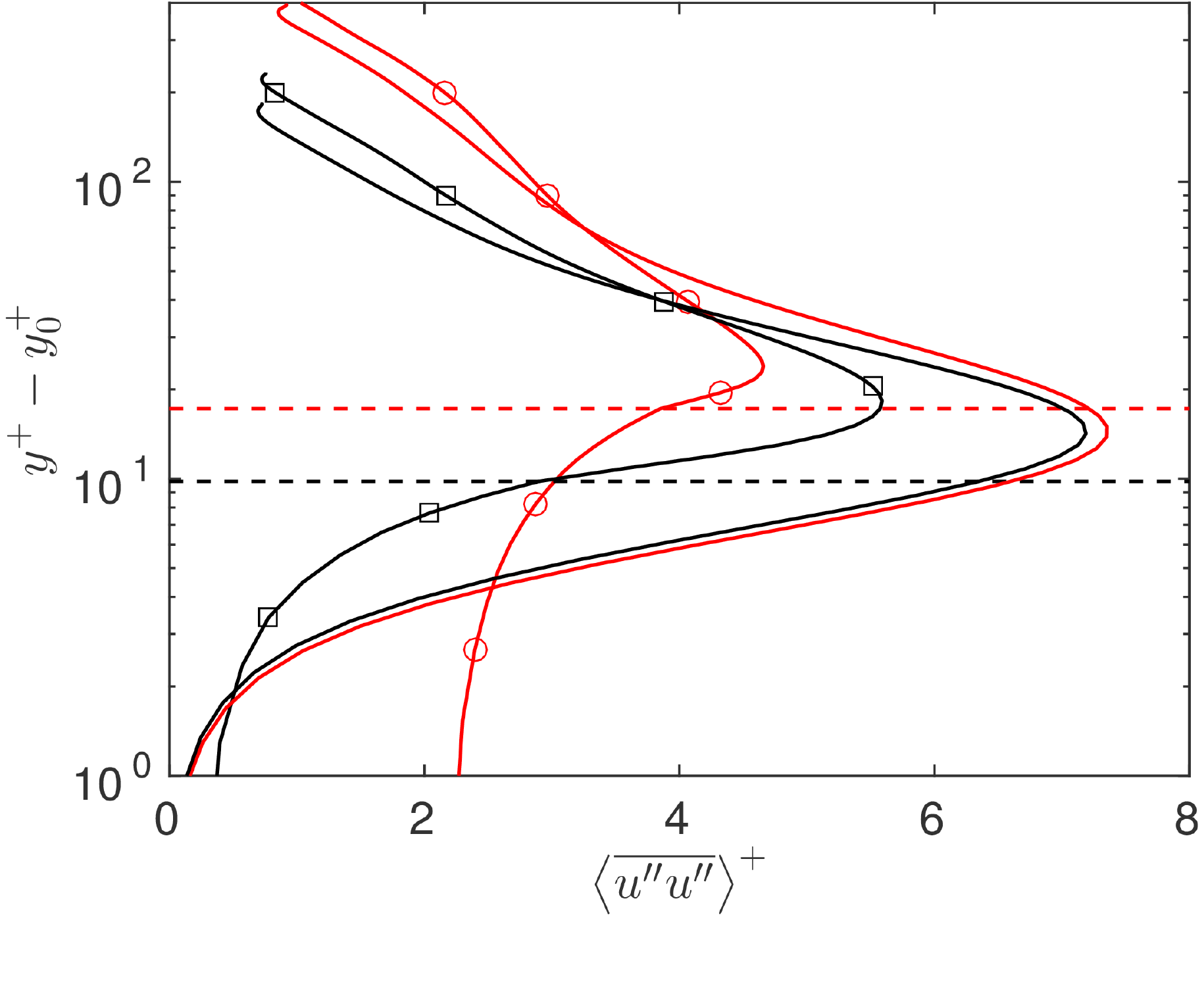}}
\put(0,132){$a)$}
\put(200,132){$b)$}
\end{picture}
\caption{Profiles of the terms of the Reynolds stress tensor $a)$ $\left\langle\overline{u'''u'''}\right\rangle^+$ and $b)$ $\left\langle \overline{u''u''}\right\rangle^+$ for $Re_{b}=2900$ (black lines) and $Re_{b}=6864$ (red lines) as function of the distance from the virtual wall. Lines ---$\square$--- and \mut{---$\bigcirc$---} indicate runs D50 and D120, respectively, while solid lines indicate the corresponding smooth wall simulations. The horizontal broken lines indicate the position of the crest of the spheres for the two runs.}
\label{fig3}
\end{figure}

\mut{
  The upper bound of the roughness sublayer was defined as the distance from the wall 
  at which the magnitude of pressure spatial disturbance $p'' -p'''$, i.e. the fluctuations
  induced by the roughness geometry, become smaller than 
  $1\%$ of its maximum value.
  This methodology to characterize the ``three-dimensionality'' of the turbulence structure
  was also adopted by \citet{chan2011}.
  Figure \ref{fig3rev1}a, shows the limit of the roughness sublayer
  compared with the wall-normal profile of stress
  defined by equation \eqref{eq11} for the streamwise velocity component and for pressure 
  (i.e. the variance of $p''-p'''$ normalized by $(\varrho u_\tau^2)^2$) for the simulations 
  D50 and D120.
  It can be noted that the decay of form-induced pressure fluctuations with the distance from 
  the crest of the spheres lies on the same line in the semi-logarithmic plot of figure 
  \ref{fig3rev1}a.
  Note that this exponential decay of the three-dimensionality of
  pressure was also observed by \citet{chan2011} for their case with
  $D^+=10.7$. 
  The curves related to the form-induced velocity fluctuations 
  do not exhibit this exponential variation with wall-distance, 
  even though the maximum variance of both velocity and pressure fluctuations is attained in 
  proximity of the crest of the spheres for both cases D50 and D120.
  The residual small variance that can be detected well above the roughness sublayer, was also 
  observed by \citet{florens2013} 
  who ascribed it 
  to time convergence error.
  However, this small residual could be associated with the roughness
  footprint that \citet{hong2011} observed on the small-scale
  turbulence, which persisted over the entire flow domain, although its
  effect on the Reynolds-stress statistics was found negligible. 
  For both the cases D50 and D120, the bound of the roughness sublayer was identified approximately
  at $y=1.8D$ (i.e. $y-y_0=0.95D$) which is in line with the experimental observations of 
  \citet{cheng2002} and of \citet{hong2011} ($y=2k$). 
  Streamwise velocity fluctuations 
  attain $1\%$ of their maximum root mean square value at $y\sim 1.45D$ which is in fair agreement 
  with the experimental results of \citet{florens2013} who 
  report 
  $y=1.5k$.
  The dispersion effect in the vicinity of the roughness elements can be also appreciated 
  by considering velocity profiles obtained at different locations with respect to the center of 
  the spheres, as shown in figure \ref{fig3rev1}b. 
  Although the streamwise velocity $\left\langle\overline{u}\right\rangle^+_B$ is remarkably
  dispersed below the crest of the spheres, %
  \pbt{%
    it rapidly converges with growing distance %
  } %
  above the crests  
  until it 
  coincides 
  with 
  the plane/time-averaged velocity in the upper part of the roughness sublayer.
}

The profiles of the mean square of streamwise velocity fluctuations were computed 
(figures \ref{fig3}a and \ref{fig3}b) by considering the definitions of fluctuations 
given by (\ref{eq3b}) and (\ref{eq3c}), respectively, in order to evaluate the 
\mut{
  effects of the roughness geometrical features on the dimensionless streamwise normal stress 
  (i.e. the streamwise turbulence intensity). 
}
While the profiles for the simulation D50 do not show significant discrepancies for $y^+>y^+_0$ by using either of the two definitions, the profiles obtained for the simulation D120 are remarkably different.
This difference is related to the particular arrangement of the spheres chosen here, which, 
\mut{
  we 
  recall, 
  is the same for the two runs, and which 
}
allows the flow to stream along streamwise intra-roughness grooves, causing a steady and 
predominantly two-dimensional spanwise-periodic pattern (of periodicity $L_B$, 
\mut{see figure \ref{sketch2}}). 
\mut{This interfacial flow} is captured by the sphere-box-average, whereas it is filtered 
out by the plane-average.
Consequently, the intensity of the groove-induced flow pattern enters the field $\boldsymbol{u}''$, 
but not $\boldsymbol{u}'''$.
In fact, the maximum of $\left\langle\overline{u'''u'''}\right\rangle^+$ is located over the crests of roughness elements for the run D50 while it falls below the crests for the run D120 (figure~\ref{fig3}a) and, in the latter case, the maximum disappears if the fluctuations $u''$ about the average flow field are considered (figure~\ref{fig3}b).
\mut{
  Larger values of the dispersive stress (of the streamwise velocity) below the crest of the spheres 
  for the case D120 can be also noted in figure \ref{fig3rev1} with respect to those attained for the 
  case D50. 
}
The destruction of the maximum normal \mut{stresses} is associated with that of the buffer 
layer and is, in our opinion the most clear effect of the fully-rough regime.
In other words, it was observed that the fully-rough regime was attained as $D^+-y^+_0$ exceeded $\sim 12$, which is almost the distance attained by the maximum of $\left\langle\overline{u'''u'''}\right\rangle^+$ from a smooth wall.

\begin{figure}
\setlength{\unitlength}{0.353mm}
\begin{picture}(0,200)(0,0)
\put(0,0){
\put(20,0){\includegraphics[trim=0cm 0cm 0cm 0cm, clip, width=.4\textwidth]{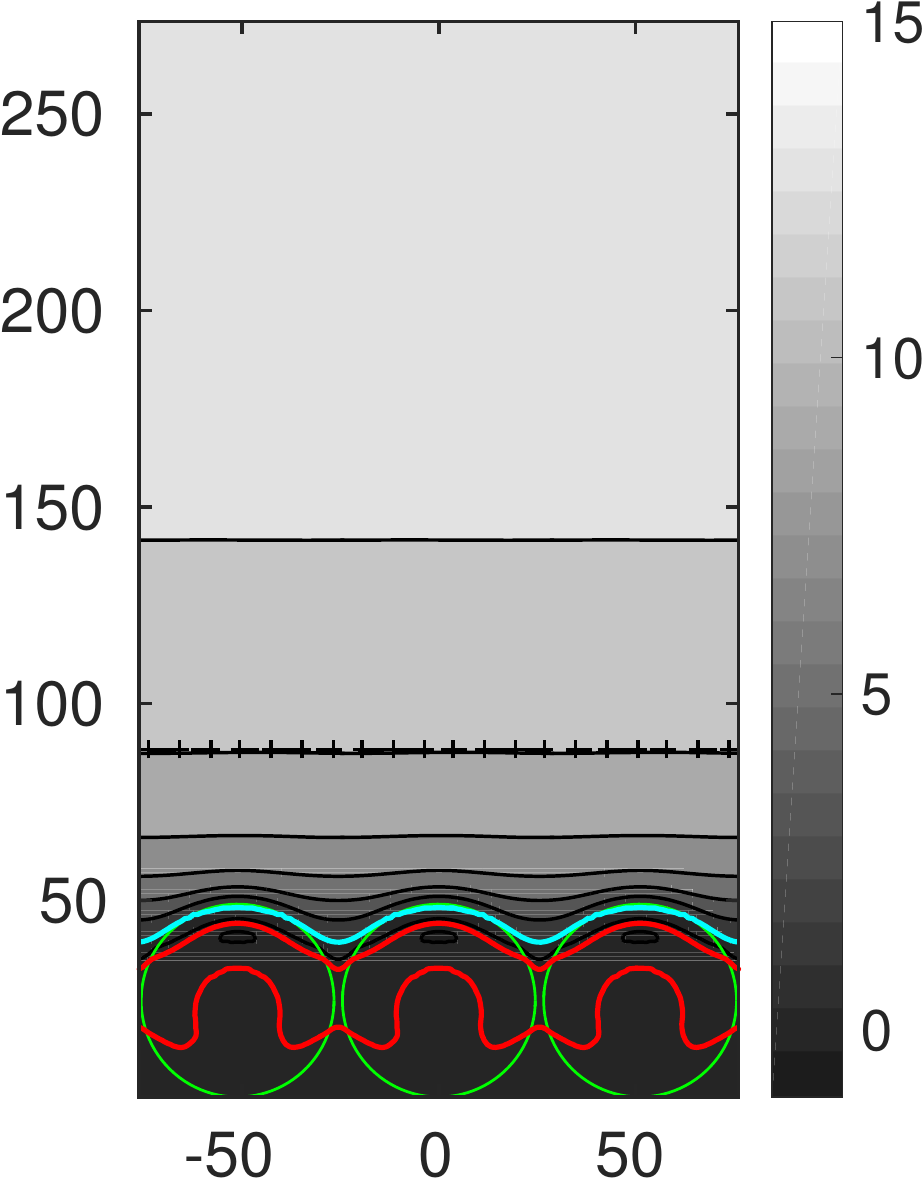}}
\put(195,0){\includegraphics[trim=0cm 0cm 0cm 0cm, clip, width=.4\textwidth]{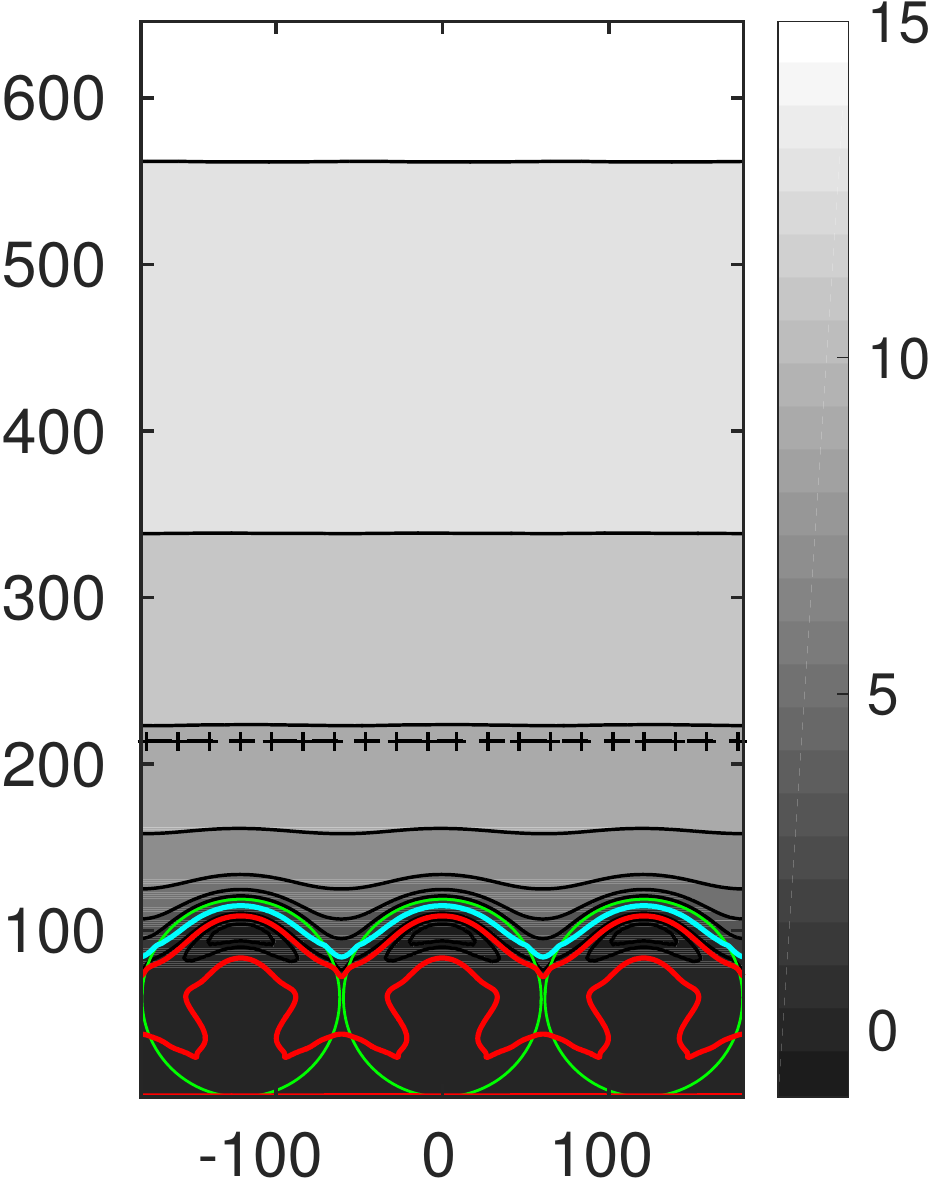}}
\put(90,-5){$z^+$}
\put(265,-5){$z^+$}
\put(5,110){$y^+$}
\put(5,190){$a)$}
\put(190,190){$b)$}
}
\end{picture}
\caption{Averaged streamwise velocity $\left\langle \overline{u}\right\rangle^+_{B,x}$ visualised by shadowed contour lines equispaced by $0.2$ and by $2$ for values smaller and larger than $1$, respectively. Red and cyan lines indicate the values $0$ and $1$. The spheres are outlined by green lines \mut{while the horizontal line $[$-{\footnotesize +}-$]$ indicates the upper limit of the roughness sublayer}. $a)$ Run D50; $b)$ run D120.}
\label{fig5}
\end{figure}
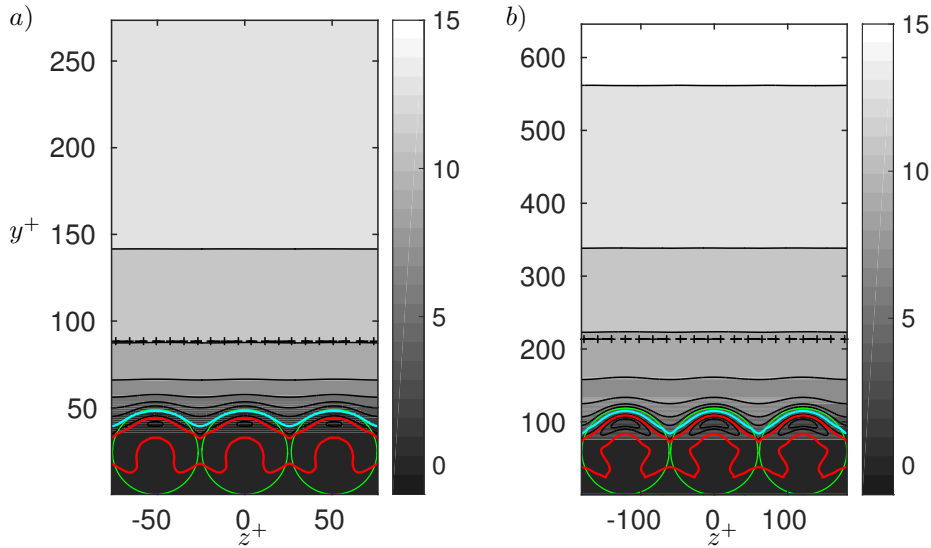
\begin{figure}
\setlength{\unitlength}{0.353mm}
\begin{picture}(0,430)(0,0)
\put(0,218){
\put(20,0){\includegraphics[trim=0cm 0cm 0cm 0cm, clip, width=.4\textwidth]{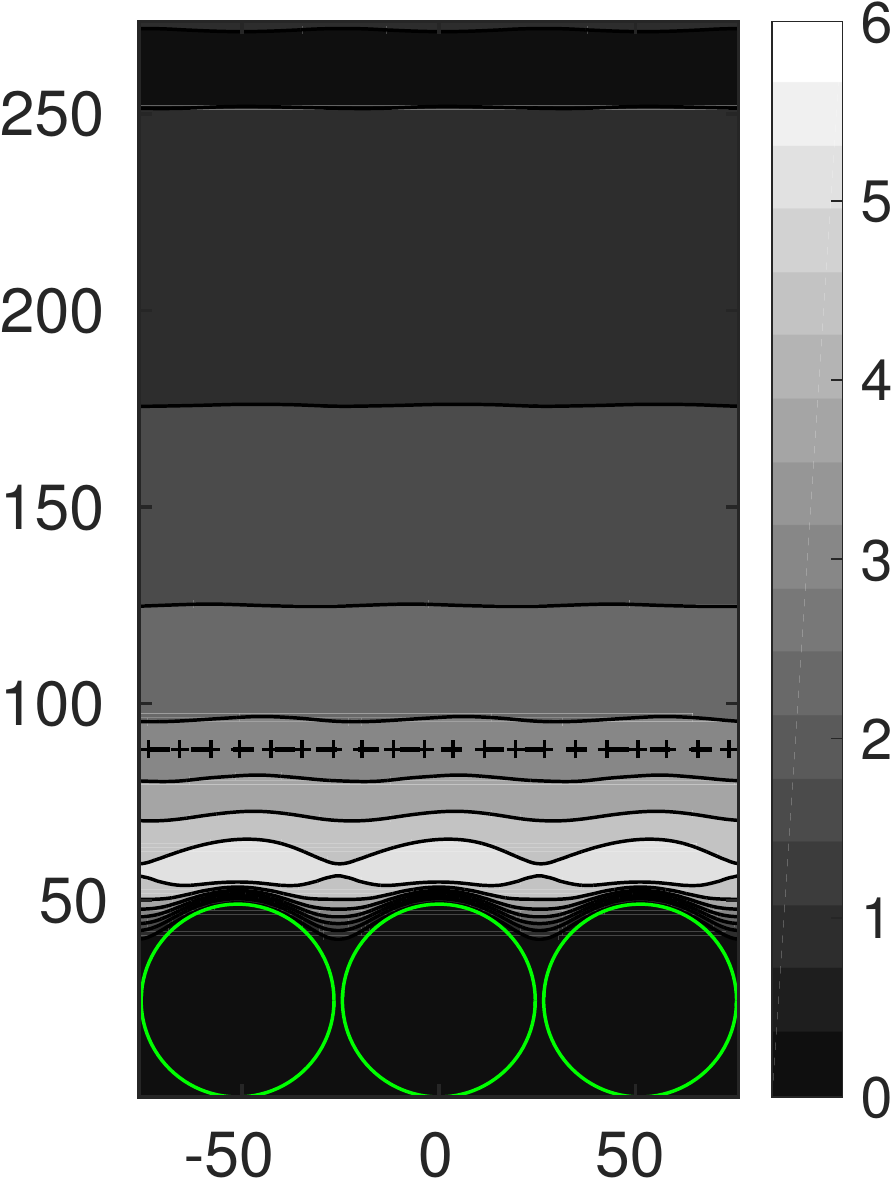}}
\put(202,0){\includegraphics[trim=0cm 0cm 0cm 0cm, clip, width=.4\textwidth]{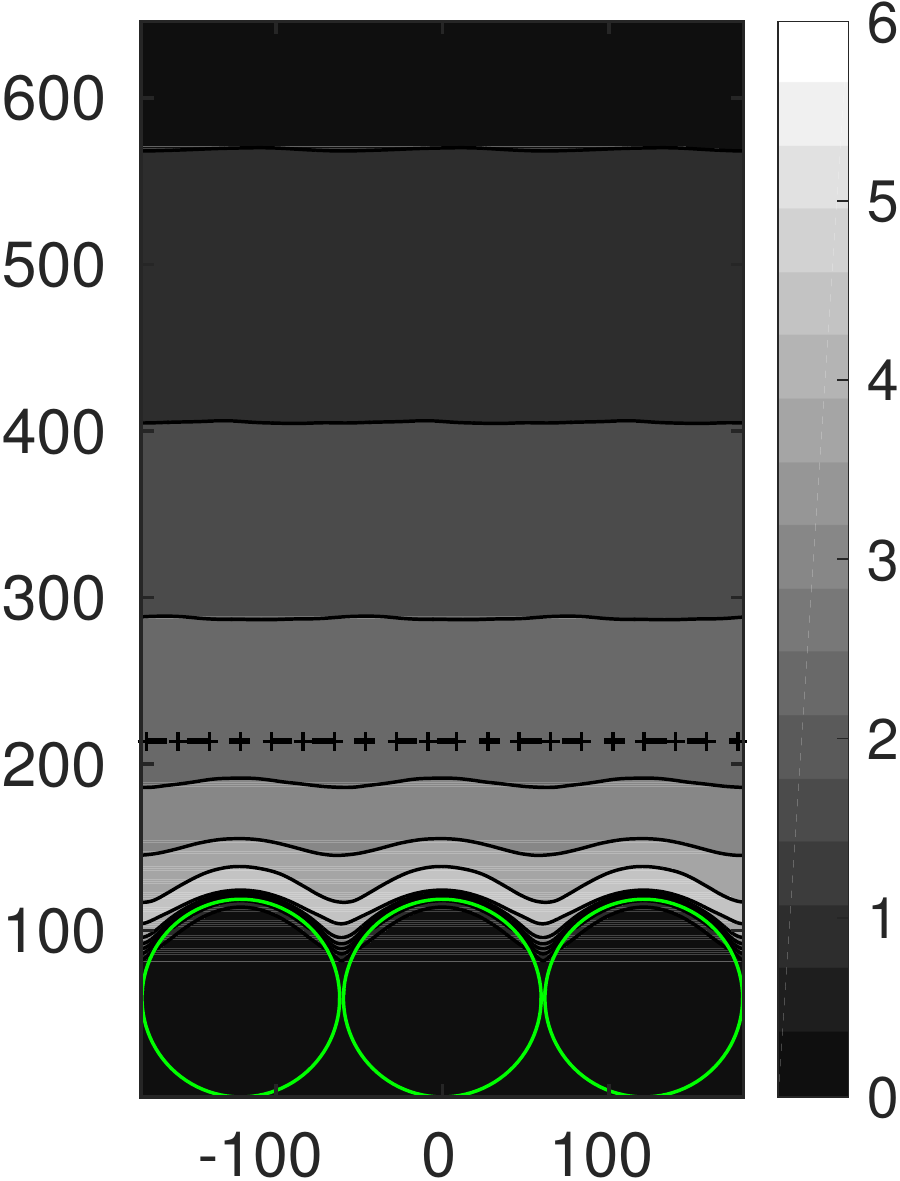}}
\put(94,-5){$z^+$}
\put(275,-5){$z^+$}
\put(5,110){$y^+$}
\put(5,190){$a)$}
\put(190,190){$b)$}
}
\put(20,0){\includegraphics[trim=0cm 0cm 0cm 0cm, clip, width=.4\textwidth]{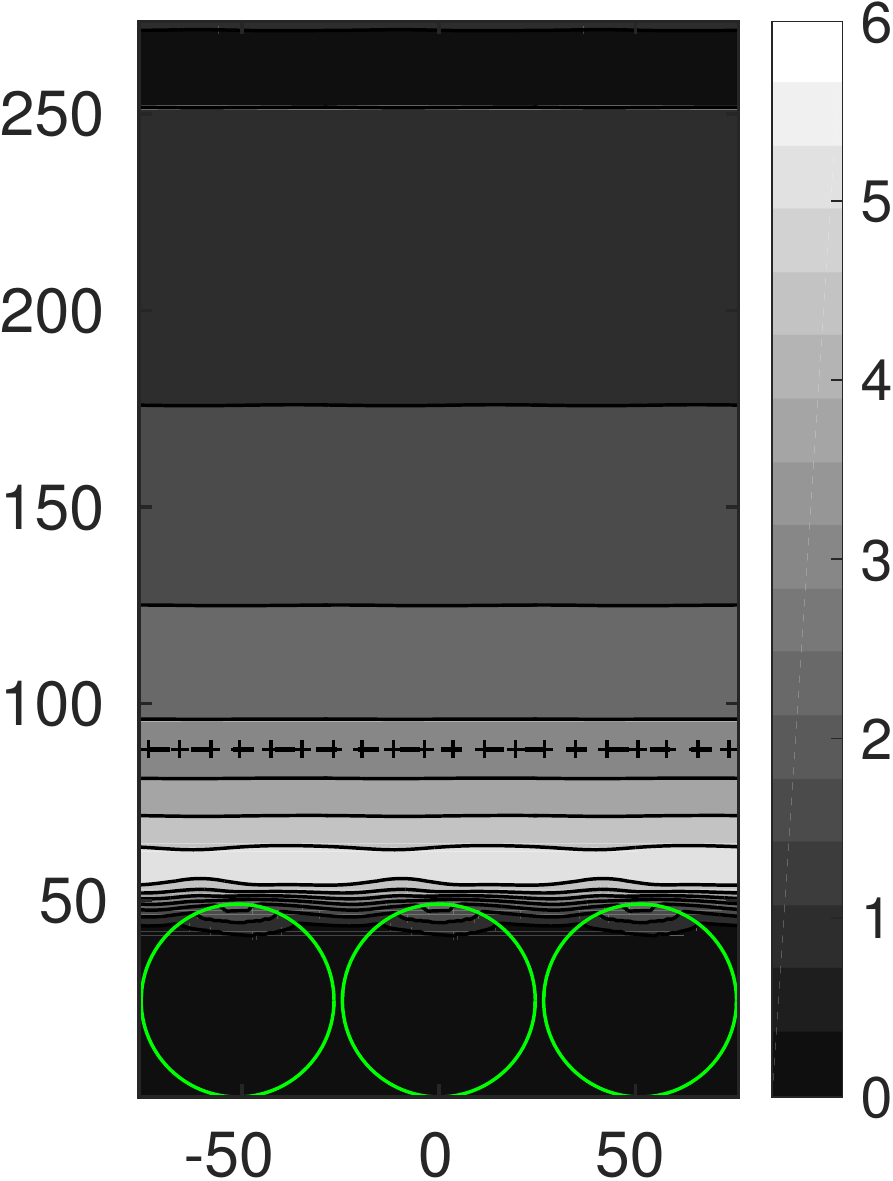}}
\put(202,0){\includegraphics[trim=0cm 0cm 0cm 0cm, clip, width=.4\textwidth]{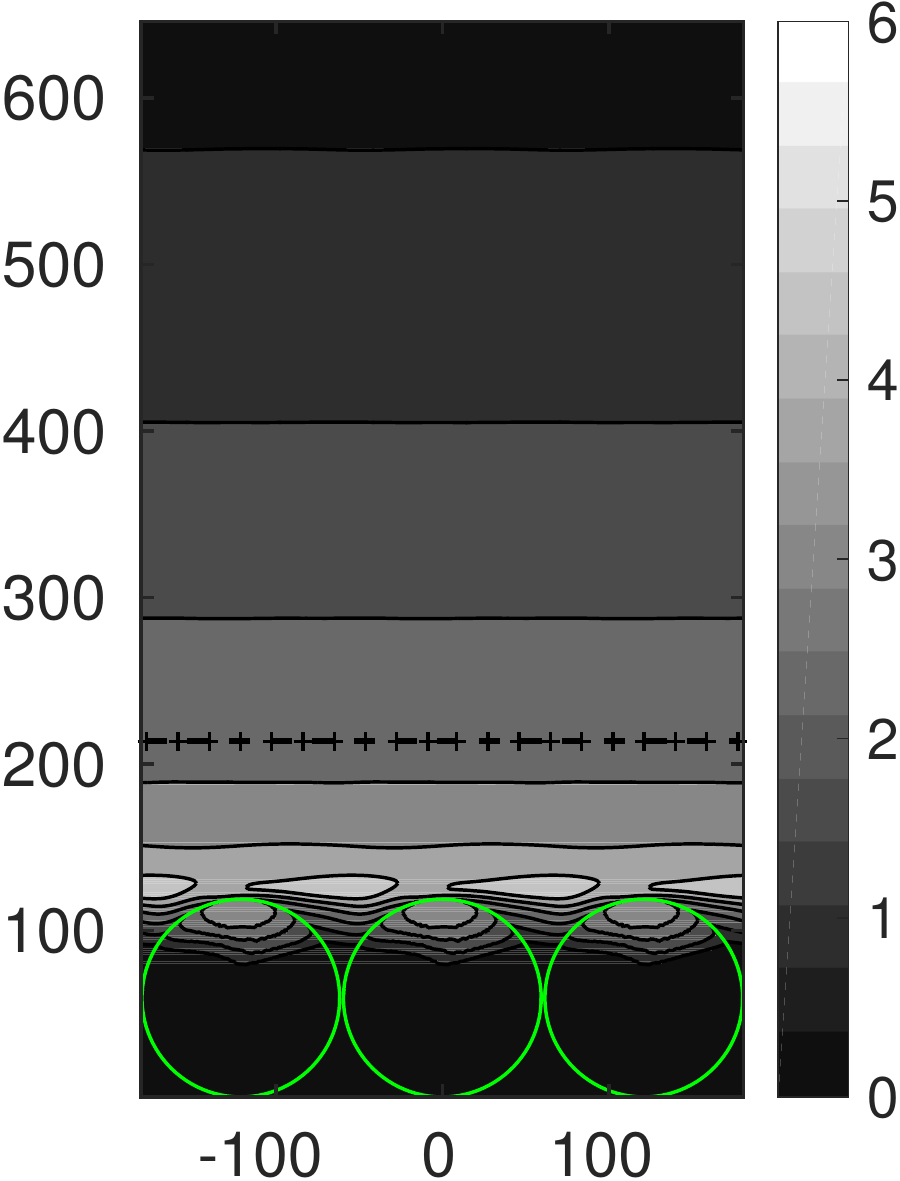}}
\put(94,-5){$z^+$}
\put(275,-5){$z^+$}
\put(5,110){$y^+$}
\put(5,190){$c)$}
\put(190,190){$d)$}
\put(81,180){\color{white!}\vector(1,0){30}}
\put(264,180){\color{white!}\vector(1,0){30}}
\end{picture}
\caption{Averaged square of the fluctuations of the averaged streamwise velocity $\left\langle \overline{u''u''}\right\rangle^+_{B,x}$ and $\left\langle \overline{u''u''}\right\rangle^+_{B,z}$ visualised in front view $(a,b)$ and side view $(c,d)$, respectively, by shadowed contour lines equispaced by \mut{$0.67$}. Cyan and red lines indicate the values $3$ and $4$, respectively. The spheres are outlined by green lines \mut{while the horizontal line $[$-{\footnotesize +}-$]$ indicates the upper limit of the roughness sublayer}. \mut{White rightward arrows indicate the direction of the flow.} $(a,c)$ Run D50; $(b,d)$ run D120.}
\label{fig6}
\end{figure}

The same picture arises also from the comparison of the average flow field computed for the simulations D50 and D120.
Indeed, the streamwise component of the streamwise-averaged velocity field, 
$\left\langle \overline{u}\right\rangle^+_{B,x}$, which is plotted in figure~\ref{fig5}, 
shows that the flow of the run D120 penetrates deeper into the crevices between the spheres.
%
However, the mean square of the corresponding fluctuations, $\left\langle \overline{u''u''}\right\rangle^+_{B,x}$, 
is more intense 
\mut{
  in the roughness sublayer 
}
over the crest of the roughness elements in the transitionally-rough case than in the fully-rough 
case (see figures \ref{fig6}a,b), although significant turbulent fluctuations are present 
\mut{
  in the roughness canopy 
}
in the latter one.
It also appears that the maximum of $\left\langle \overline{u''u''}\right\rangle^+_{B}$ is 
localised over the top of the spheres, where the interaction between the shear layers 
produced by neighbouring spheres is smaller, and not elsewhere.
This can be also deduced from figure~\ref{fig6}c which shows the distribution of 
$\left\langle \overline{u''u''}\right\rangle^+_{B,z}$, where the peaks of 
figure~\ref{fig6}a are filtered out through the spanwise-average.
%

\begin{figure}
\setlength{\unitlength}{0.353mm}
\begin{picture}(0,150)(0,0)
\put(0,0){\includegraphics[trim=1cm 0cm 1.5cm 0cm, clip, width=.5\textwidth]{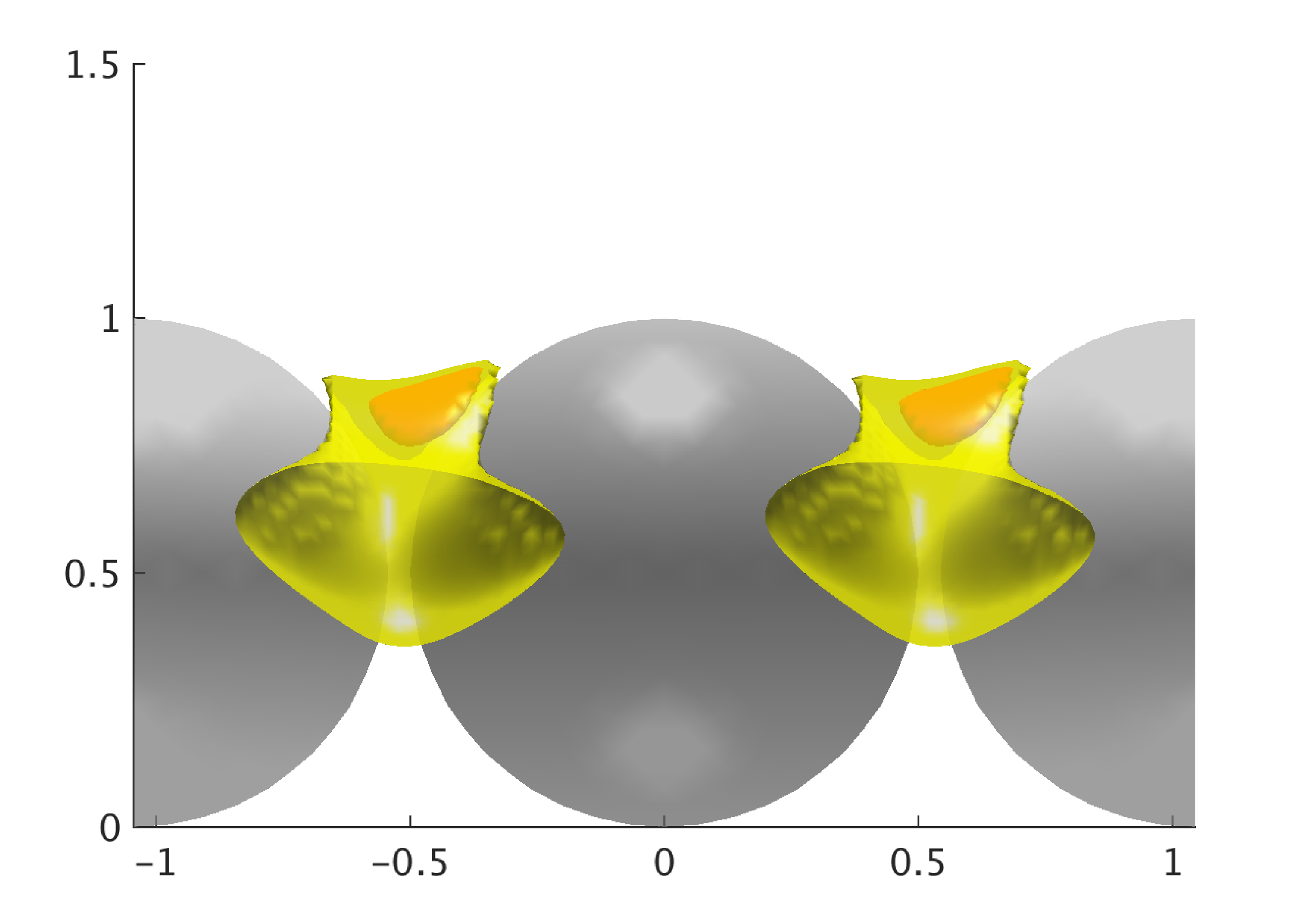}}
\put(190,0){\includegraphics[trim=1cm 0cm 1.5cm 0cm, clip, width=.5\textwidth]{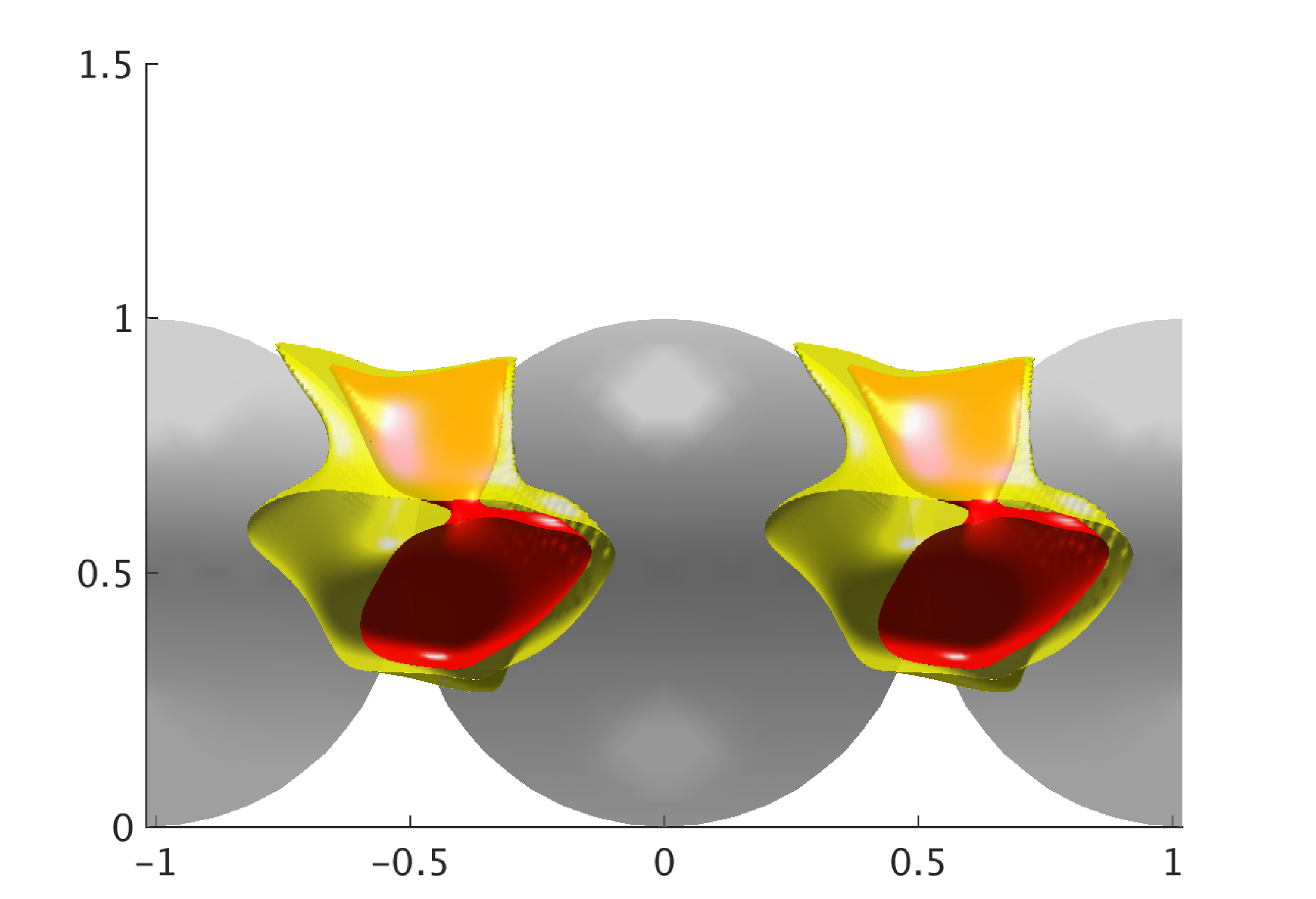}}
\put(-5,140){$a)$}
\put(190,140){$b)$}
\put(90,-5){${x}/D$}
\put(280,-5){${x}/D$}
\put(-5,75){$\dfrac{y}{D}$}
\put(85,110){\vector(1,0){30}}
\put(275,110){\vector(1,0){30}}
\end{picture}
\caption{Separation regions where the mean flow reverses are visualised by contour surfaces of the average streamwise velocity $\left\langle \overline{u}\right\rangle^+_{B}$ at the values $-0.02$ (yellow) and $-0.2$ (red). \mut{Rightward arrows indicate the direction of the flow.} $a)$ Run D50; $b)$ run D120.}
\label{fig4}
\end{figure}
\begin{figure}
\setlength{\unitlength}{0.353mm}
\begin{picture}(0,295)(0,0)
\put(65,5){
\put(14,150){\includegraphics[trim=0cm 0cm 0cm 0cm, clip, width=.64\textwidth]{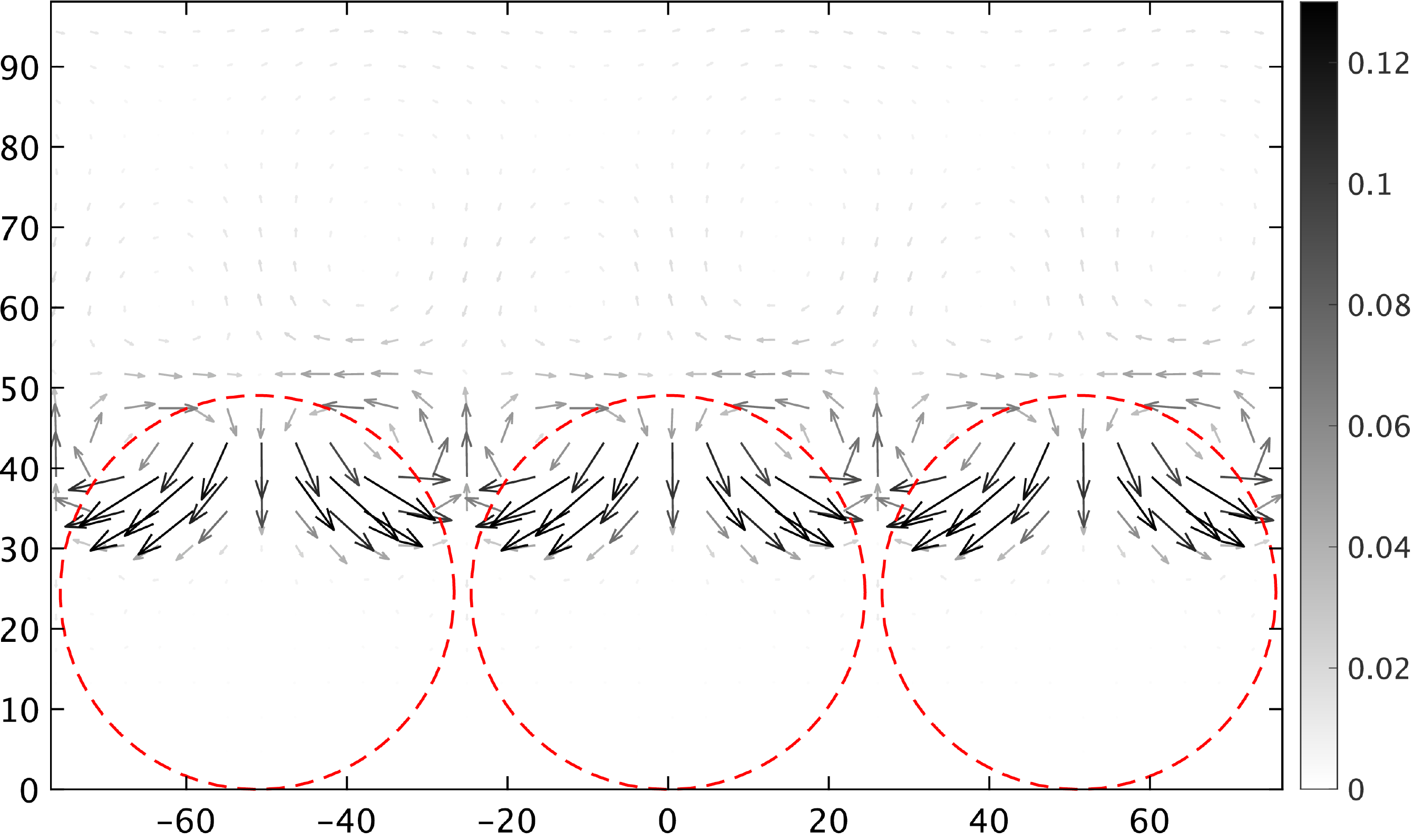}}
\put(10,-2){\includegraphics[trim=0cm 0cm 0cm 0cm, clip, width=.65\textwidth]{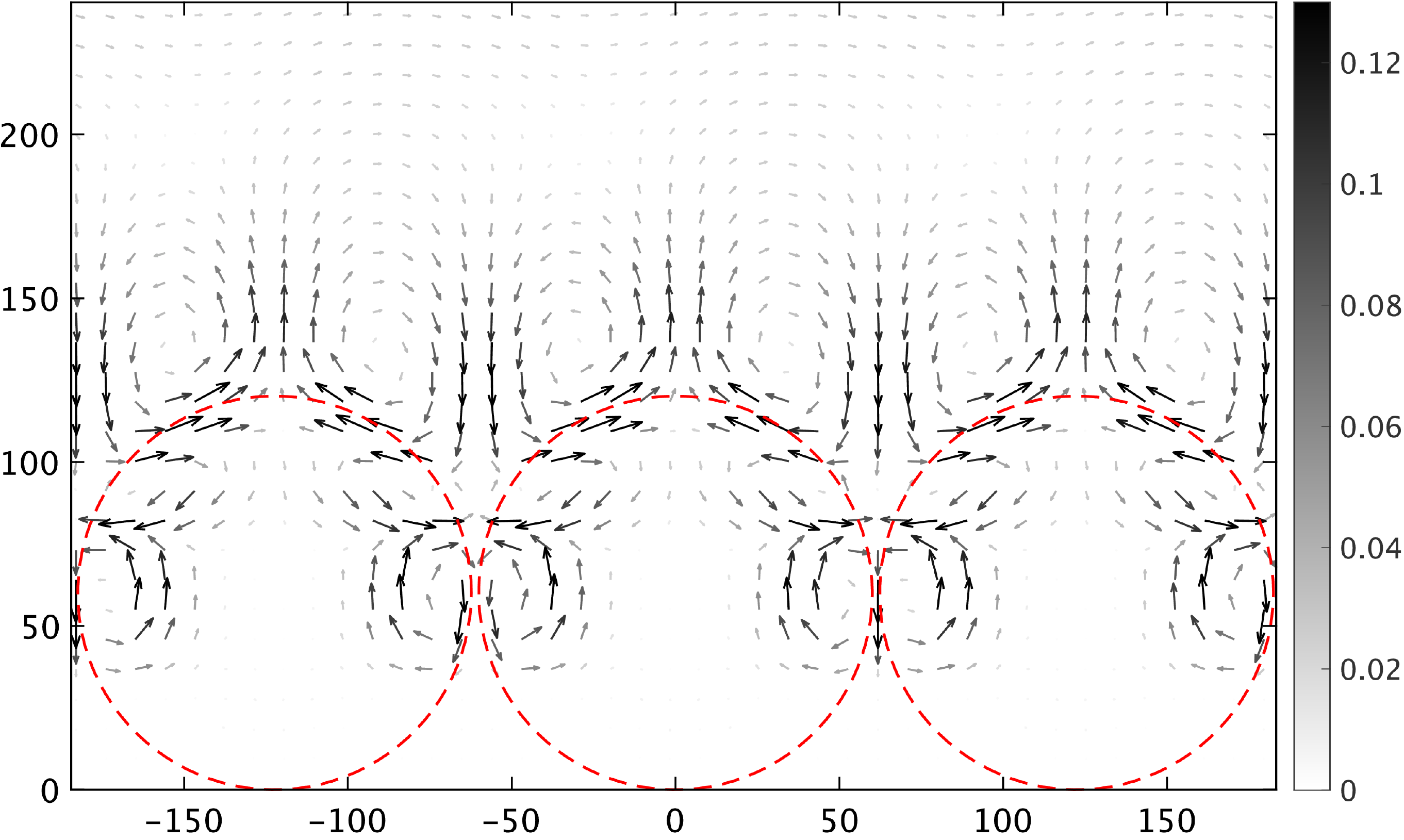}}
\put(125,-10){${z}^+$}
\put(0,75){$y^+$}
\put(0,225){$y^+$}
\put(2,278){$a)$}
\put(2,133){$b)$}
\put(60,218){A}
\put(92,56){C}
\put(95,228){B}
\put(57,62){D}
}
\end{picture}
\caption{Secondary flow visualised by velocity vectors 
  $\left\langle \overline{(w,v)}\right\rangle^+_{B,x}$ and 
  shaded by their modulus. The position of the spheres is
  outlined by red broken lines. $a)$ Run D50; $b)$ run D120.}  
\label{fig7}
\end{figure}

Flow separation was observed, both for the simulations D50 and D120, downstream of 
the roughness elements and over the spherical caps placed directly adjacent to the wall.
Figure~\ref{fig4} shows that the regions where the average streamwise velocity, 
$\left\langle \overline{u}\right\rangle^+_{B}$, is negative, increase in size and 
intensity for increasing values of the Reynolds number.
These regions are paired with as many spanwise-oriented recirculation cells from which fluid 
particles can escape flowing either along the upstream side of the spheres or the downstream 
side of the spherical caps. 
A mean secondary motion arises in both the transitionally- and fully-rough regime simulations. %
\pbt{%
  Here we use the term ``secondary motion/flow'' without implying 
  any statement on the origin of the
  motion, which we believe to be due to the combined effects of the
  three-dimensional geometry (curvature), inertia (breaking of
  fore/aft symmetry) and turbulence. Note that the latter effect is
  expected to be relatively weak, since turbulence intensity decays
  rapidly when entering the interstitial below the inter-particle
  grooves (i.e.\ for $y<y_0/2$). %
}%
Vectors in figure~\ref{fig7} highlight the direction of the secondary flow projected on the 
$(y,z)$-plane: while for the run D50 (in terms of streamwise-averaged flow) the fluid is directed 
towards the regions characterised by negative streamwise velocity in the zone denoted by ``A'' and 
it moves away from the roughness through B, the opposite picture appears for the run D120, since 
the secondary motion converges at C and diverges from D.
This can be explained by observing that the mean flow tends to penetrate deeper into the 
roughness along with the secondary recirculation cells (the migration of their center of 
rotation can be noted in figure~\ref{fig7}), also encountering the spherical caps mounted 
on the wall.
A pair of recirculation cells appears in figure~\ref{fig7}b which is absent in the 
transitionally-rough simulation, and which is possibly associated with the flow 
interaction with the underlying spherical caps.
It is interesting to note that the direction of rotation of the recirculation cells over 
the spherical caps is opposite with respect to those in the vinicity of the crest of the 
(complete) spheres, while the average streamwise velocity has also opposite sign, i.e. 
being negative over the caps and positive over the spheres.
This picture is confirmed by the following analysis of the average
vorticity field $\left\langle
  \overline{\boldsymbol{\omega}}\right\rangle^+_{B}$. 
%
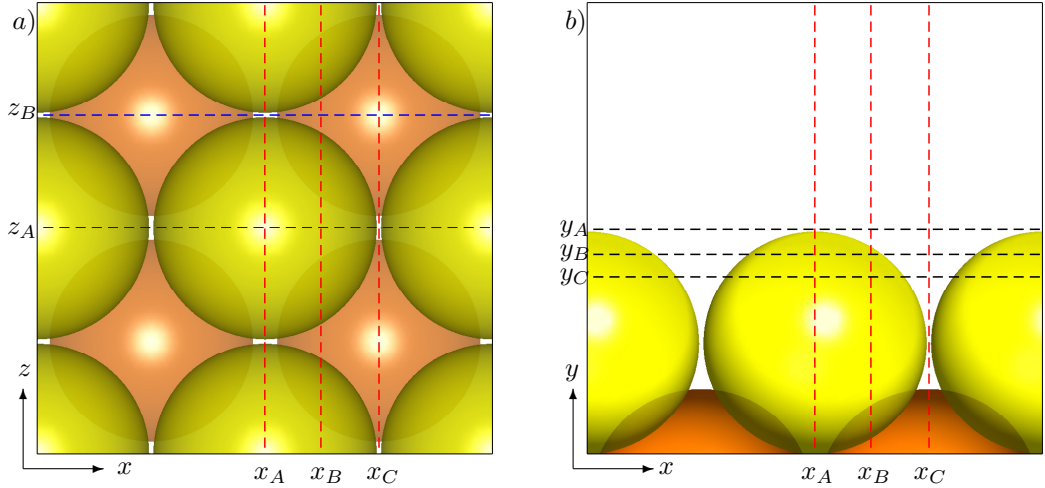
\begin{figure}
\setlength{\unitlength}{0.353mm}
\begin{picture}(0,175)(0,0)
\put(0,5){
\put(5,0){\includegraphics[trim=0cm 0cm 0cm 0cm, clip, width=.45\textwidth]{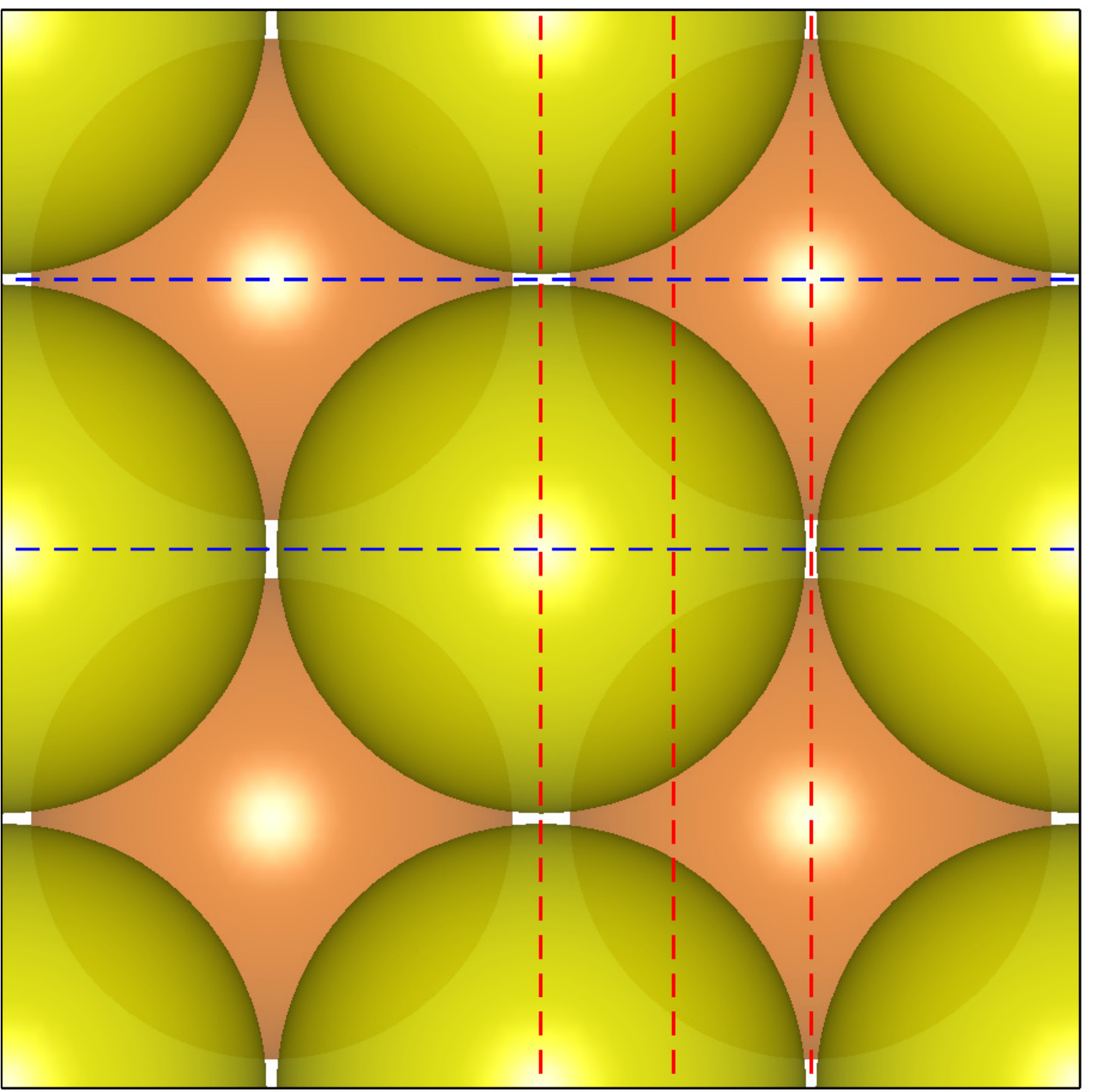}}
\put(210,0){\includegraphics[trim=0cm 0cm 0cm 0cm, clip, width=.45\textwidth]{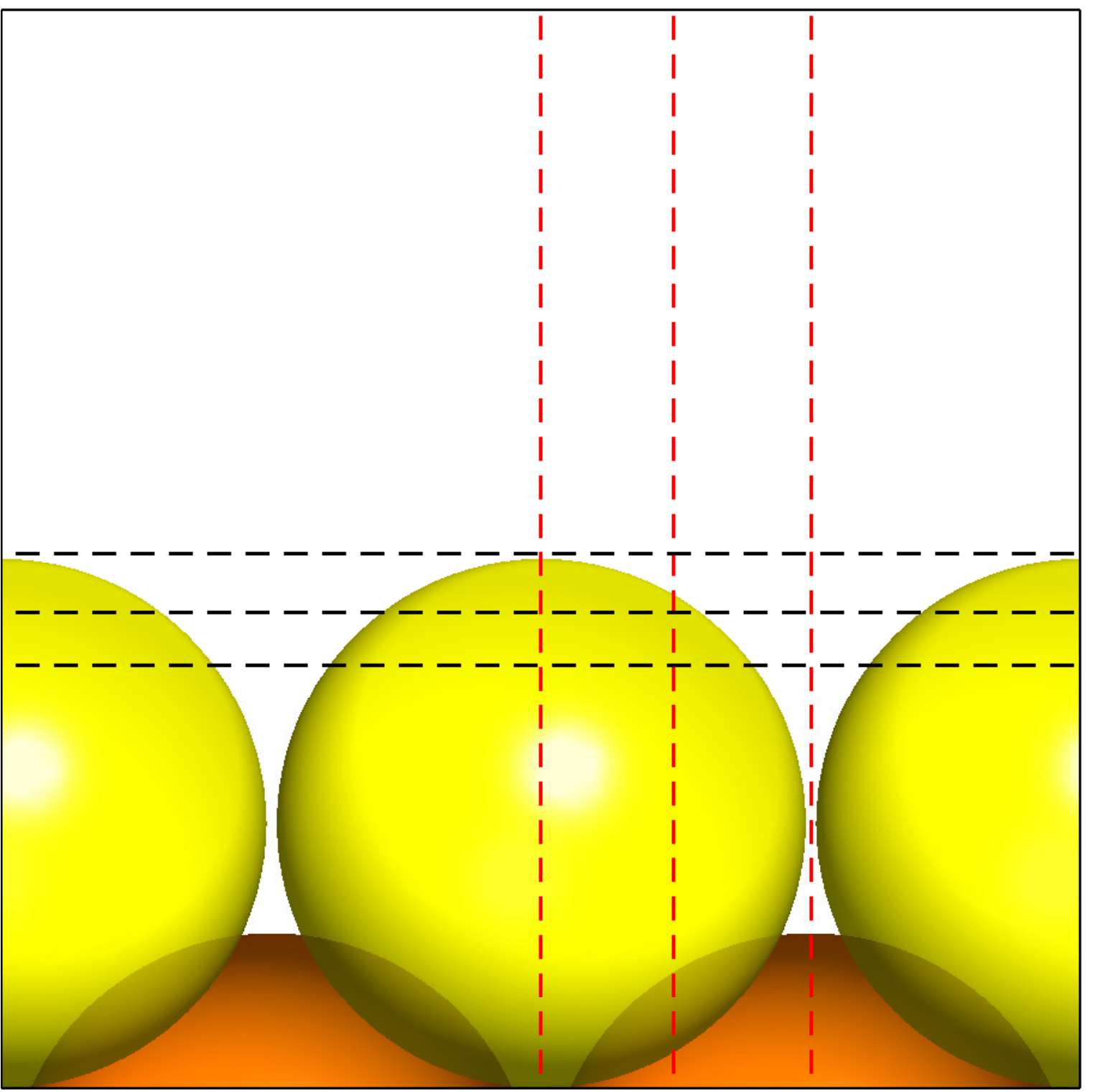}}
\put(202,30){$y$}
\put(-2,30){$z$}
\put(237,-7){$x$}
\put(35,-7){$x$}
\put(-4,160){$a)$}
\put(202,160){$b)$}
\put(0,-5){\vector(1,0){30}}
\put(0,-5){\vector(0,1){30}}
\put(205,-5){\vector(1,0){30}}
\put(205,-5){\vector(0,1){30}}
\put(86,-9){$x_A$}
\put(107,-9){$x_B$}
\put(128,-9){$x_C$}
\put(290,-9){$x_A$}
\put(311,-9){$x_B$}
\put(332,-9){$x_C$}
\put(-6,83){$z_A$}
\put(-6,128){$z_B$}
\put(200,84){$y_A$}
\put(200,75){$y_B$}
\put(200,66){$y_C$}
}
\end{picture}
\caption{Sketch of the top $(a)$ and side $(b)$ view of the region (sphere-box) over which the flow field was averaged. Broken lines indicate the position of the cross sections at which the streamwise (red broken lines), spanwise (blue broken lines) and wall-normal (black broken lines) vorticity components where computed and shown in figures \ref{fig8}, \ref{fig9} and \ref{fig10}, respectively.}
\label{fig20}
\end{figure}
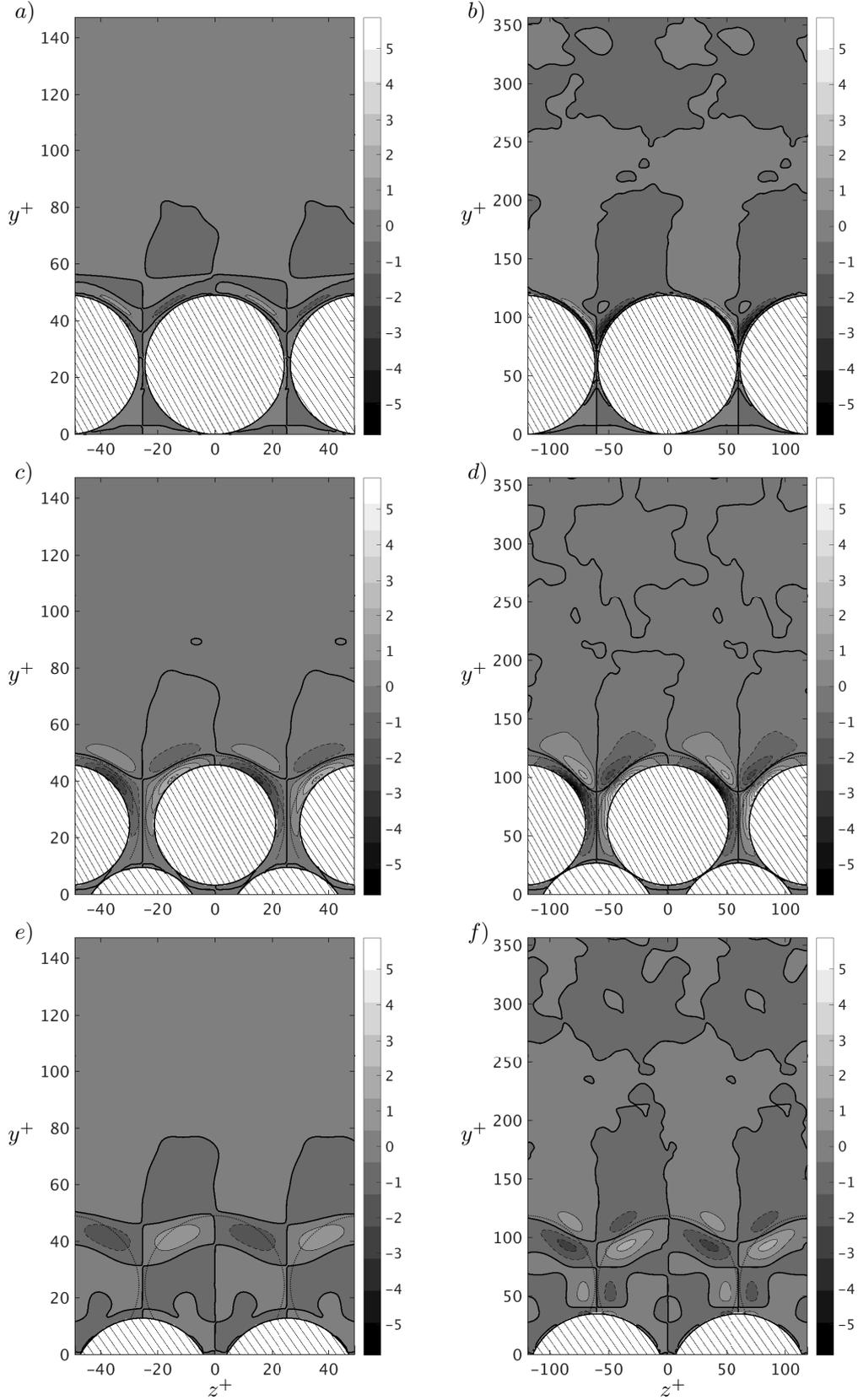
\begin{figure}
\setlength{\unitlength}{0.353mm}
\begin{picture}(0,620)(0,0)
\put(0,410){
\put(0,0){\includegraphics[trim=0cm 0cm 0cm 0cm, clip, width=.45\textwidth]{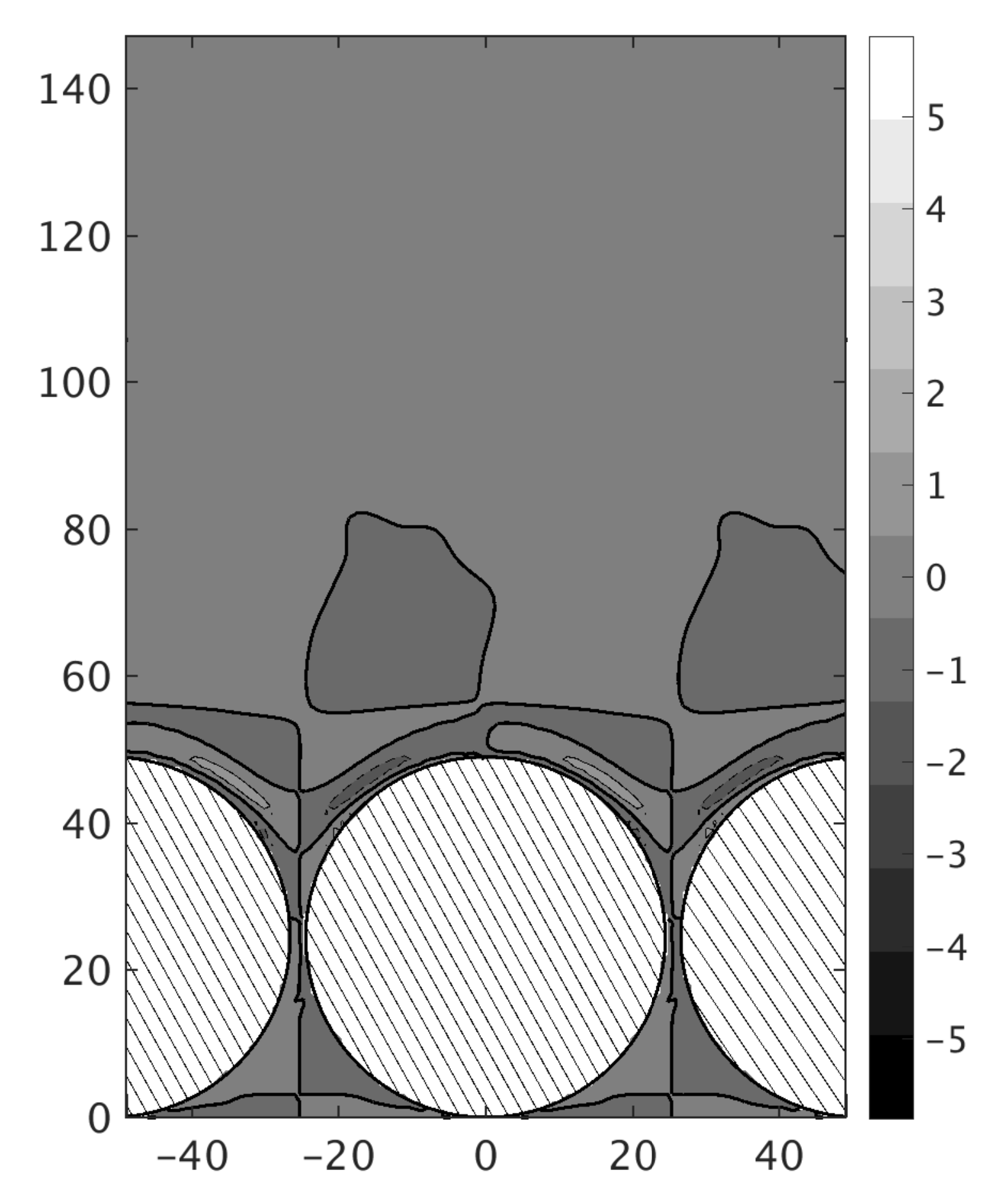}}
\put(-8,110){$y^+$}
}
\put(200,410){
\put(0,0){\includegraphics[trim=0cm 0cm 0cm 0cm, clip, width=.45\textwidth]{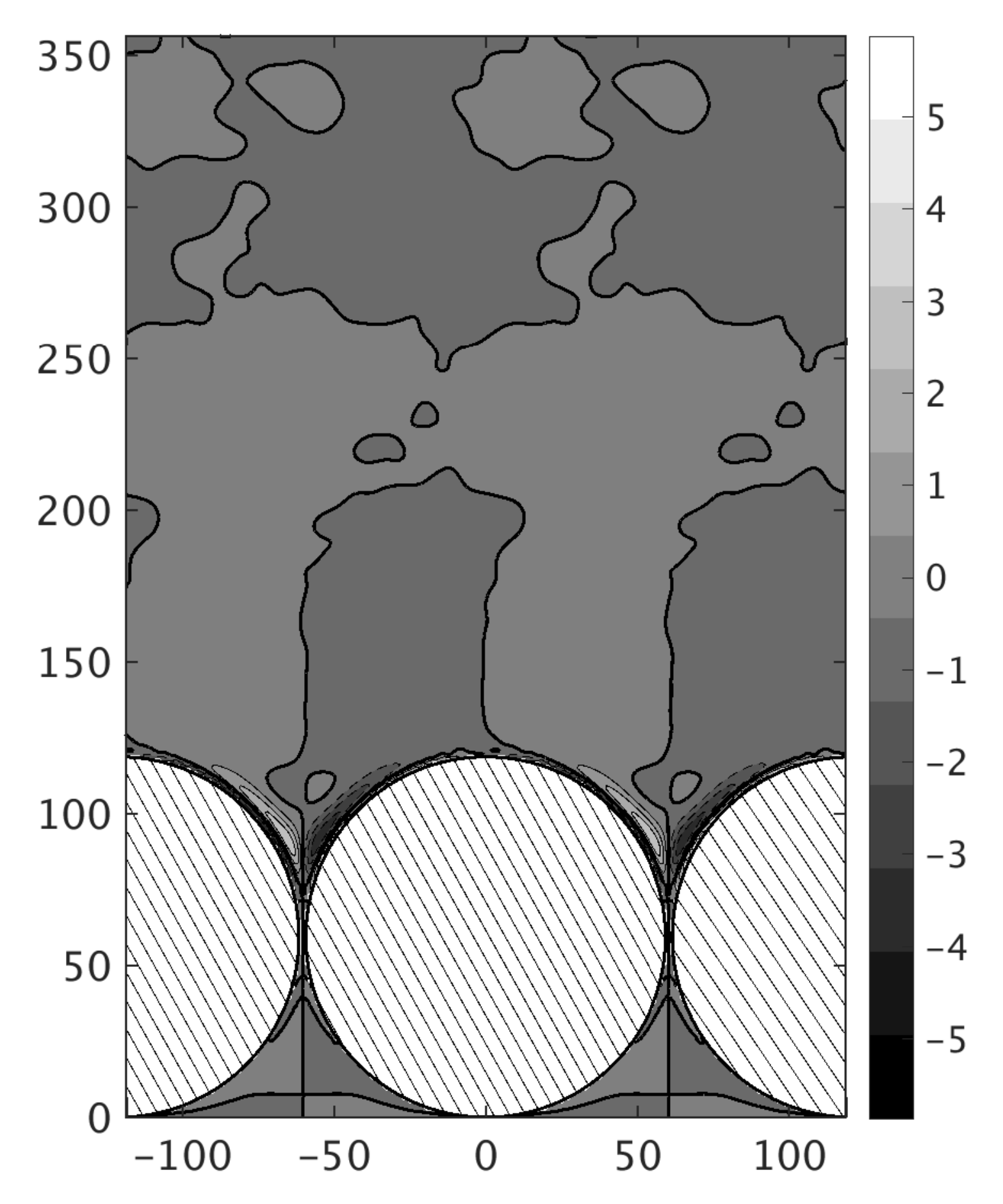}}
\put(-8,110){$y^+$}
}
\put(0,205){
\put(0,0){\includegraphics[trim=0cm 0cm 0cm 0cm, clip, width=.45\textwidth]{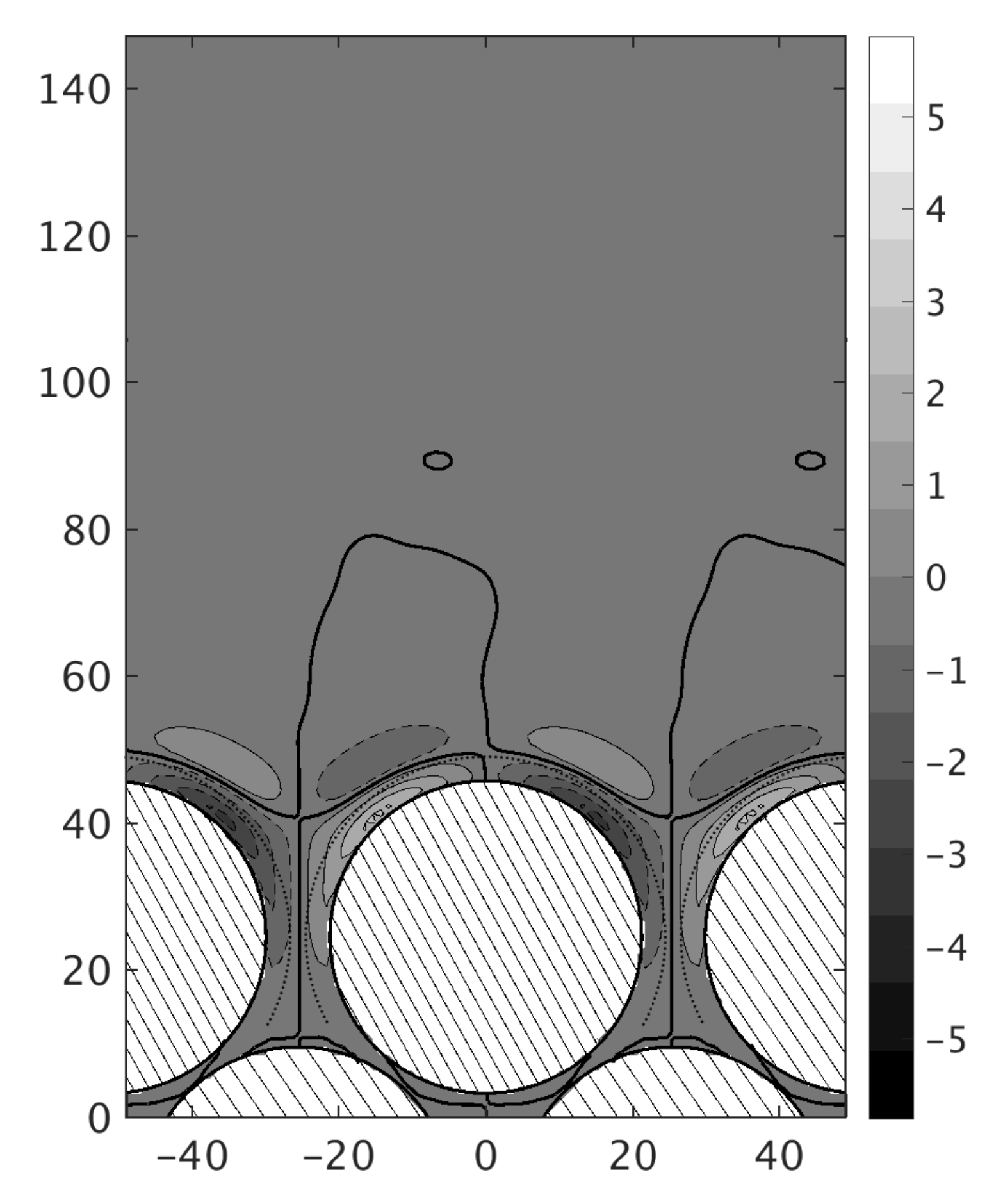}}
\put(-8,110){$y^+$}
}
\put(200,205){
\put(0,0){\includegraphics[trim=0cm 0cm 0cm 0cm, clip, width=.45\textwidth]{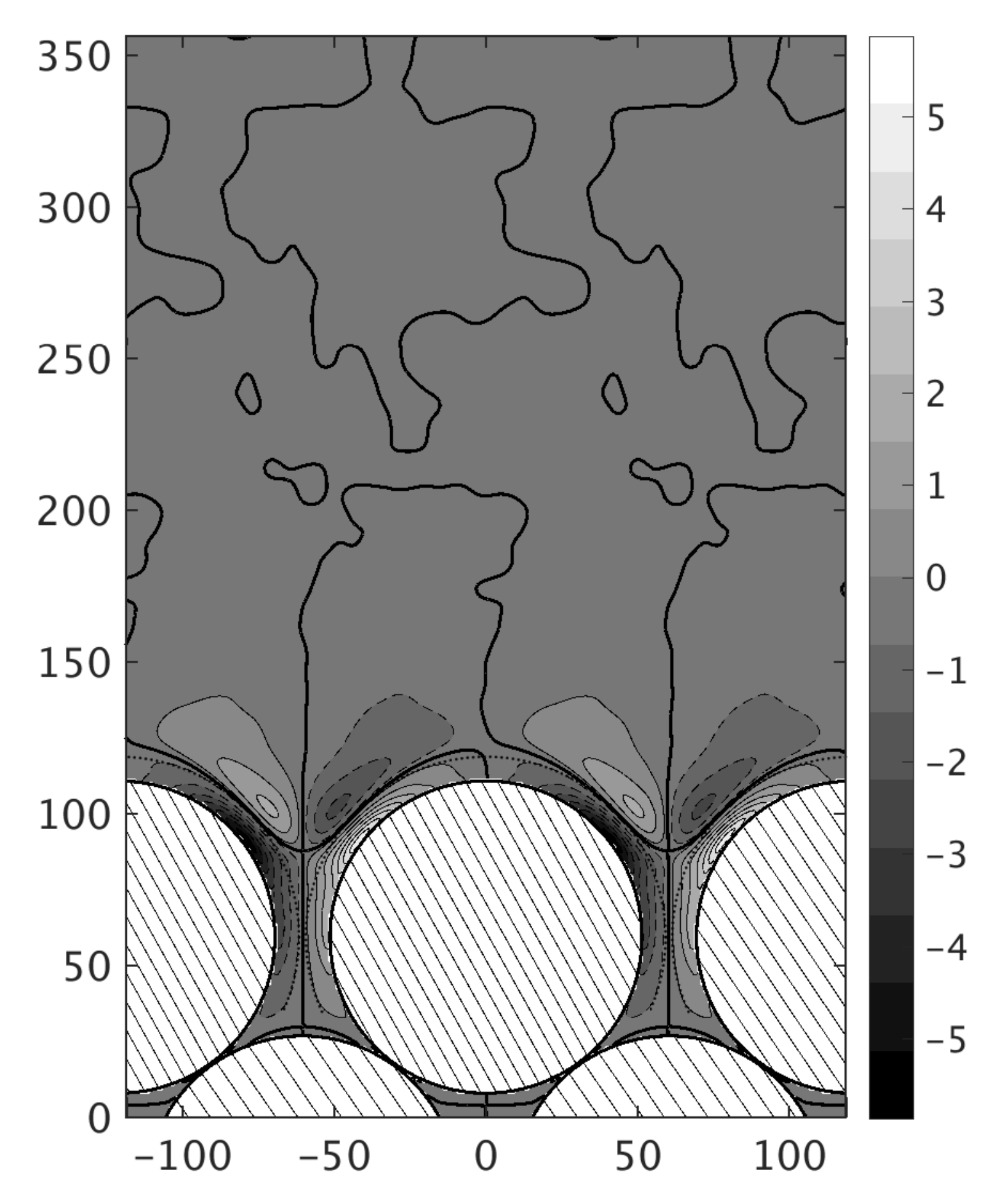}}
\put(-8,110){$y^+$}
}
\put(0,0){
\put(0,0){\includegraphics[trim=0cm 0cm 0cm 0cm, clip, width=.45\textwidth]{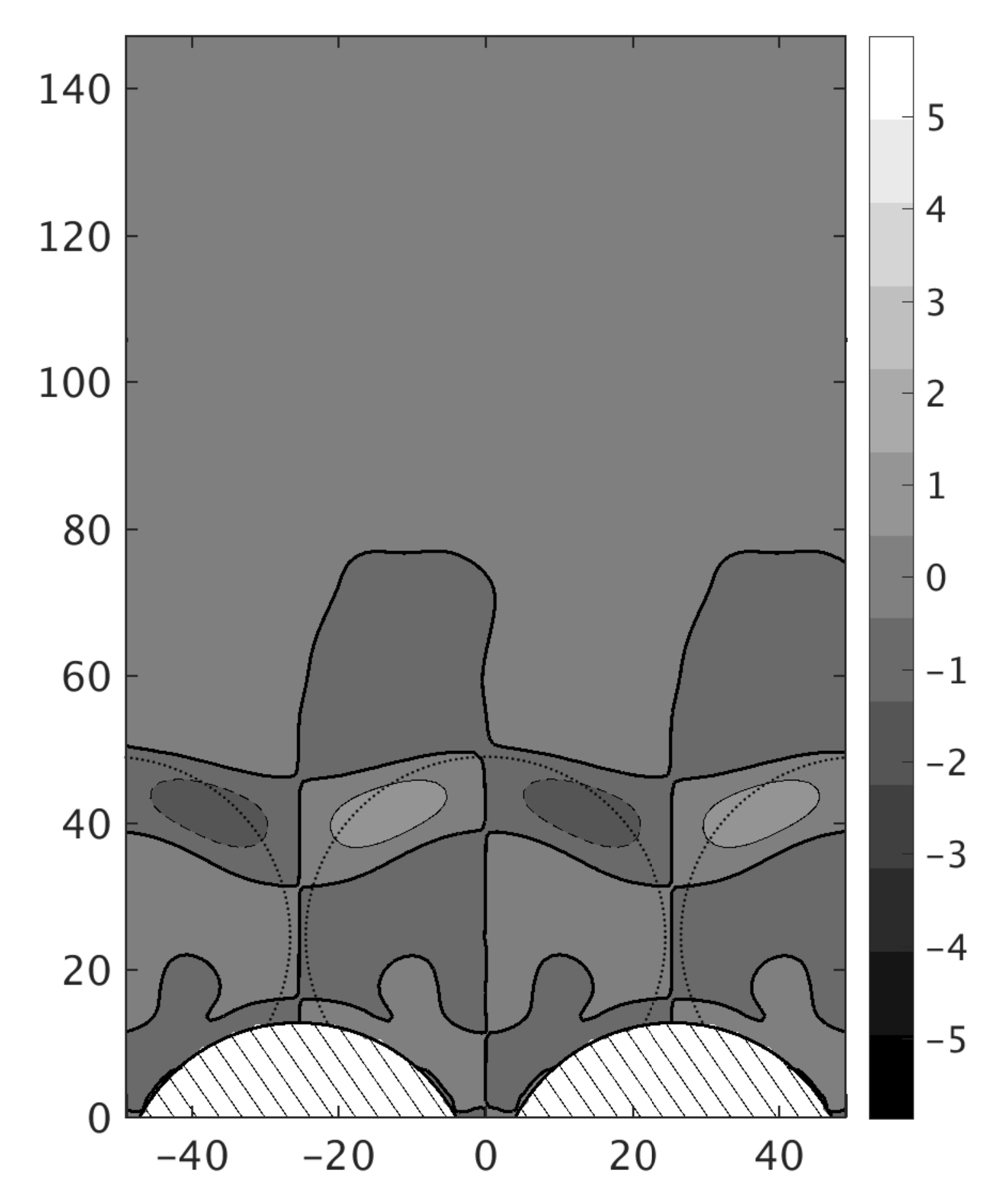}}
\put(-8,110){$y^+$}
\put(80,-4){${z}^+$}
}
\put(200,0){
\put(0,0){\includegraphics[trim=0cm 0cm 0cm 0cm, clip, width=.45\textwidth]{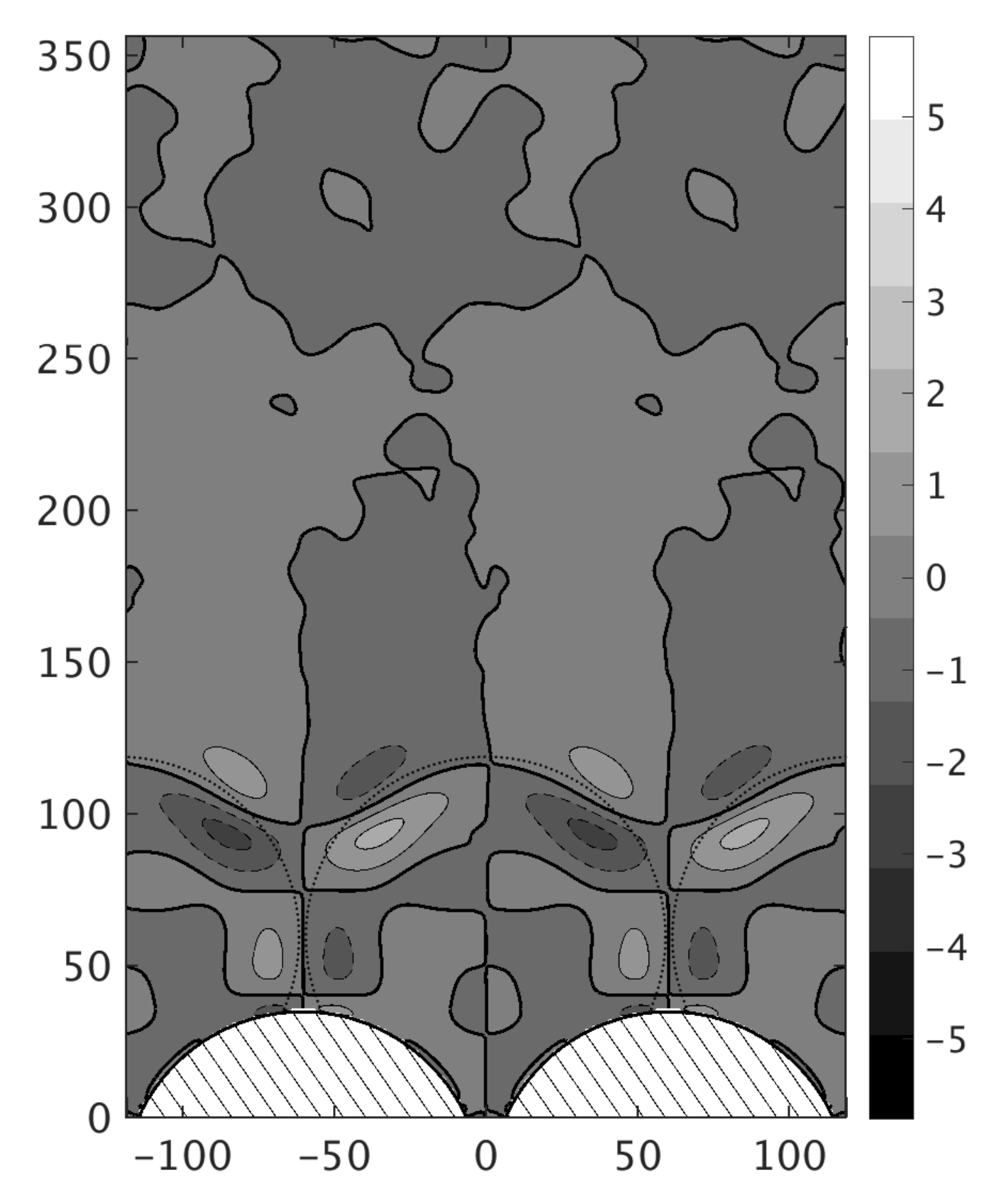}}
\put(-8,110){$y^+$}
\put(80,-4){${z}^+$}
}
\put(-5,610){$a)$}
\put(195,610){$b)$}
\put(-5,405){$c)$}
\put(195,405){$d)$}
\put(-5,200){$e)$}
\put(195,200){$f)$}
\end{picture}
\caption{Streamwise component 
  $\left\langle\overline{\omega_x}\right\rangle^+_{B}$ 
  of the sphere-box/time-averaged vorticity field visualised by
  filled contours $(a,b)$ in 
  the vertical plane through the center
  of the sphere $\tilde{x}=0$,  $\tilde{x}$ being the streamwise
  coordinate in the frame of 
  reference $\mathcal{B}$, $(c,d)$ at
  $\tilde{x}=(D+\Delta_B)/4$ and (e-f) at $\tilde{x}=(D+\Delta_B)/2$,
  i.e.\ at the slices $x_A$, $x_B$ and $x_C$ of figure~\ref{fig20},
  respectively. 
  The value 
  $\left\langle\overline{\omega_x}\right\rangle^+_{B}=0$ is shown as a
  thick line. 
  $(a,c,e)$ Run D50; $(b,d,f)$ run D120.} 
\label{fig8}
\end{figure}
\begin{figure}
\setlength{\unitlength}{0.353mm}
\begin{picture}(0,420)(0,0)
\put(0,210){
\put(0,0){\includegraphics[trim=0cm 0cm 0cm 0cm, clip, width=.45\textwidth]{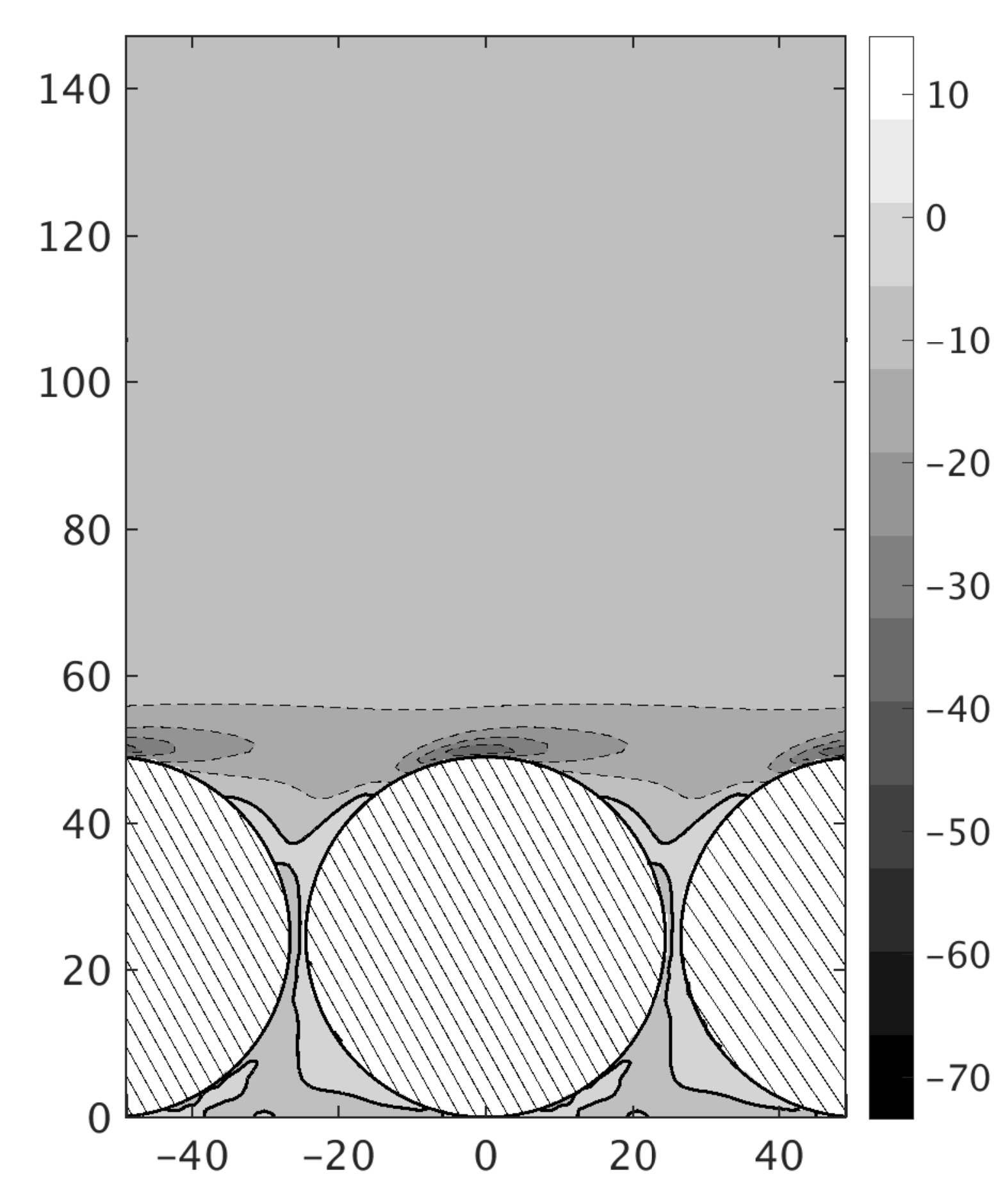}}
\put(-8,110){$y^+$}
\put(70,180){\vector(1,0){30}}
}
\put(200,210){
\put(0,0){\includegraphics[trim=0cm 0cm 0cm 0cm, clip, width=.45\textwidth]{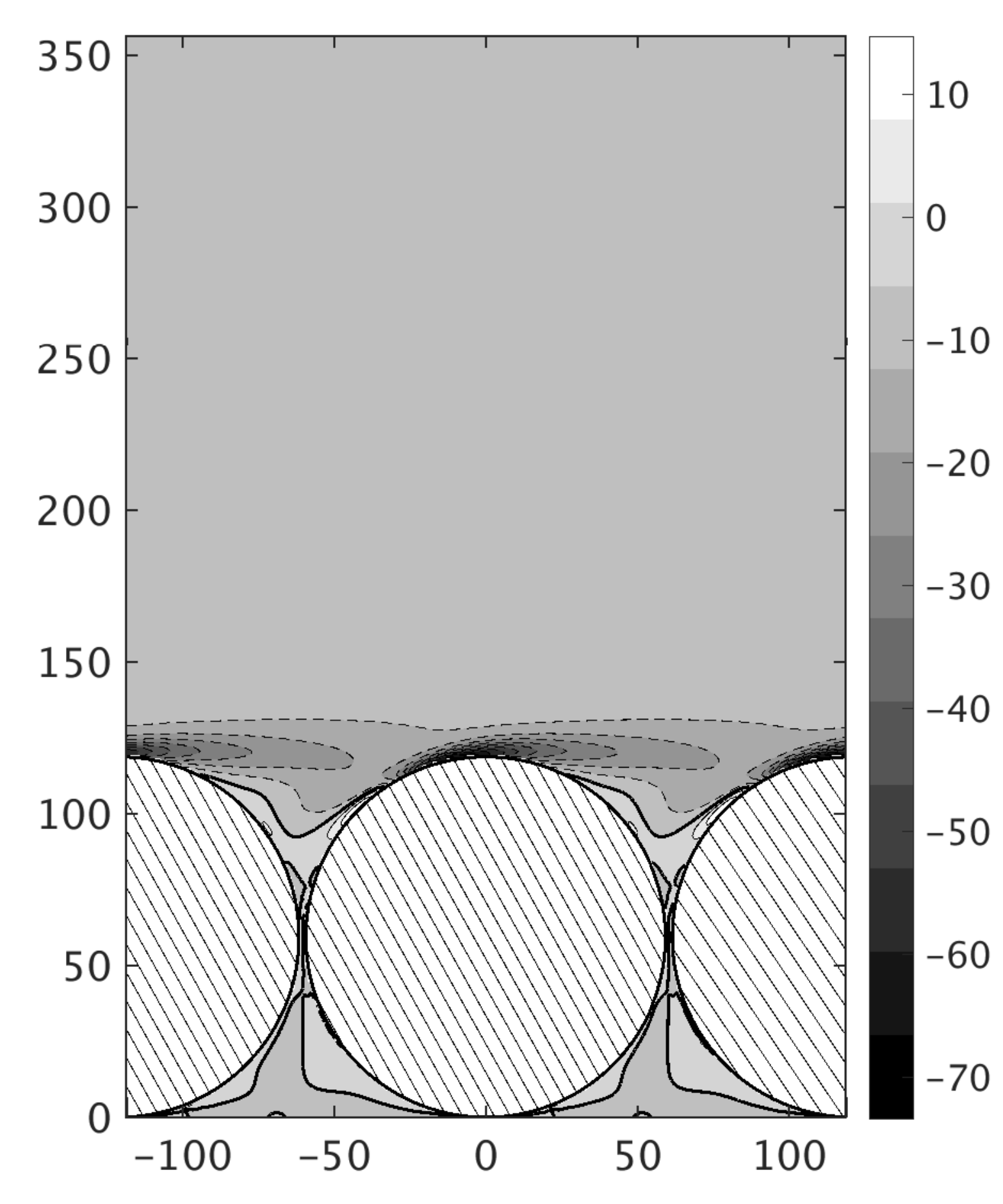}}
\put(-8,110){$y^+$}
\put(70,180){\vector(1,0){30}}
}
\put(0,0){
\put(0,0){\includegraphics[trim=0cm 0cm 0cm 0cm, clip, width=.45\textwidth]{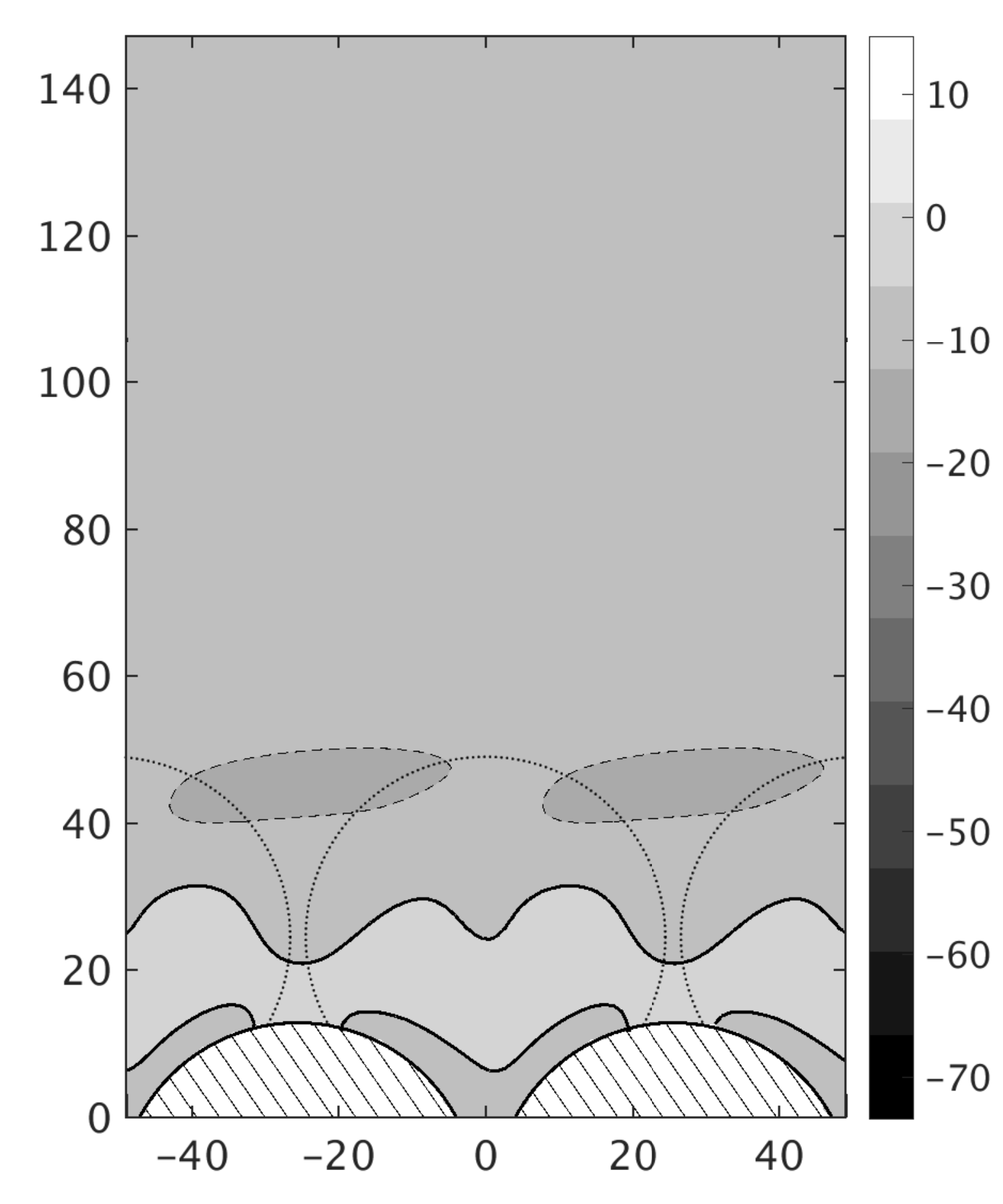}}
\put(-8,110){$y^+$}
\put(80,-4){${x}^+$}
\put(70,180){\vector(1,0){30}}
}
\put(200,0){
\put(0,0){\includegraphics[trim=0cm 0cm 0cm 0cm, clip, width=.45\textwidth]{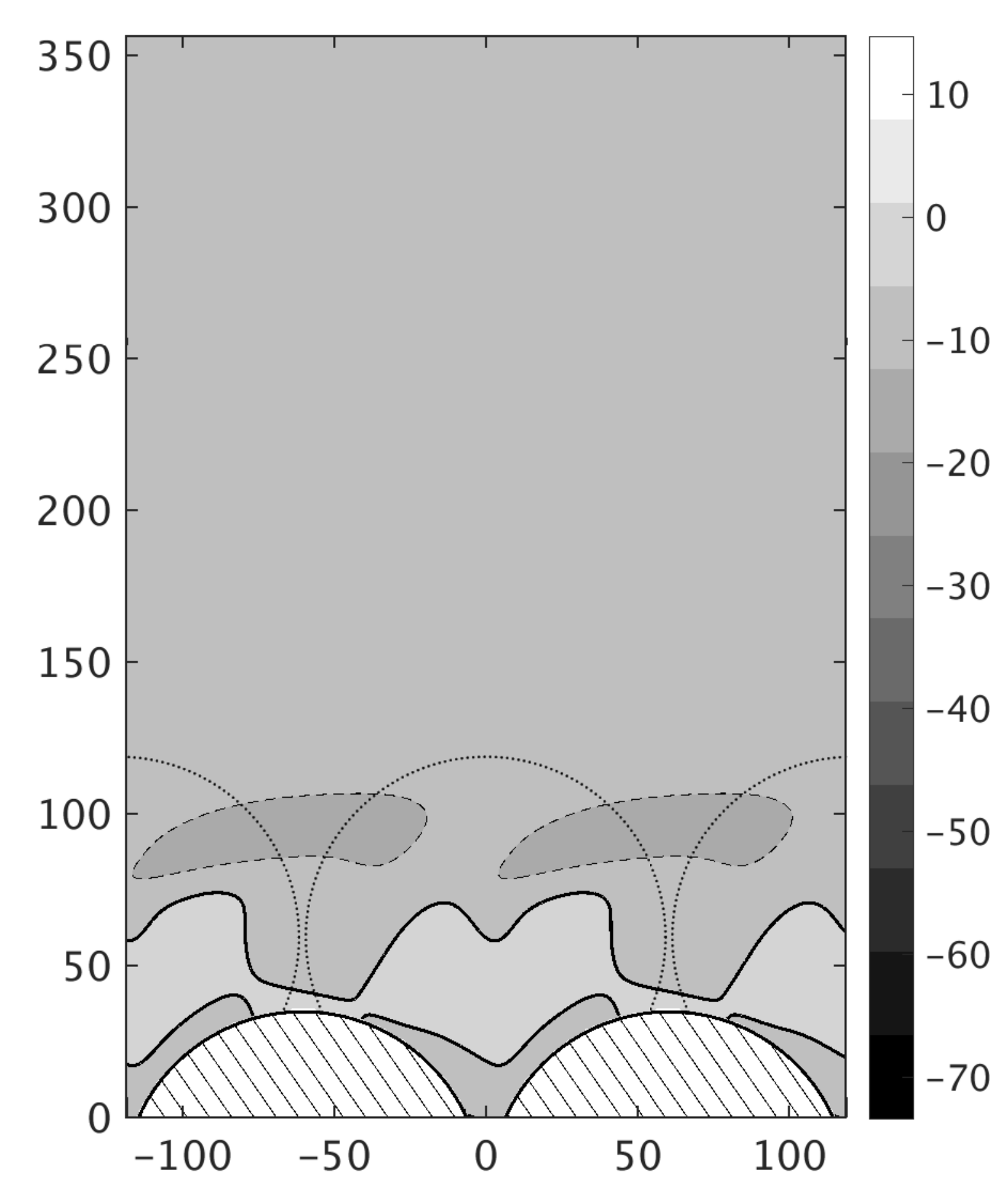}}
\put(-8,110){$y^+$}
\put(80,-4){${x}^+$}
\put(70,180){\vector(1,0){30}}
}
\put(-5,400){$a)$}
\put(195,400){$b)$}
\put(-5,190){$c)$}
\put(195,190){$d)$}
\end{picture}
\caption{Spanwise component 
  $\left\langle\overline{\omega_z}\right\rangle^+_{B}$ 
  of the sphere-box/time-averaged vorticity field visualised by filled 
  contours $(a,b)$ at the vertical plane through the center of the
  sphere $\tilde{z}=0$, $\tilde{z}$ being the spanwise coordinate in
  the frame of reference $\mathcal{B}$, $(c,d)$ at $\tilde{z}=(D+\Delta_B)/2$,
  i.e.\ at the slices $z_A$ and $z_B$ of figure~\ref{fig20},
  respectively. 
  The value $\left\langle\overline{\omega_z}\right\rangle^+_{B}=0$ is shown as a
  thick line. 
  \mut{The principal flow direction is indicated by the arrows.}
  $(a-c)$ Run D50; $(b-d)$ run D120.} 
\label{fig9}
\end{figure}
\begin{figure}
\setlength{\unitlength}{0.353mm}
\begin{picture}(0,480)(0,0)
\put(0,320){
\put(0,0){\includegraphics[trim=0cm 0cm 0cm 0cm, clip, width=.45\textwidth]{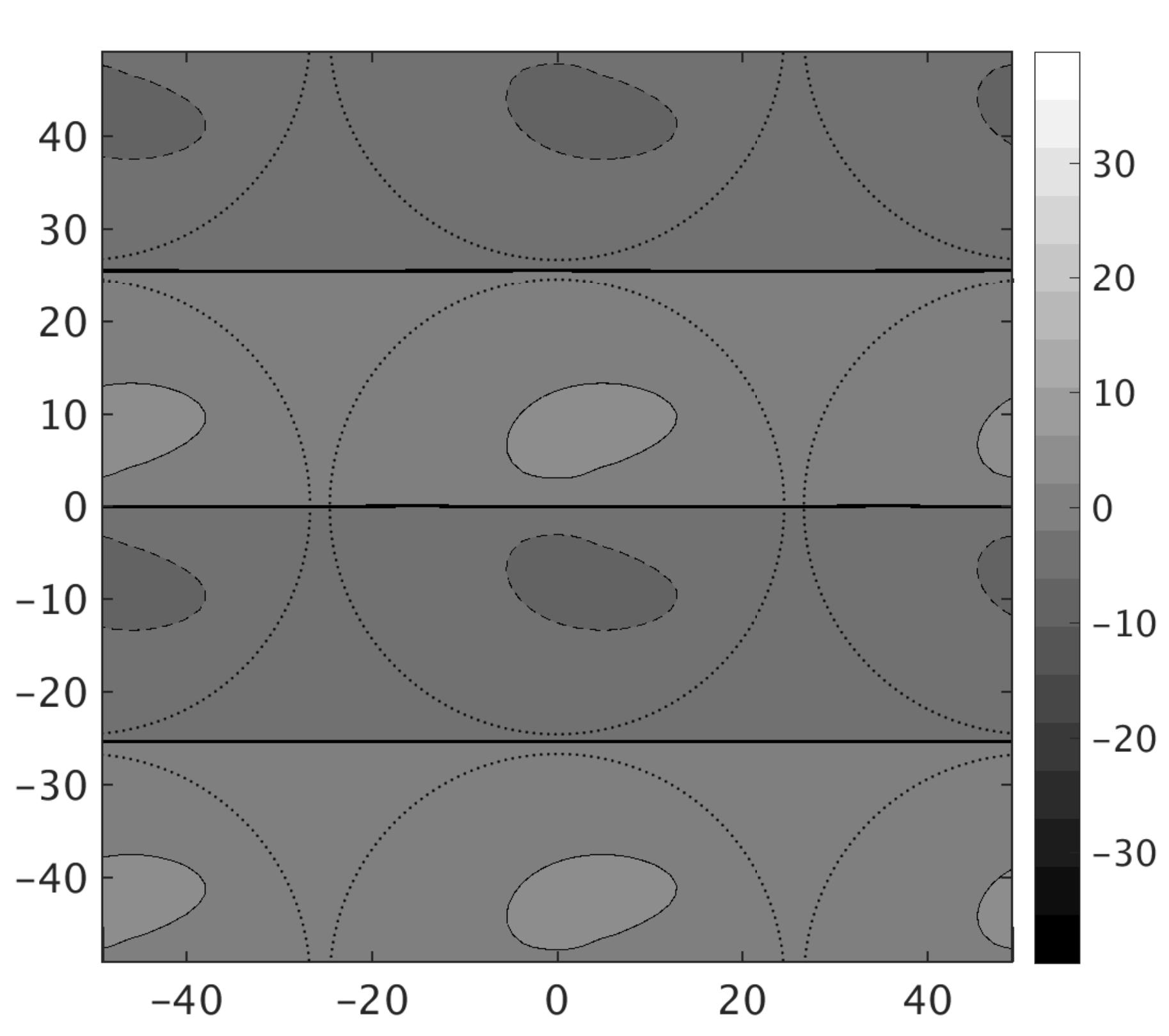}}
\put(-8,78){${z}^+$}
}
\put(200,320){
\put(0,0){\includegraphics[trim=0cm 0cm 0cm 0cm, clip, width=.45\textwidth]{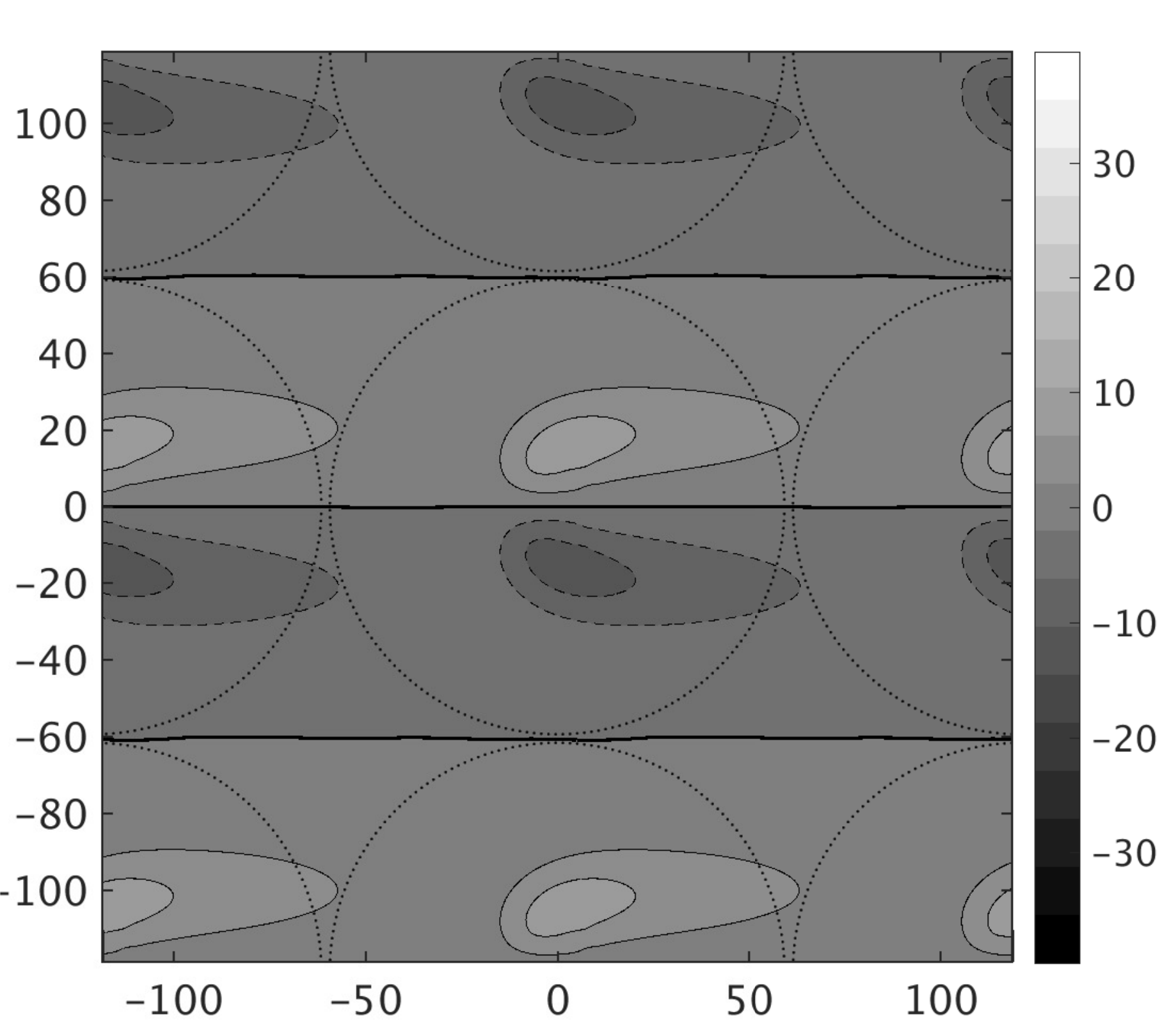}}
\put(-8,78){${z}^+$}
}
\put(0,160){
\put(0,0){\includegraphics[trim=0cm 0cm 0cm 0cm, clip, width=.45\textwidth]{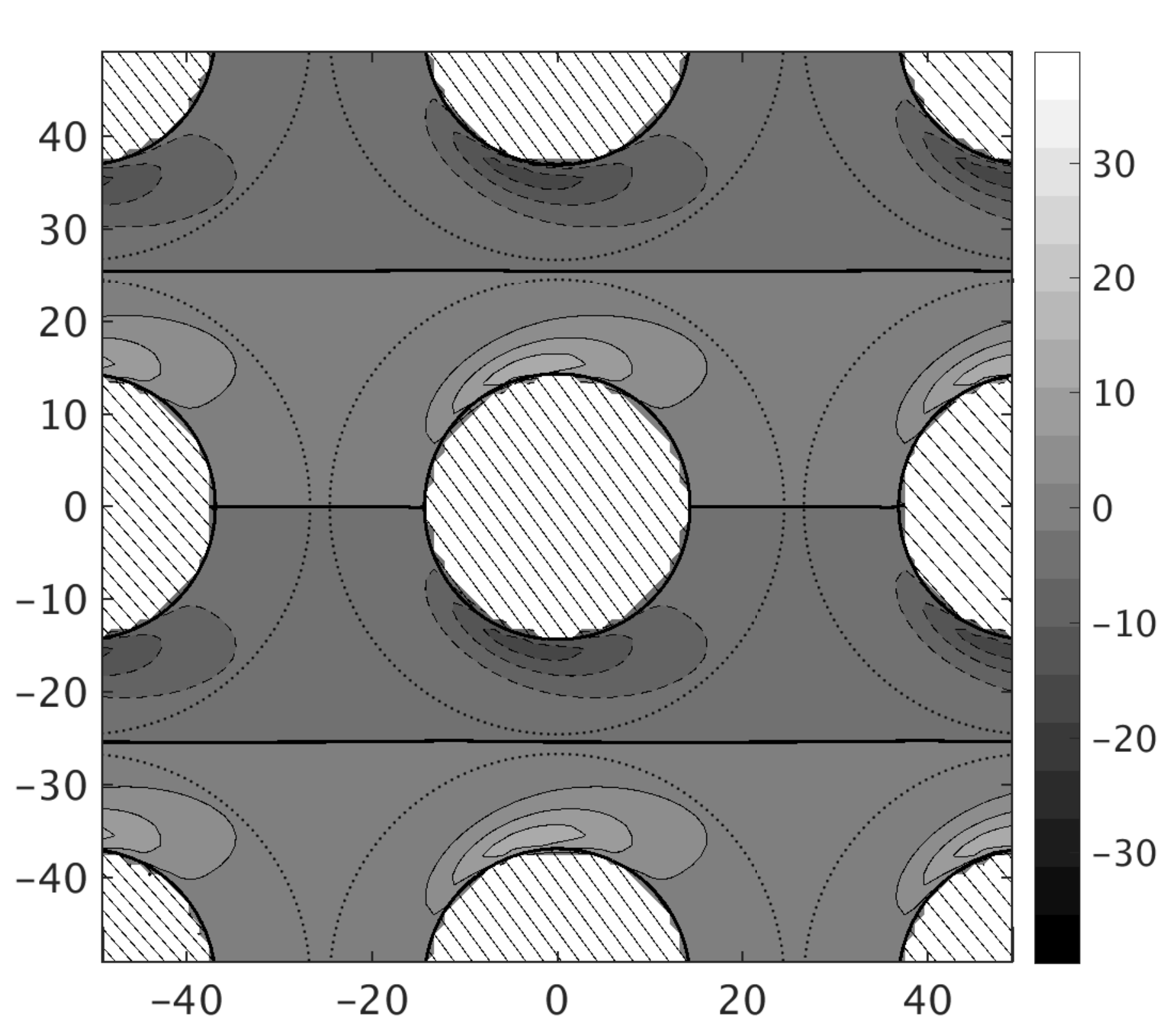}}
\put(-8,78){${z}^+$}
}
\put(200,160){
\put(0,0){\includegraphics[trim=0cm 0cm 0cm 0cm, clip, width=.45\textwidth]{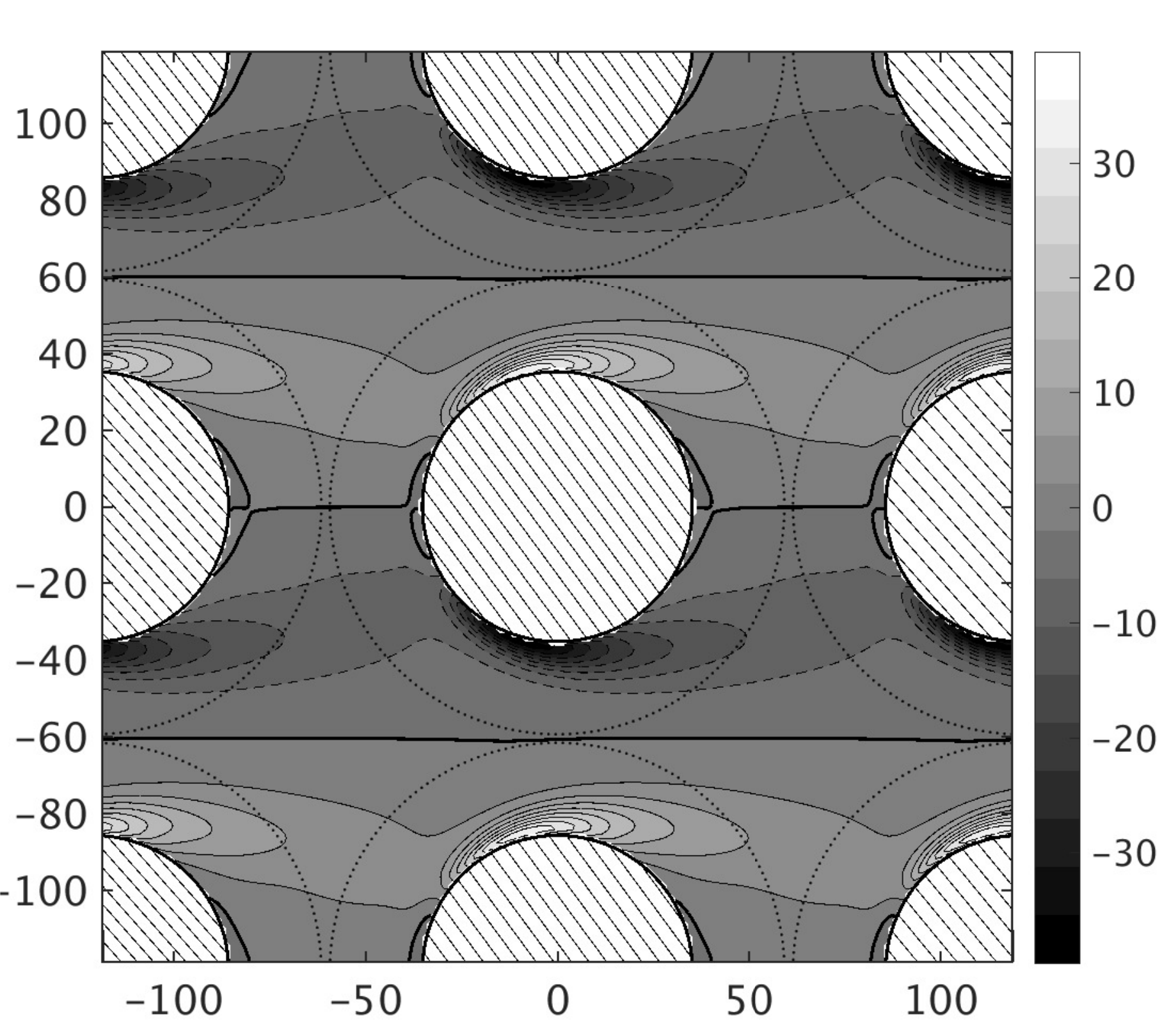}}
\put(-8,78){${z}^+$}
}
\put(0,0){
\put(0,0){\includegraphics[trim=0cm 0cm 0cm 0cm, clip, width=.45\textwidth]{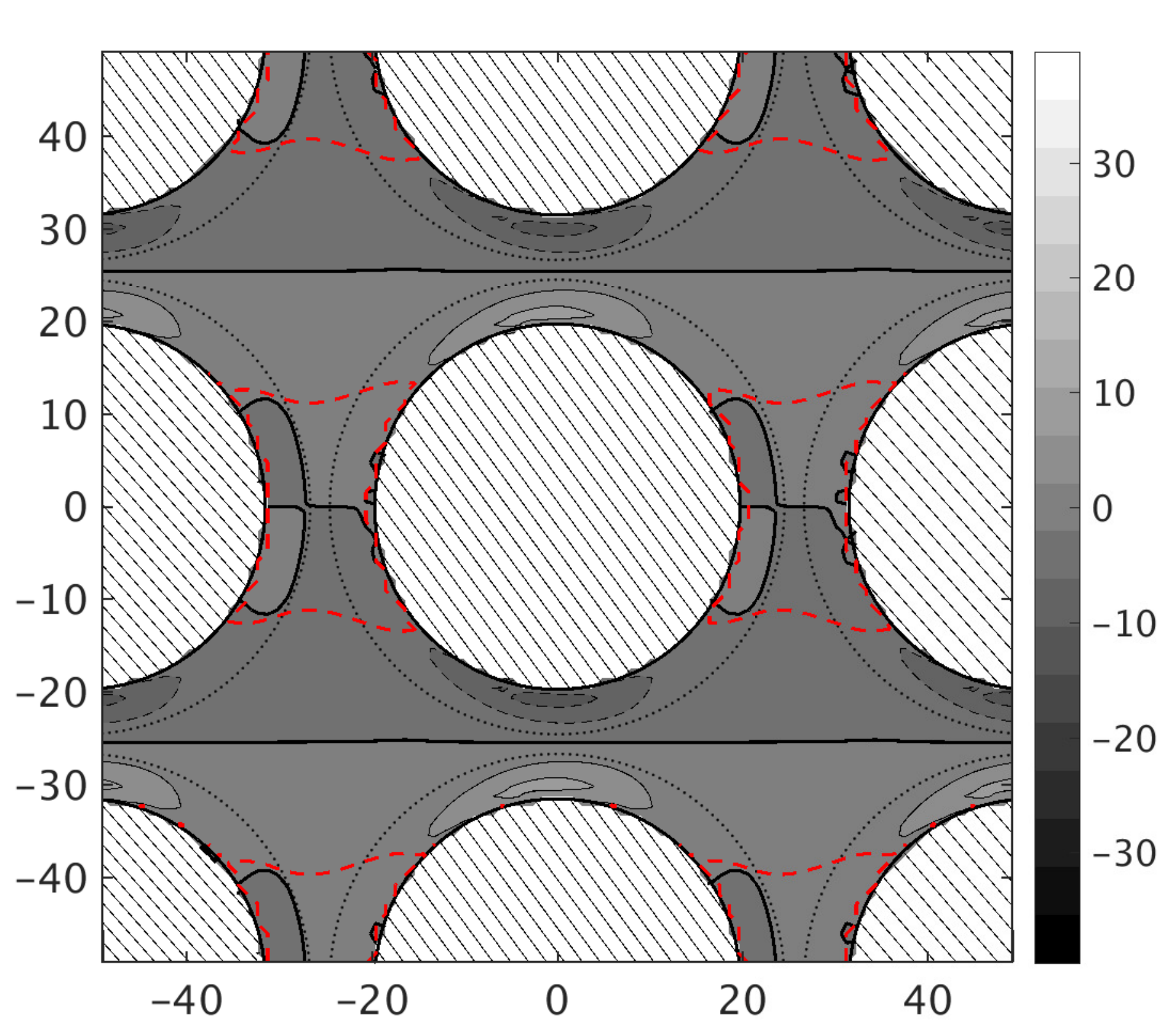}}
\put(-8,78){$z^+$}
\put(80,-7){${x}^+$}
}
\put(200,0){
\put(0,0){\includegraphics[trim=0cm 0cm 0cm 0cm, clip, width=.45\textwidth]{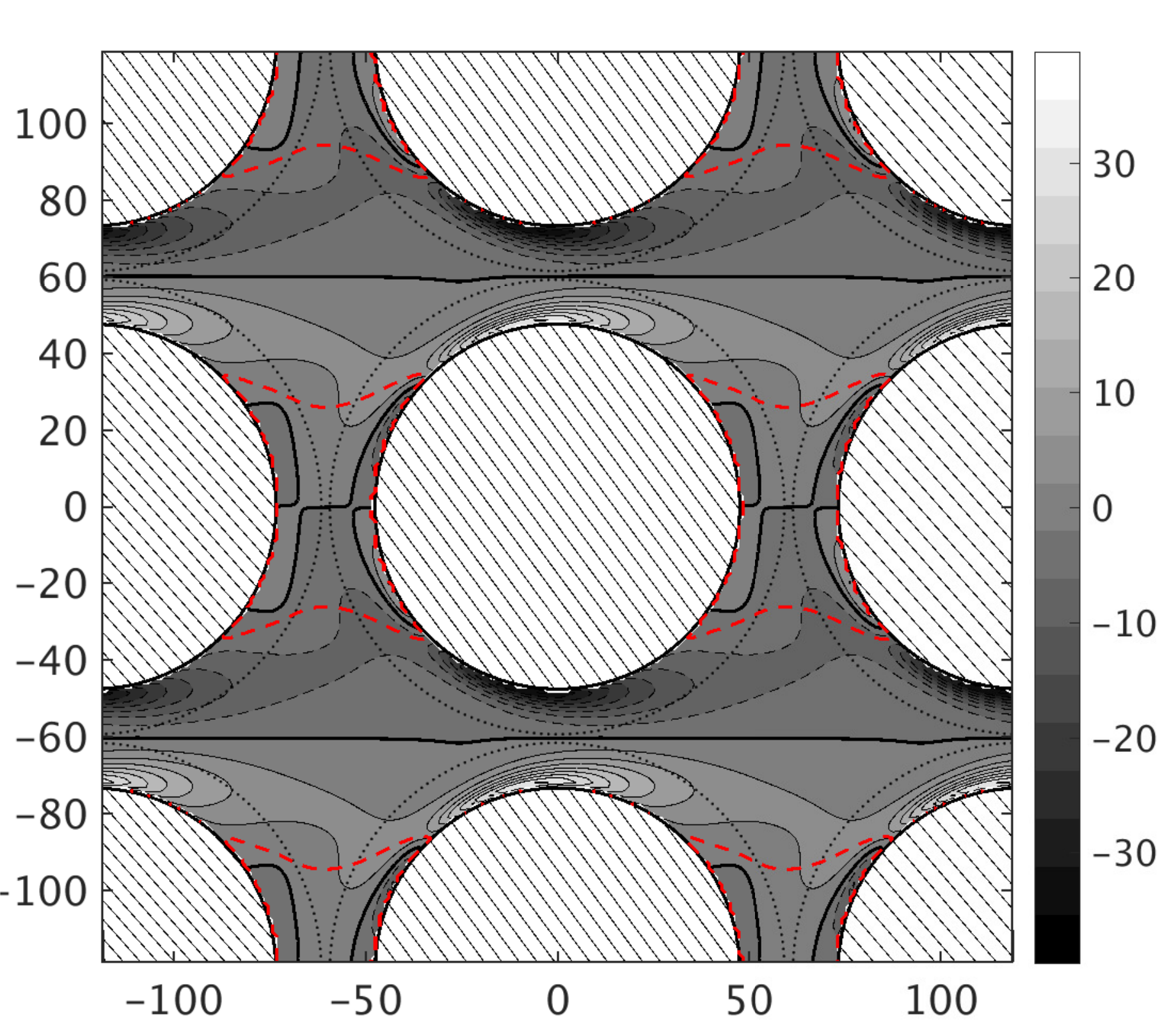}}
\put(-8,78){$z^+$}
\put(80,-7){${x}^+$}
}
\put(-5,465){$a)$}
\put(195,465){$b)$}
\put(-5,305){$c)$}
\put(195,305){$d)$}
\put(-5,145){$e)$}
\put(195,145){$f)$}
\end{picture}
\caption{Wall-normal component
  $\left\langle\overline{\omega_y}\right\rangle^+_{B}$ 
  of the sphere-box/time-averaged vorticity field visualised by filled
  contours $(a,b)$ at the wall-parallel plane $y=D+dx$ just over the
  crest of the spheres, $(c,d)$ at $y=0.9D$ and $(e,f)$ at $y=y_0$,
  i.e.\ at the slices $y_A$, $y_B$ and $y_C$ of figure~\ref{fig20},
  respectively. 
  The value $\left\langle\overline{\omega_y}\right\rangle^+_{B}=0$ 
  is shown as a thick line, 
  while, in panels e and f, red broken contour lines indicate
  $\left\langle \overline{u}\right\rangle^+_{B}=0$.
  \mut{The principal flow direction is from left to right.}
  $(a,c,e)$ Run D50; $(b,d,f)$ run D120.} 
\label{fig10}
\end{figure}
%

The vorticity related to the average flow field, 
$\left\langle\overline{\boldsymbol{\omega}}\right\rangle^+_{B}$, 
is shown in
  figures~\ref{fig8}-\ref{fig10} in planes which are orthogonal to the
  respective vorticity components, as sketched in figure~\ref{fig20}.  
  For both simulations D50 and D120, the distribution of the
  streamwise vorticity component, $\left\langle\overline{\omega_x}\right\rangle^+_{B}$, 
  in figure~\ref{fig8} shows the presence of four 
  streamwise-oriented vortical structures 
  per streamwise-orientated inter-particle gap. 
  In figure~\ref{fig8}a,b, which corresponds to a plane orthogonal to the $x$-axis and 
  crossing the center of the spheres 
  \mut{
     (referred to as plane $x=x_A$ in figure \ref{fig20}),
  }
  these structures can be seen as thin sheets. 
  Then, as we move to the next cross-section further downstream
  (figure~\ref{fig8}c,d
  \mut{
    related to the plane $x=x_B$ of figure \ref{fig20}),
  }
  these structures remain practically attached
  to the spheres while the upper pair weakens and the lower one
  intensifies. 
  Finally, 
  at the vertical plane passing through the 
  mid-plane 
  between 
  two adjacent 
  spheres
  (figure~\ref{fig8}e,f
  \mut{
    related to the plane $x=x_C$ of figure \ref{fig20}),
  } 
  the vortical structures still extend along the shear layer which forms
  over the recirculation region in the ``wake'' of the reference
  sphere. Here we observe that while in case D50 the upper pair of
  vortices practically disappears, it is still present in case D120. 
  It is also noteworthy that in the fully-rough case, 
  another quadruplet
  of streamwise-vorticity 
  structures similar to that previously described, but much weaker and
  of opposite sign, appears over the wall-mounted spherical caps 
  (figure~\ref{fig8}f). The reversal of vorticity direction
  confirms the picture previously given for the recirculation cells
  related to the secondary flow (cf.\ figure~\ref{fig7}$b$). 

  The spanwise vorticity component, $\left\langle\overline{\omega_z}\right\rangle^+_{B}$, 
  is the most intense, 
  since it is associated with the primary mean flow. 
  The most intense regions of $\left\langle\overline{\omega_z}\right\rangle^+_{B}$ 
  are located at and around the top of the spheres (figure~\ref{fig9}a,b 
  \mut{
    corresponding to the plane $z=z_A$ of figure \ref{fig20}a);
  } 
  this high-vorticity zone extends clearly much further downstream along the shear 
  layer in the fully-rough case than at lower roughness Reynolds number. 
  On the other hand, very small levels of 
  $\left\langle\overline{\omega_z}\right\rangle^+_{B}$ 
  are attained in the center plane between the spheres (figure~\ref{fig9}c,d, 
  \mut{
    plane $z=z_B$ of figure \ref{fig20}a).
  }  

  Finally, let us turn to the wall-normal component 
  $\left\langle\overline{\omega_y}\right\rangle^+_{B}$ which is
  shown in figure~\ref{fig10} at the wall-parallel planes indicated
  in figure~\ref{fig20}$b$. 
  Due to the deflection of the flow around the spheres, two regions characterised 
  by high absolute values of the wall-normal vorticity component appear 
  (with opposite sign) on either side of the upper part of the spheres. 
  Again, the principal difference between the transitionally-rough and 
  the fully-rough case is the extent to which the vorticity patches emanating 
  from the spheres reach downstream.
  In the latter case this extent is significantly larger; 
  e.g. at a wall-normal distance equal to the virtual origin (figure~\ref{fig10}d), 
  the zones with intense values of the mean wall-normal vorticity reach all the way 
  to the adjacent sphere. 

  The
  statistical 
  footprint of the interaction of
  roughness-induced 
  high-vorticity structures with the
  surface of the spheres can be detected 
  in the distribution of the
  stress on the roughness elements which is
  discussed 
  in section \ref{sec2}.
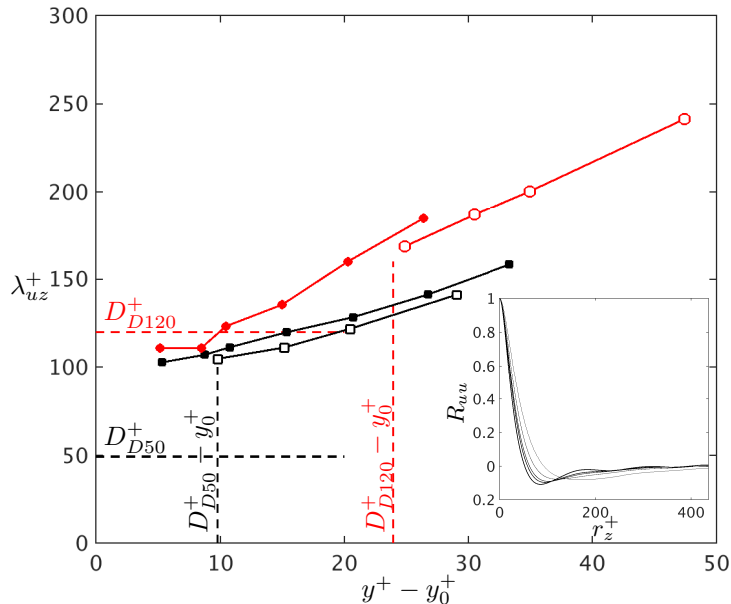
\begin{figure}
\setlength{\unitlength}{0.353mm}
\begin{picture}(0,220)(0,0)
\put(50,0){
\put(-4,-2){\includegraphics[trim=0cm 0cm 0cm 0cm, clip, height=.59\textwidth]{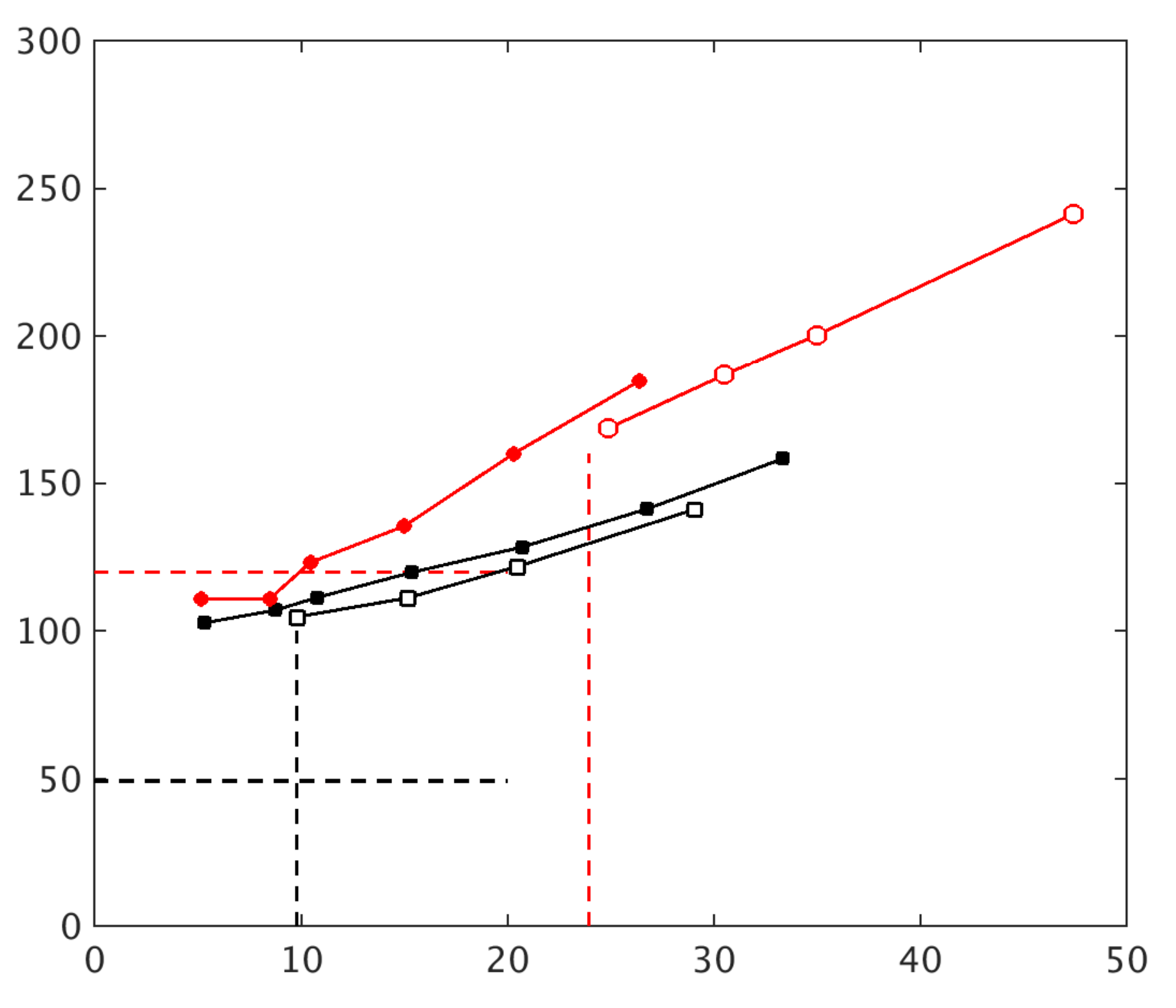}}
\put(157.4,26){\includegraphics[trim=0cm 0cm .2cm .3cm, clip, width=.23\textwidth]{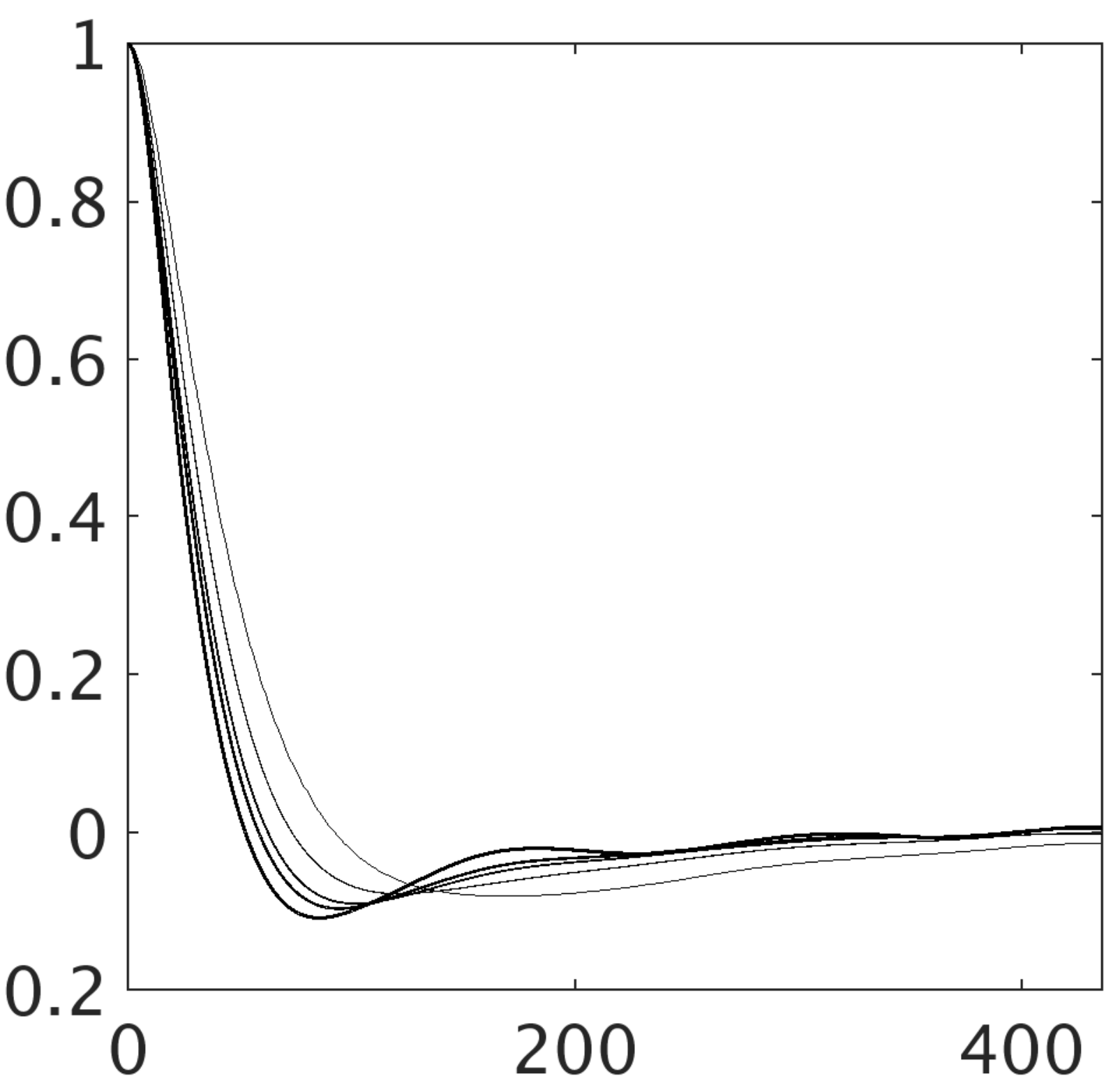}}
\put(116,-4){$y^+-y^+_0$}
\put(-15,110){$\lambda^+_{uz}$}
\put(202,19){\small $r_z^+$}
\put(148,61){\small \rotatebox{90}{$R_{uu}$}}
\put(20,100){\color{red!}$D^+_{D120}$}
\put(116,20){\rotatebox{90}{\color{red!}$D^+_{D120}-y^+_0$}}
\put(20,53){$D^+_{D50}$}
\put(50,20){\rotatebox{90}{$D^+_{D50}-y^+_0$}}
}
\end{picture}
\caption{
Distance between two adjacent low-(or high-)speed streaks $\lambda^+_{uz}$ 
plotted as a function of the distance form the virtual wall, for the runs D50 
(black line, empty circles) and D120 (red line, empty circles).
The first point is located just above the crest of the spheres.
$\lambda^+_{uz}$ was estimated 
as twice the location of the minimum of the correlation function
$R_{uu}(r_z)$ of the fluctuations $u'$ of the streamwise velocity
component (small inset). 
Solid lines with filled symbols show the trend of $\lambda^+_{uz}$ for the respective 
smooth-wall cases S1 and S2.
Vertical broken lines indicate the distance of the crest 
of the spheres from $y^+_0$, while horizontal broken lines indicate $D^+$.
}
\label{fig22}
\end{figure}
\begin{figure}
\setlength{\unitlength}{0.353mm}
\begin{picture}(0,170)(0,0)
\put(0,2){\includegraphics[trim=0cm 0cm 0cm 0cm, clip, width=.5\textwidth]{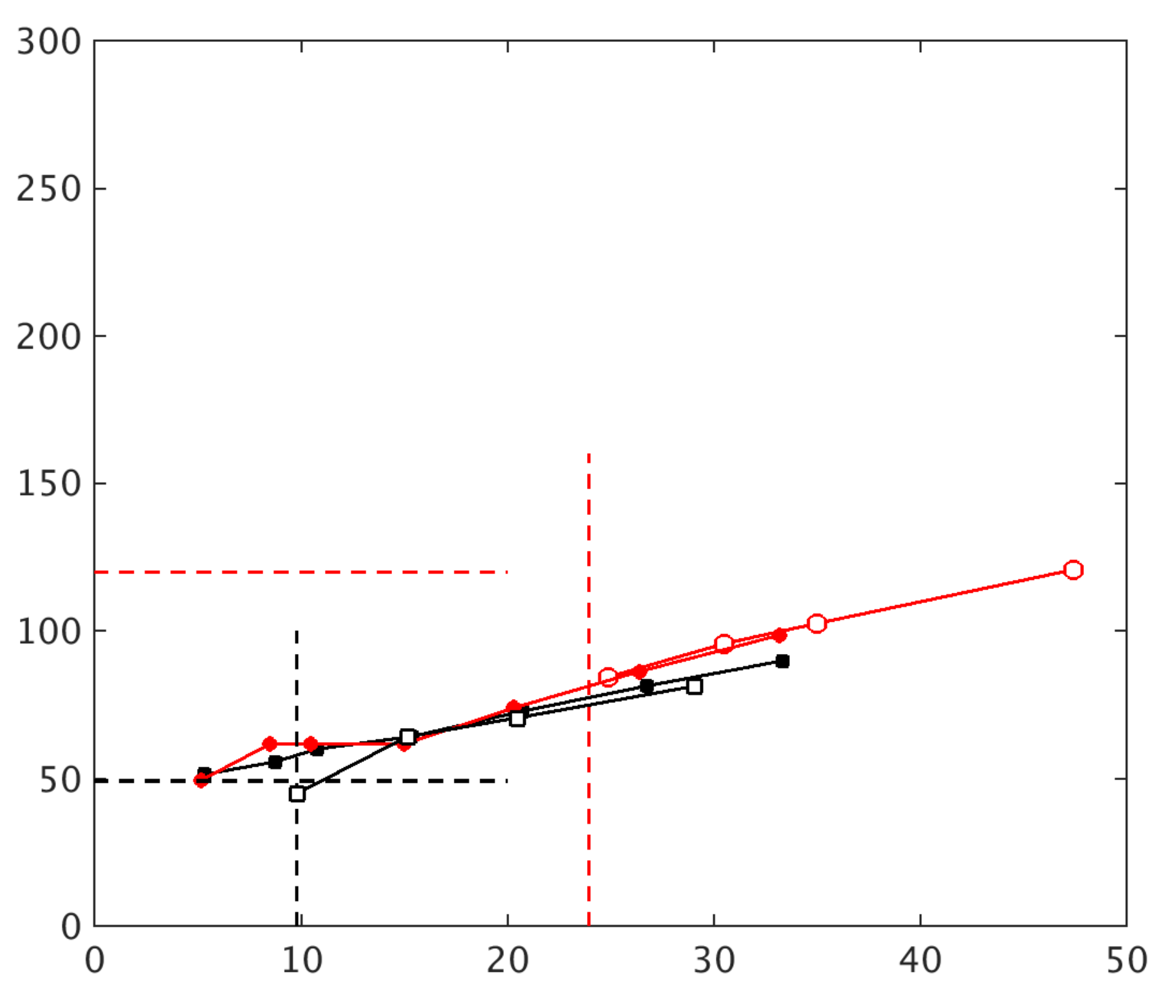}}
\put(20,148){$(a)$}
\put(84,-2){$y^+-y^+_0$}
\put(-10,80){$\lambda^+_{vz}$}
\put(60,77){\scriptsize\color{red!}$D^+_{D120}$}
\put(84,55){\scriptsize\rotatebox{90}{\color{red!}$D^+_{D120}-y^+_0$}}
\put(60,31){\scriptsize $D^+_{D50}$}
\put(37,50){\scriptsize \rotatebox{90}{$D^+_{D50}-y^+_0$}}
\put(200,0){
\put(0,2){\includegraphics[trim=0cm 0cm 0cm 0cm, clip, width=.5\textwidth]{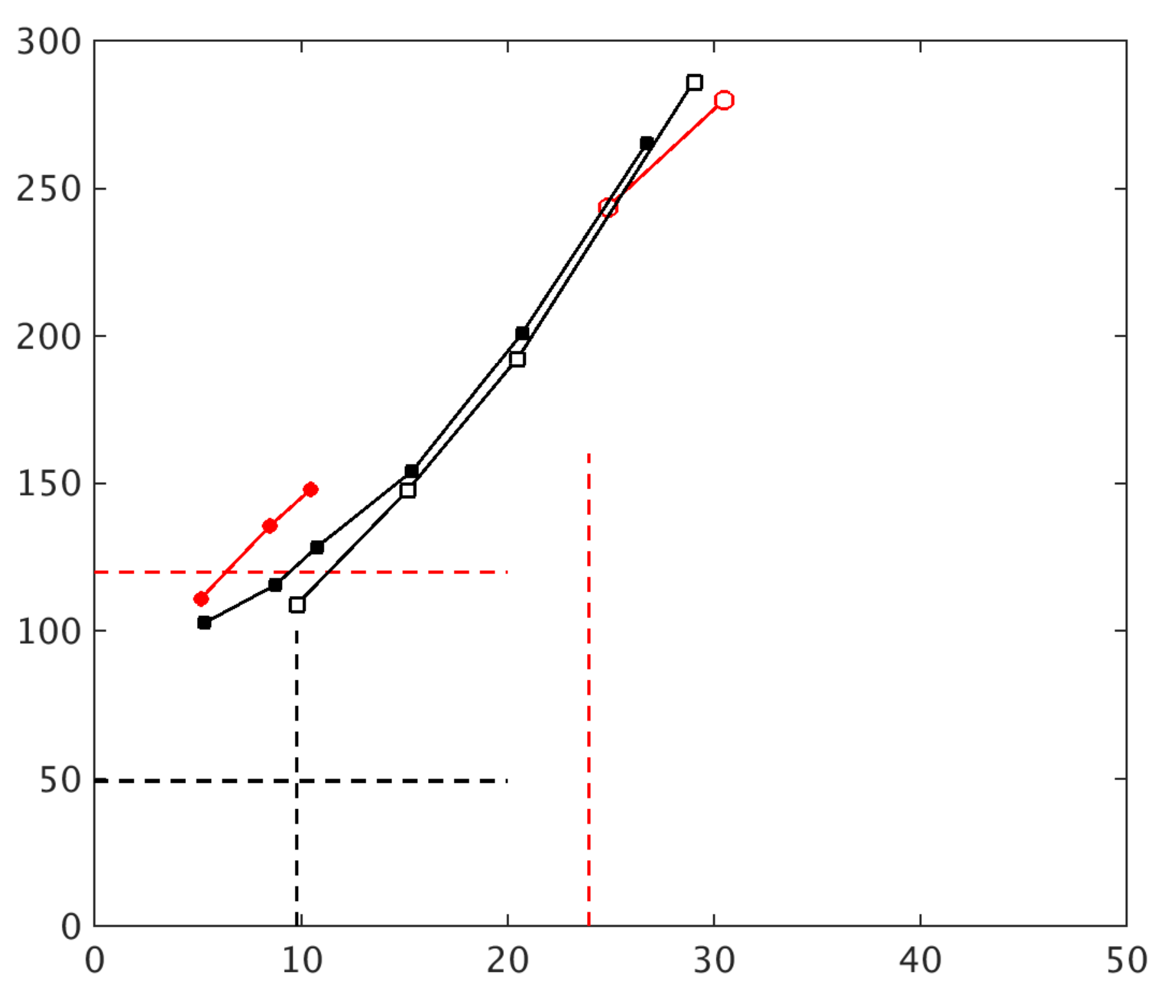}}
\put(20,148){$(b)$}
\put(84,-2){$y^+-y^+_0$}
\put(-10,80){$\lambda^+_{wz}$}
\put(60,65){\scriptsize\color{red!}$D^+_{D120}$}
\put(84,19){\scriptsize\rotatebox{90}{\color{red!}$D^+_{D120}-y^+_0$}}
\put(60,43){\scriptsize $D^+_{D50}$}
\put(37,19){\scriptsize \rotatebox{90}{$D^+_{D50}-y^+_0$}}
}
\end{picture}
\caption{
Panels $(a)$ and $(b)$ show the trends of $\lambda^+_{vz}$ and $\lambda^+_{wz}$ 
obtained from $R_{vv}(z)$ and $R_{ww}(z)$, 
in the same manner as $\lambda^+_{uz}$ in figure~\ref{fig22}. 
Symbols and lines are the same as those indicated in the caption of figure~\ref{fig22}.
}
\label{fig22a}
\end{figure}
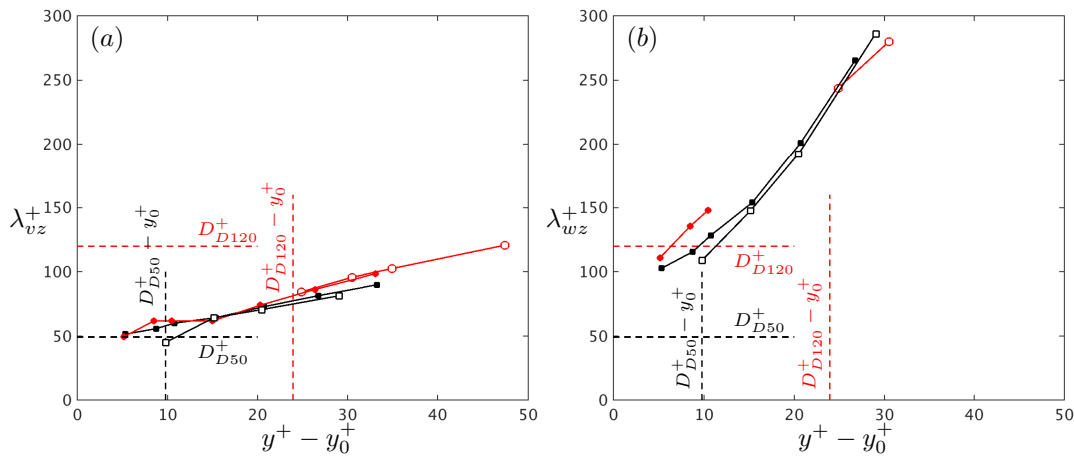

\begin{figure}
\setlength{\unitlength}{0.353mm}
\begin{picture}(0,215)(0,0)
\put(10,0){
\put(0,0){\includegraphics[trim=3cm 0cm 6cm 1cm, clip, width=.95\textwidth]{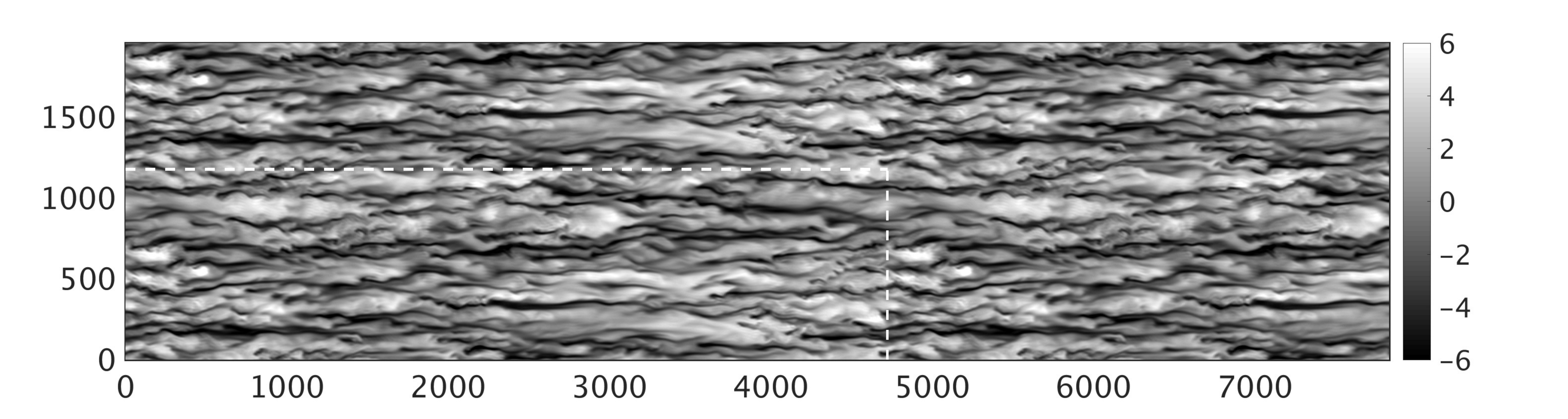}}
\put(-10,55){$z^+$}
\put(180,-6){$x^+$}
\put(-5,95){$b)$}
\put(362,52){$\displaystyle\frac{u'}{u_\tau}$ }
}
\put(10,110){
\put(0,0){\includegraphics[trim=3cm 0cm 6cm 1cm, clip, width=.95\textwidth]{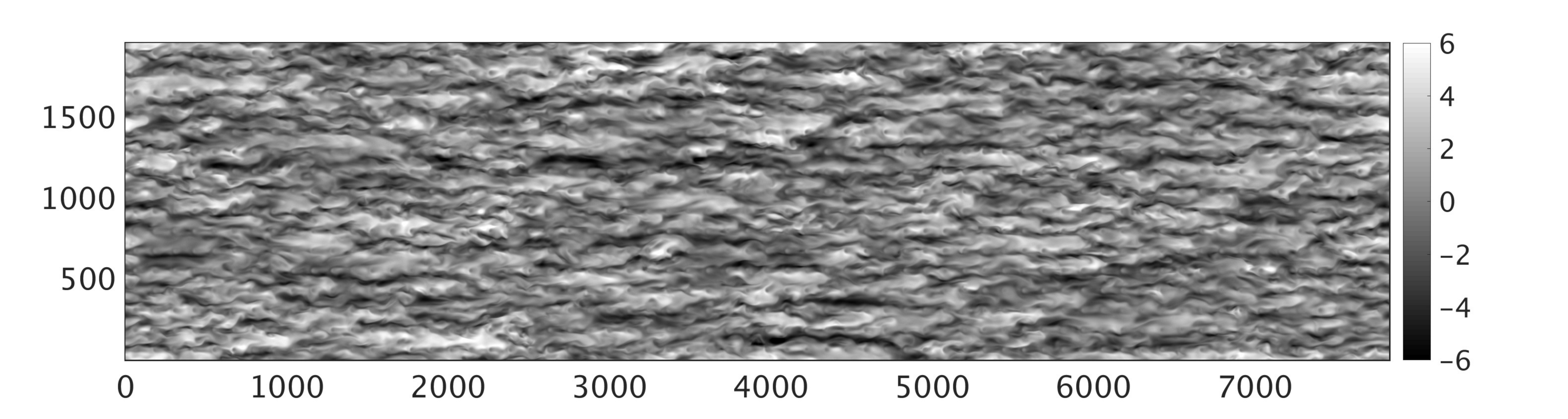}}
\put(-10,55){$z^+$}
\put(180,-6){$x^+$}
\put(-5,95){$a)$}
\put(362,52){$\displaystyle\frac{u'}{u_\tau}$ }
}
\end{picture}
\caption{
  Instantaneous visualisation 
  of velocity fluctuations $u'$ for
  $(a)$ the simulation D120 at $y^+-y^+_0=20$ and 
  $(b)$ the smooth-wall simulation  with 
  $\mut{Re_{bH}}=6900$ at $y^+=20$. 
  White broken lines delimit the original computational
  domain, extended periodically in the streamwise and spanwise
  direction up to the dimensions of the domain of the simulation D120
  in order to facilitate the visual comparison. 
  \mut{
    The principal flow direction is from left to right.
  }
}
\label{fig30}
\end{figure}
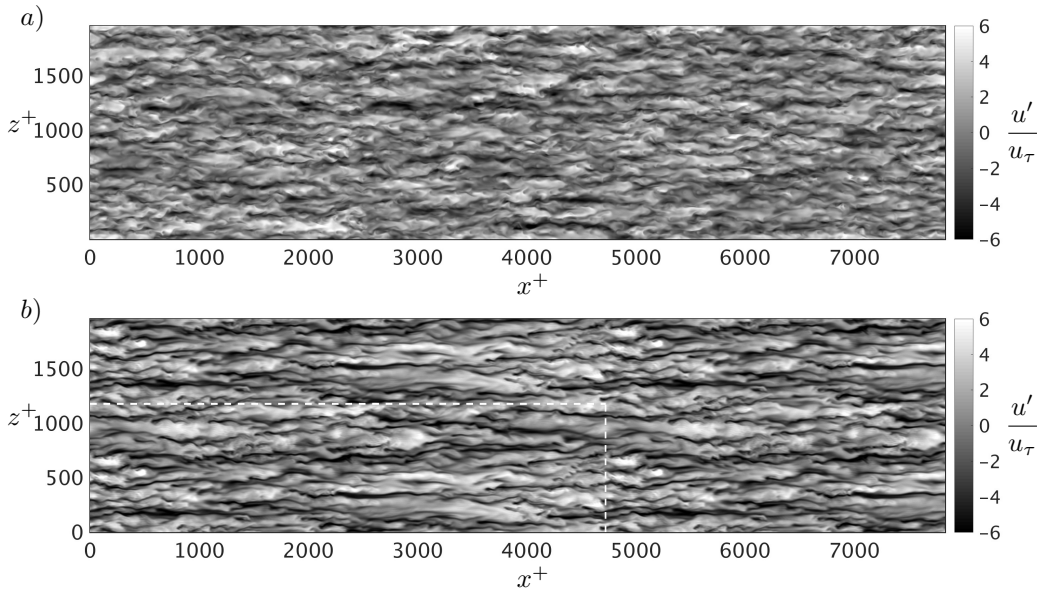

With the aim of determining the influence of the roughness on the
turbulence structure, the distribution of the 
temporal fluctuations, 
($\mathbf{u}^\prime$), 
is investigated both for the transitionally- and 
fully-rough simulations.
Typically, the presence of low-/high-speed streaks is associated 
with that of streamwise vortices. 
Indeed, the characteristics of velocity streaks are an effective 
indicator of the structure of turbulence in wall bounded flows.
It is well known that, in the smooth-wall case, the distance 
in the spanwise direction between two adjacent low- or high-speed 
streaks is approximately $\lambda_{uz}^+=100$ immediately over the 
viscous sublayer ($y^+\sim 5$), and grows at a constant rate with 
the distance from the wall until $y^+\sim 30$ \citep{KMM1987}.
The value of $\lambda_{uz}^+$ can be estimated as twice the separation
distance at which the spanwise two-point correlation of the streamwise
velocity fluctuations, $R_{uu}$, 
attains the maximum negative value (see the small panel of figure~\ref{fig22}), 
as long as the minimum is negative.
On the basis of this definition, \citet{KMM1987} extended the idea of 
a spatial scale $\lambda^+_{uz}$ less intuitively to the other components 
of the velocity and found that $\lambda^+_{wz}\simeq\lambda^+_{uz}$ and 
$\lambda^+_{vz}\simeq 2\lambda^+_{uz}$.
Presently, a similar approach was adopted for both the smooth- and rough-wall 
simulations (see figure~\ref{fig22a}).
As shown in figure~\ref{fig22}, figure~\ref{fig22a}a and figure~\ref{fig22a}b, respectively, 
the values of $\lambda^+_{uz}$, $\lambda^+_{vz}$ and $\lambda^+_{wz}$, at 
$y^+=D^+$ for the simulations D50 and D120 
were \mut{roughly} the same as those attained for the %
\pbt{%
  respective smooth-wall cases %
}%
at the
same distance from the virtual wall, i.e.\ at $y^+_{smooth}=(D^+-y_0^+)_{rough}$. 
\mut{
  Figure~\ref{fig30}
  shows the distribution of instantaneous fluctuations of the
  streamwise velocity, ${u}'$, 
  in the wall-parallel plane located at $y^+\simeq D^+$ for the run D120
  (panel $(a)$) and  at $y^+\simeq D^+-y^+_0$ for the smooth wall
  simulation at the same bulk Reynolds number (panel $(b)$). 
  Visually, the length and width of low-speed streaks in those planes
  is remarkably similar, as is their intensity.
}
The value of $\lambda^{+}_{wz}$, extrapolated at 
$y^+_{smooth}=(D^+-y^+_0)_{D120}$ 
for the smooth-wall simulation at $\mut{Re_{bH}}=6900$, is almost equal
to that of $\lambda^{+}_{wz}$ obtained for the fully-rough simulation
D120 (see figure~\ref{fig22}c). %
Hence, the relationships observed by \citet{KMM1987} between 
$\lambda^+_{uz}$ and $\lambda^+_{vz}$ and between $\lambda^+_{uz}$ and 
$\lambda^+_{wz}$ still hold both in the transitionally- and fully-rough 
regimes for the present shape and arrangement of the roughness elements, 
and an increase of the Reynolds number $D^+$ results in the increase of 
the slope of $\lambda^+_{uz}$ as a function of $y^+$. 
However, the methodology described above does not allow us to estimate 
the value of $\lambda^+_{z}$ below the crest of roughness elements where 
velocity fluctuations are not defined continuously on wall-parallel planes.
%
%

It emerges from this picture that the effect of roughness is not to change sharply the structure of turbulence above the roughness elements, but to select a particular scale related to the roughness which then appears 
more pronounced 
than in an equivalent turbulent open-channel flow developing over a smooth wall at the same bulk Reynolds number.
In the fully-rough simulation, the distance between low- and
high-speed streaks, $\lambda^+_{uz}/2$, at 
$y^+=D^+$, 
was equal to $85$.
\subsection{Force and torque acting on the roughness elements}\label{sec2}
The stress acting on each roughness element was calculated on the 
basis of the volume forces associated with the immersed boundary
method \citep{chan2011}. 
The square arrangement of the spheres allows us to refer to them with
the indices $(i,j)$ as in a two-dimensional array, where
$i=1,\ldots,n_c$ and $j=1,\ldots,n_r$ enumerate the 
streamwise indices (columns) and the spanwise indices 
(rows) of the array, respectively. 
For the present simulations $n_c=64$ and $n_r=16$. 
  We apply the sphere-box/time-average operator described in
  section~\ref{sec1}   
  to the total stress tensor $\boldsymbol{\tau}$ 
  (containing both the viscous and pressure contributions, as defined
  in equation~\ref{equ-define-total-stress-tensor}).
  At the sphere surface, 
  the projection of the average stress tensor upon the outward facing
  normal vector $\widetilde{\boldsymbol{n}}$ yields the average stress
  vector, viz.
  \begin{equation}\label{equ-define-stress-tensor}
    \widetilde{\boldsymbol{\tau}}_n
    =
    \widetilde{\left\langle\overline{\boldsymbol{\tau}}
    \right\rangle}_B \cdot\widetilde{\boldsymbol{n}}
    \,.
  \end{equation}
Then the sphere-box/time-averaged hydrodynamic force is defined as:
\begin{equation}
\boldsymbol{F}=\displaystyle\int_{\widetilde{\mathcal{S}}} \widetilde{\boldsymbol{\tau}}_n\, dS
\label{eq5a}
\end{equation}
while the time-average of the force acting on the $(i,j)$-th sphere is: 
\begin{equation}
\boldsymbol{F}^{(i,j)}=\displaystyle\int_{\mathcal{S}^{(i,j)}}\overline{\boldsymbol{\tau}} \cdot\boldsymbol{n}^{(i,j)}\, dS\:\: ,
\label{eq5b}
\end{equation}
where $\widetilde{\mathcal{S}}$ denotes the surface of the sphere contained in 
the sphere-box $\mathcal{B}$ (see figure~\ref{sketch2} and the definition 
\eqref{eqA5} in the Appendix), $\mathcal{S}^{(i,j)}$ indicates the surface 
of the $(i,j)$-th sphere centered in $\boldsymbol{x}^{(i,j)}_c$ and $\boldsymbol{n}$ 
the surface-normal unit vector. 

The fluctuations $\boldsymbol{F}''$ 
around $\boldsymbol{F}$ can be defined in a similar way as those of
the velocity given in (\ref{eq3b}). 
The components $(F_x,F_y,F_z)$ of the force $\boldsymbol{F}$ are the mean 
drag, vertical lift and lateral lift forces acting on roughness
elements, respectively, which can be normalised by the reference force
$F_R=\varrho u_\tau^2 L^2_B$. 
\setlength{\tabcolsep}{0.35em} 
{\renewcommand{\arraystretch}{1.5}%
\begin{table}
    \begin{center}
    \begin{tabular}{l l l l l l l l l l}
    Run & $ F_x/F_R $ & $ F_y/F_R $ & $ F_z/F_R $ &
          $ \sigma_{F_x}/F_R $ & $ \sigma_{F_y}/F_R $  & $ \sigma_{F_z}/F_R $ & $ \alpha_F $ &
          \mut{$ F_x^{(\nu)}/F_x $} & \mut{$ F_y^{(\nu)}/F_y $} \vspace*{.05cm}
          \\

    \hline
    D50  & $1.14$ & $0.37$ & $0.0002$ & $1.31$ & $0.65$ & $1.26$ & $18.1^\circ$ &
    \mut{$0.55$} & \mut{$0.06$} \\
    D120 & $1.17$ & $0.48$ & $0.002$  & $1.57$ & $0.76$ & $1.40$ & $22.2^\circ$ & 
    \mut{$0.42$} & \mut{$0.02$} \\
    \hline
        & $ \mathcal{S}_{F_x} $ & $ \mathcal{S}_{F_y} $ & $ \mathcal{S}_{F_z} $ &
          $ \mathcal{K}_{F_x} $ & $ \mathcal{K}_{F_y} $ & $ \mathcal{K}_{F_z} $ & & &\\
    \hline
    D50  & $0.056$ & $0.27$ & -$0.014$ & $5.07$ & $5.76$ & $4.32$ & & & \\
    D120 & $0.09$  & $0.13$ & -$0.012$ & $3.64$ & $4.17$ & $3.91$ & & & \\
    \hline
    \end{tabular}
    \caption{Statistics of 
      the hydrodynamic force acting on the top-layer particles in 
      runs D50 and D120. $F_i/F_R$ denotes the normalised mean force
      component in the $x_i$-direction,
      $\alpha_F=\arctan{(\mathcal{C}_{F_y}/\mathcal{C}_{F_x})}$
      denotes the angle of the resulting force with respect to the
      $x$-axis, $\sigma_{F_i}$, $\mathcal{S}_{F_i}$ and
      $\mathcal{K}_{F_i}$ are the standard deviation, the skewness and
      the kurtosis of the $i$-th component of the force,
      respectively. 
      The reference force is defined as $F_R=\varrho u^2_{\tau}L_B^2$.
    }
    \label{tab3}
    \end{center}
\end{table}
}
The values of the mean dimensionless force components for the simulations 
D50 and D120 are provided in table~\ref{tab3} along with their second, 
third and forth moment statistics, 
\mut{
  the angle $\alpha_F$ between the average drag and lift forces, and 
  the values of the ratios of the viscous components of drag and lift 
  forces, $F_x^{(\nu)}$ and $ F_y^{(\nu)}$, to the respective total forces.
}
The mean dimensionless drag force remains essentially unchanged between the 
simulations D50 and D120 as well as the value of $\mut{U_{bh}}/u_\tau$ (see table \ref{tab1}). 
This can be explained in terms of Darcy-Weisbach's friction factor, 
since $H/D$ (which is the inverse of the relative roughness) had nearly 
the same value in the two simulations, while the values of bulk Reynolds 
number were sufficiently close and high to keep the friction factor almost constant.
Contrarily, the mean vertical lift experiences a significant increase, resulting 
in the increase of the angle $\alpha_F$. %
\citet{chan2011} observed that 
$F_y/F_R$ increased as an effect of increasing $D^+$, but
since the relative submergence $H/D$ was simultaneously changed,
they could not conclude on a pure scaling with particle Reynolds
number $D^+$. 
\mut{
  Despite the fact that the dimensionless drag remains essentially
  unchanged from D50 to D120, the pressure contribution to the mean
  drag force significantly increases (table~\ref{tab3}).
  To a lesser extent this is also true for mean wall-normal lift. 
}
The kurtosis of the three components of $\boldsymbol{F}''$ decreases for 
increasing values of $D^+$, tending towards the value $3$ expected for a 
Gaussian distribution. %
Indeed, fluctuations of small length scale, at large values of $D^+$, are 
filtered out through the surface integral, thereby reducing the probability 
of values far from the mean and consequently the value of $\mathcal{K}$.

The average torque acting on the roughness elements is defined as follows: 
\begin{equation}
\boldsymbol{T}=\displaystyle\int_{\widetilde{\mathcal{S}}}\widetilde{\boldsymbol{r}_c}\times \widetilde{\boldsymbol{\tau}}_n\, dS
\label{eq6}
\end{equation}

where the distance vector is denoted by 
$\widetilde{\boldsymbol{r}_c}=(\widetilde{\boldsymbol{x}}-\widetilde{\boldsymbol{x}}_c)$, 
with $\widetilde{\boldsymbol{x}}_c=(0,D/2,0)$. 
\setlength{\tabcolsep}{0.35em} 
{\renewcommand{\arraystretch}{1.5}%
\begin{table}
        \begin{center}
    \begin{tabular}{l l l l l l l}
    Run & $ T_x/T_R $ & $ T_y/T_R $ & $ T_z/T_R $ &
          $ \sigma_{T_x}/T_R $ & $ \sigma_{T_y}/T_R $  & $ \sigma_{T_z}/T_R $ \\
    \hline
    F50  & -$0.0006$ & -$0.00043$ & -$0.72$ & $0.17$ & $0.11$ & $0.27$ \\
    D120 & -$0.0011$ & $0.0005$   & -$0.44$ & $0.10$ & $0.10$ & $0.11$ \\
    \hline
        & $ \mathcal{S}_{T_x} $ & $ \mathcal{S}_{T_y} $ & $ \mathcal{S}_{T_z} $ &
          $ \mathcal{K}_{T_x} $ & $ \mathcal{K}_{T_y} $ & $ \mathcal{K}_{T_z} $ \\
    \hline
    F50  & -$0.010$ & -$0.0082$ & -$0.75$ & $3.78$ & $4.92$ & $3.35$ \\
    D120 & $0.009$  & -$0.001$  & -$0.52$ & $3.48$ & $3.51$ & $3.38$ \vspace*{.05cm}
    \\
    \hline
    \end{tabular}
    \caption{Statistics of torque acting on the roughness elements for cases 
      D50 and D120. $T_i/F_R$ denotes the normalised mean torque component in 
      the $x_i$-direction, $\sigma_{T_i}$, $\mathcal{S}_{F_i}$ and $\mathcal{K}_{T_i}$ 
      are the standard deviation, the skewness and the kurtosis of the $i$-th component 
      of the torque, respectively.
    }
    \label{tab4}
    \end{center}
\end{table}
}
For the sake of completeness, the values of $\boldsymbol{T}$ normalised by the 
reference torque $T_R=F_R (y_0-D/2)$, as well as the statistics of torque fluctuations 
$\boldsymbol{T}''$ were computed and reported in table \ref{tab4} for the simulation D120. %
The values of the standard deviation of torque fluctuations in the three directions, 
$(\sigma_{T_x}, \sigma_{T_y}, \sigma_{T_z})/T_R$, are nearly the same whereas the 
counterparts for force fluctuations are significantly different. %
\mut{
  This suggests that the fluctuations of the shear stress
  (associated with viscous effects) are less anisotropic than pressure
  fluctuations, which dominate the hydrodynamic force and which do not 
  contribute to the torque.
  This is true despite the fact that only the most exposed part of the
  sphere surface contributes significantly to the surface shear
  stress, i.e.\ this is essentially the upstream-facing part of the upper
  hemisphere (figure not shown).
}
%

\begin{figure}
\setlength{\unitlength}{0.353mm}
\begin{picture}(0,300)(0,0)
\put(5,145){
\put(0,2){\includegraphics[trim=0cm 0cm 0cm 0cm, clip, width=.438\textwidth]{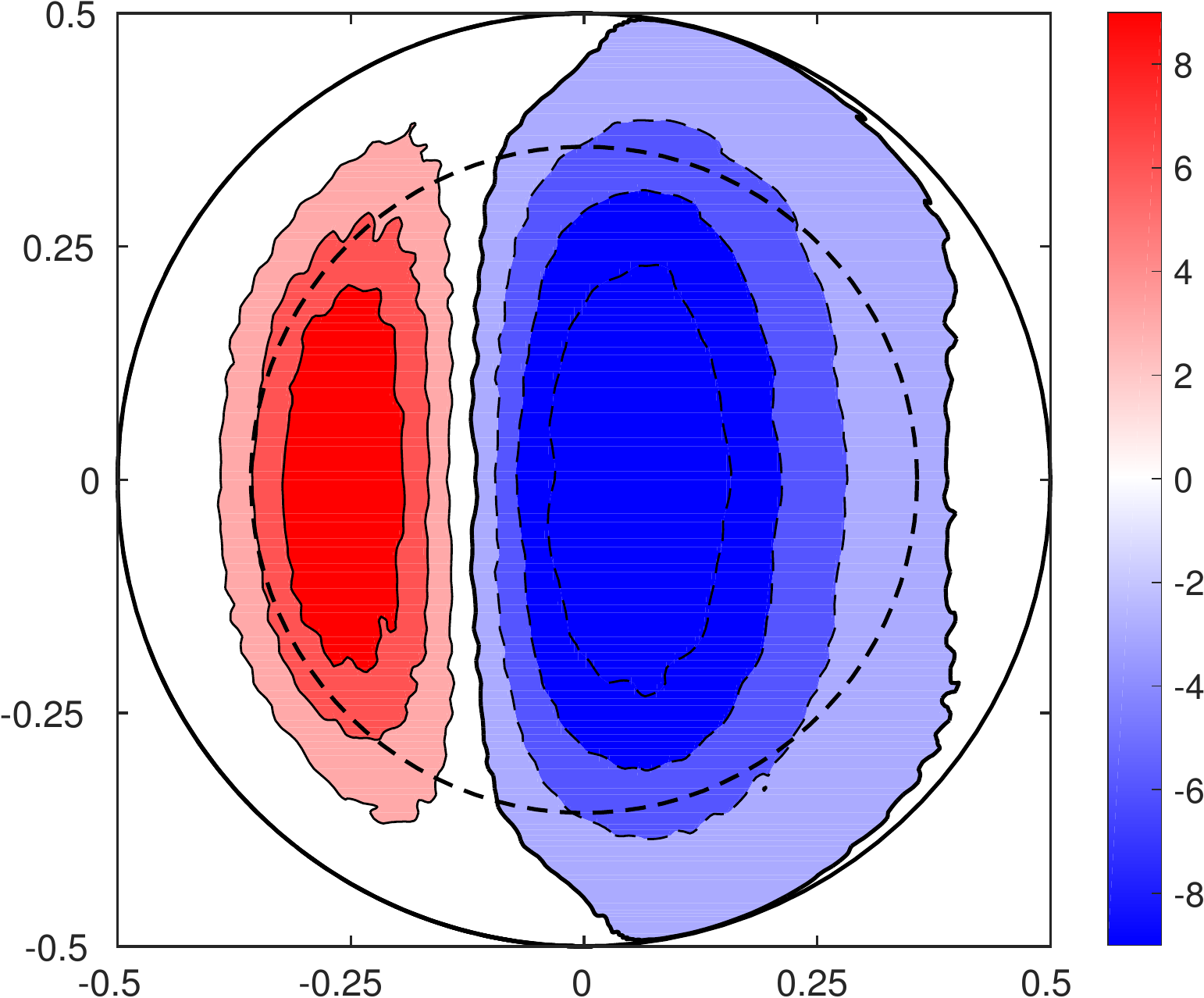}}
\put(-7,73){$\dfrac{\widetilde{z}}{D}$}
\put(-7,135){$a)$}
}
\put(200,145){
\put(0,2){\includegraphics[trim=0cm 0cm 0cm 0cm, clip, width=.43\textwidth]{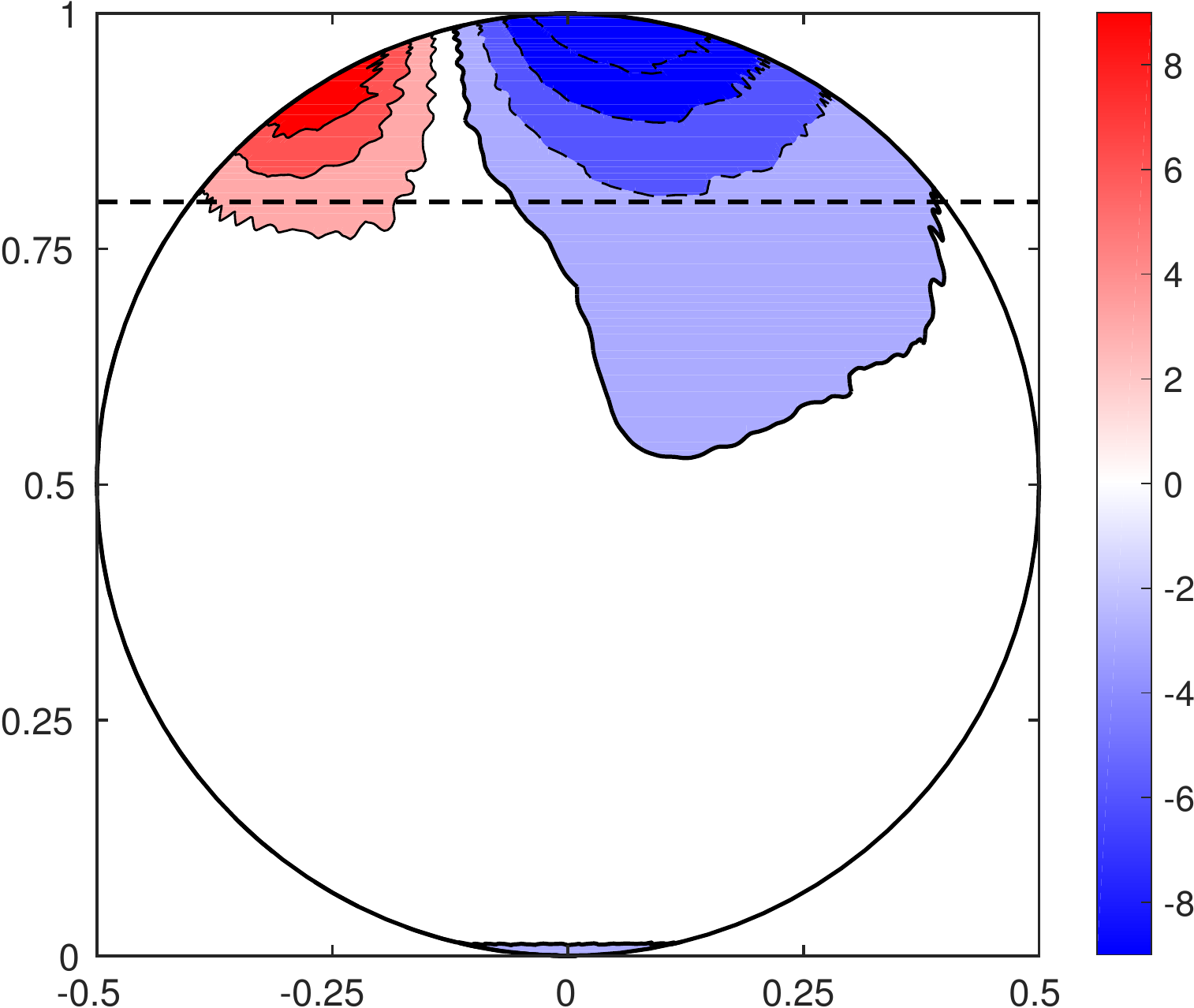}}
\put(-10,73){$\dfrac{y}{D}$}
\put(-8,119){\scriptsize $\dfrac{y_0}{D}$}
\put(3,118.5){\line(2,-1){10}}
\put(-7,135){$b)$}
}
\put(5,0){
\put(0,2){\includegraphics[trim=0cm 0cm 0cm 0cm, clip, width=.438\textwidth]{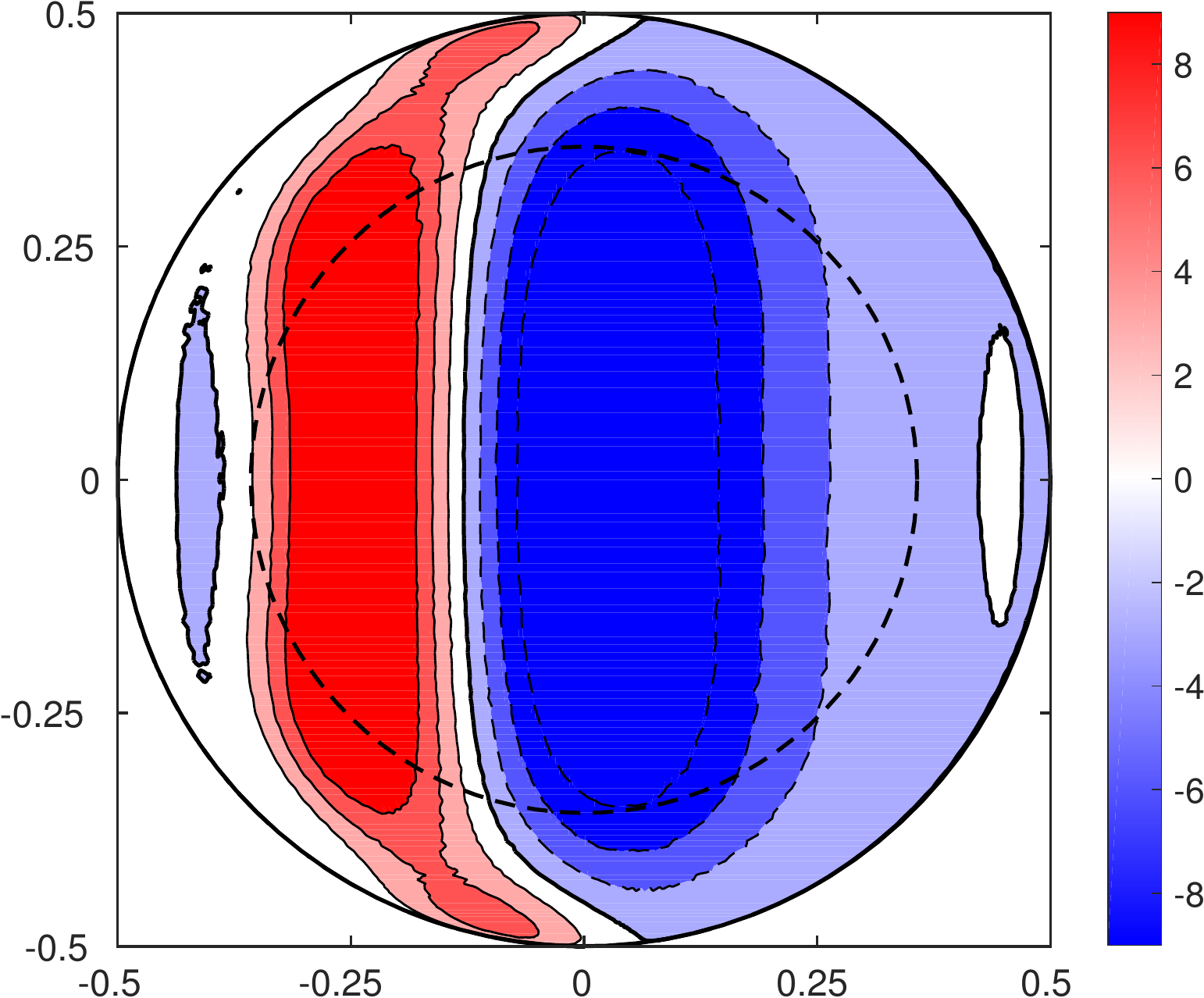}}
\put(70,-5){$\widetilde{x}/D$}
\put(-7,73){$\dfrac{\widetilde{z}}{D}$}
\put(-7,135){$c)$}
}
\put(200,0){
\put(0,2){\includegraphics[trim=0cm 0cm 0cm 0cm, clip, width=.43\textwidth]{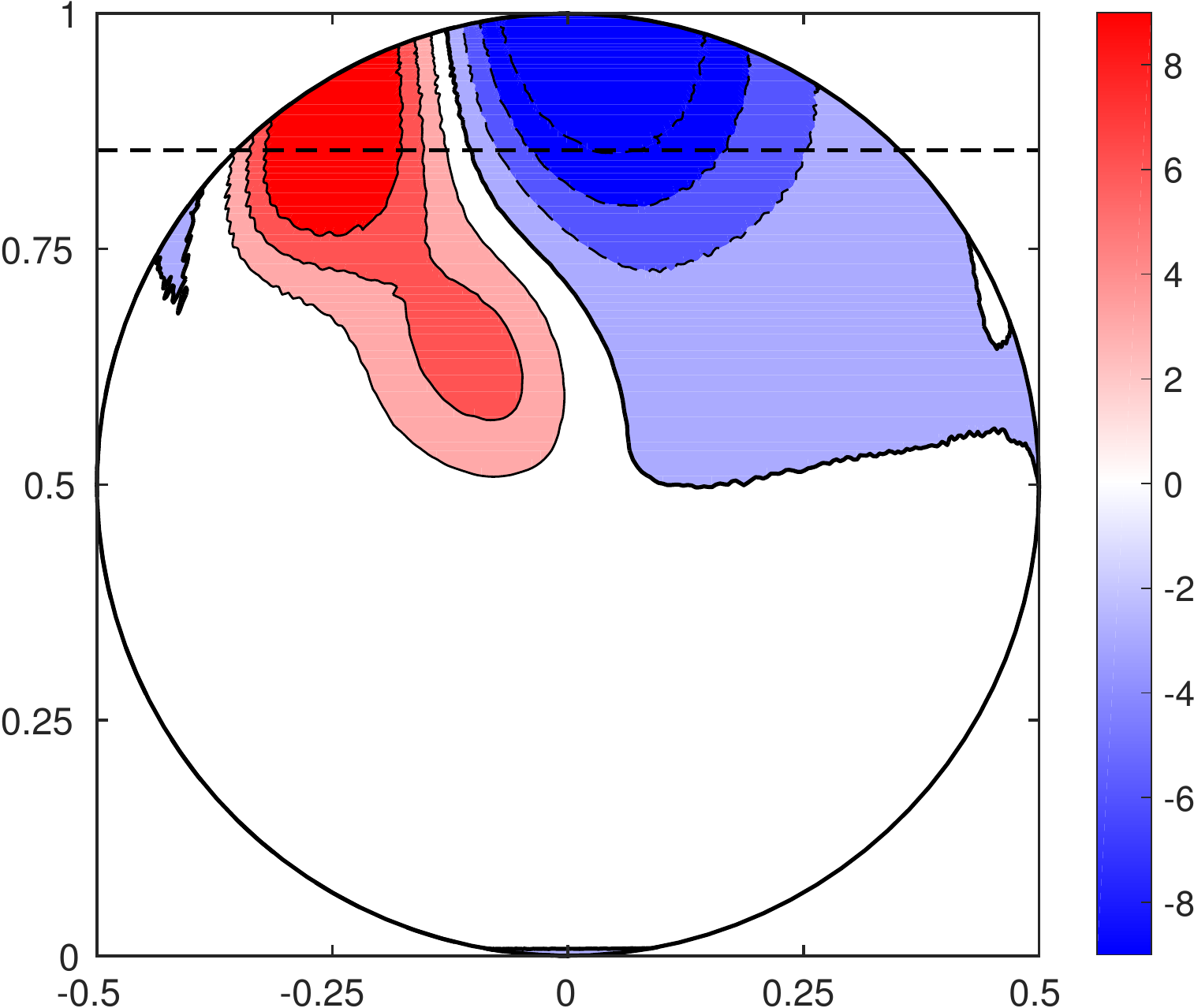}}
\put(70,-5){$\widetilde{x}/D$}
\put(-10,73){$\dfrac{y}{D}$}
\put(-8,123.5){\scriptsize $\dfrac{y_0}{D}$}
\put(3,124.5){\line(2,-1){10}}
\put(-7,135){$d)$}
}
\end{picture}
\caption{%
  Distribution of the mean pressure, $\langle\overline{p}_{tot}\rangle_{B}$, 
  evaluated at the sphere surface and normalised by $F_R/A_{sph}$. 
  Panels $(a,c)$ show the top view while panels $(b,d)$ show the side
  view of the sphere. $(a,b)$ run D50; $(c,d)$ run D120. Thin contour
  lines at values $\pm[3,\, 6,\, 9]$. The thick contour line indicates
  the value $0$.} 
\label{fig14}
\end{figure}
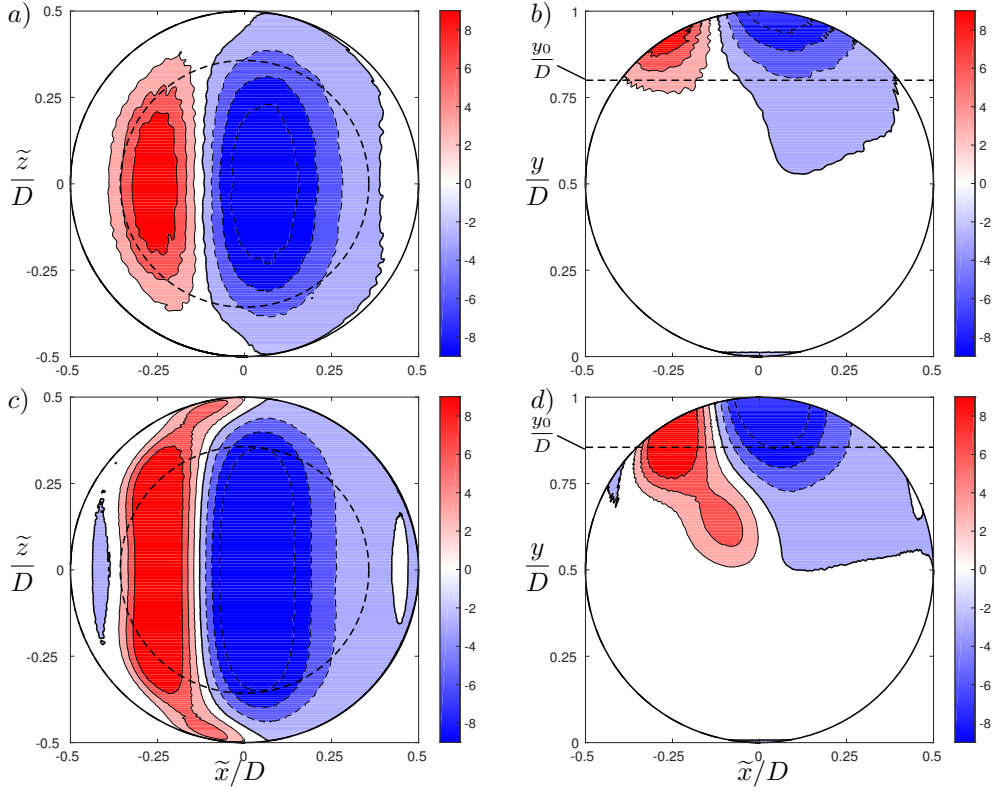
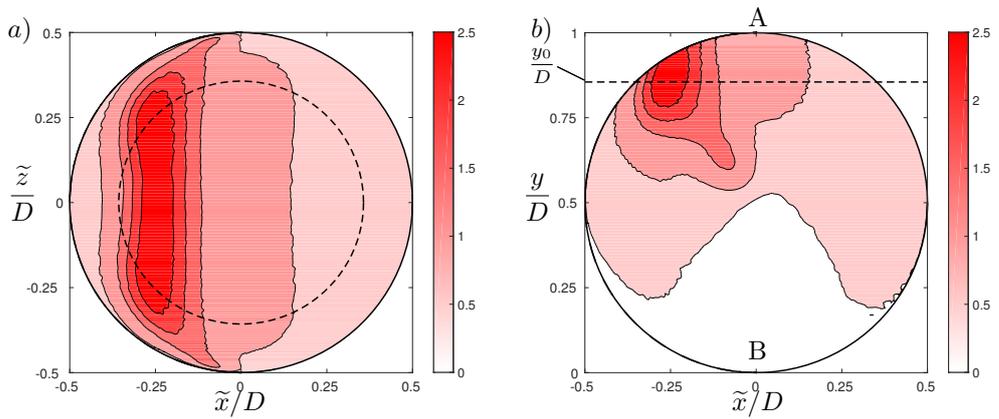
\begin{figure}
\setlength{\unitlength}{0.353mm}
\begin{picture}(0,150)(0,0)
\put(5,0){
\put(0,2){\includegraphics[trim=0cm 0cm 0cm 0cm, clip, width=.438\textwidth]{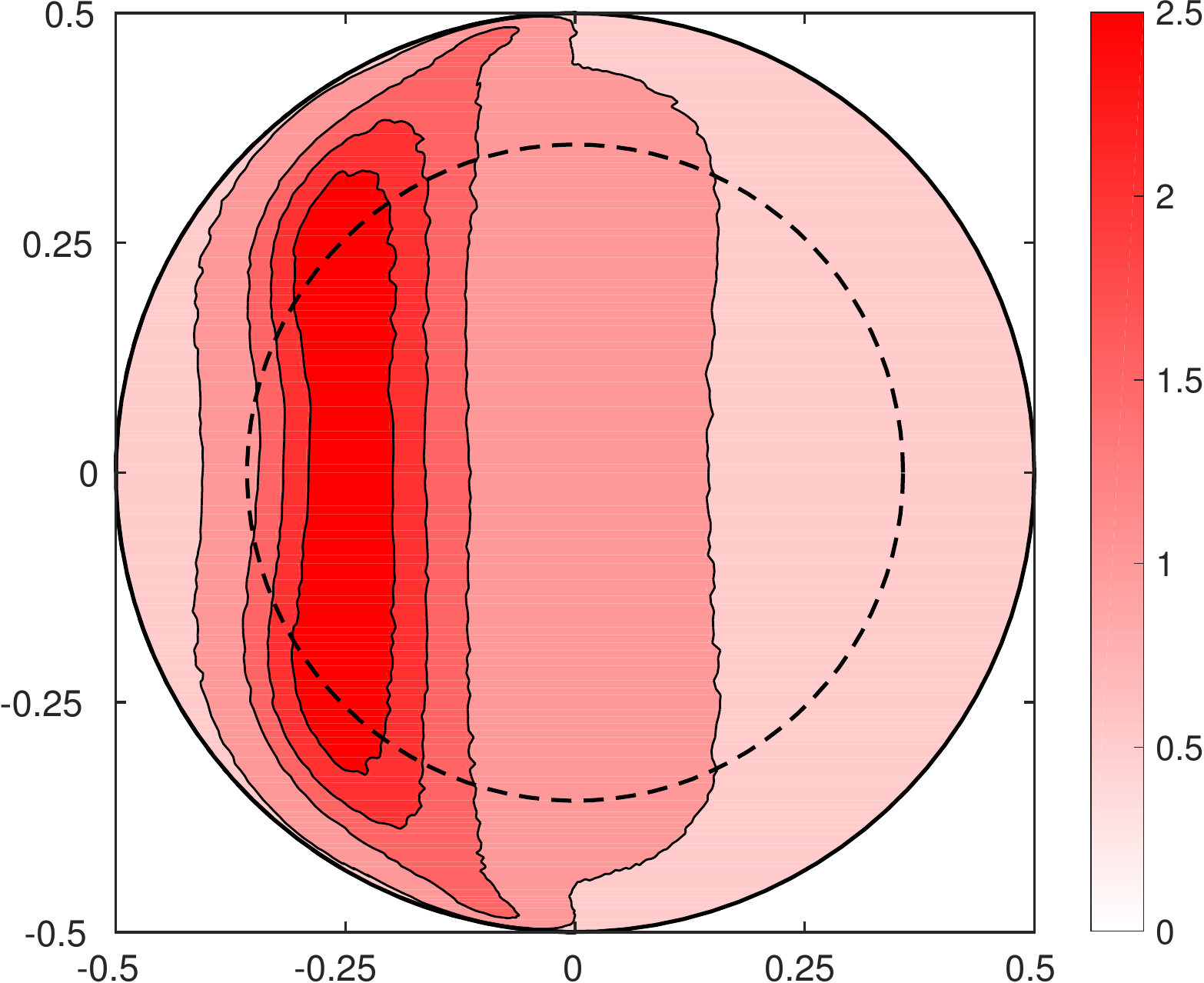}}
\put(70,-5){$\widetilde{x}/D$}
\put(-7,73){$\dfrac{\widetilde{z}}{D}$}
\put(-7,135){$a)$}
}
\put(200,0){
\put(0,2){\includegraphics[trim=0cm 0cm 0cm 0cm, clip, width=.43\textwidth]{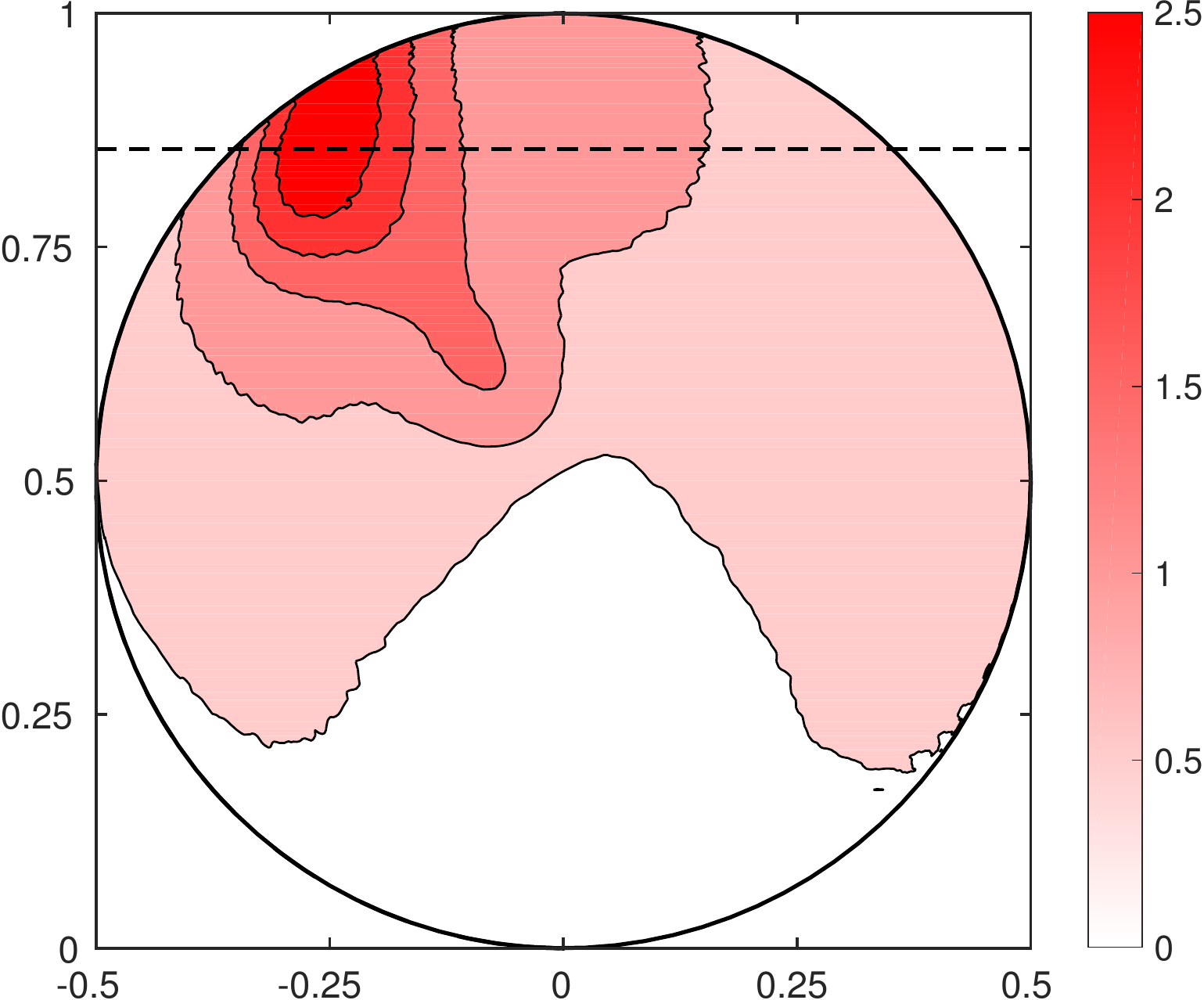}}
\put(68,-5){$\widetilde{x}/D$}
\put(-10,73){$\dfrac{y}{D}$}
\put(-8,123){\scriptsize $\dfrac{y_0}{D}$}
\put(3,124){\line(2,-1){10}}
\put(-7,135){$b)$}
\put(74,139){A}
\put(74,14){B}
}
\end{picture}
\caption{%
  Root mean square of the pressure fluctuations
  $p_{tot}^{\prime\prime}$ at the sphere surface, 
  normalised by $F_R/A_{sph}$. Panels $a)$ and $b)$ show the top and side views of the sphere for the run D120. Contour lines are equispaced by $0.5$.} 
\label{fig24}
\end{figure}
The distribution of the 
average surface pressure $\langle\overline{p}_{tot}\rangle_{B}$ 
(i.e.\ the surface-normal component of the stress vector
$\widetilde{\boldsymbol{\tau}}_n$ with the sign reversed), 
normalised by $F_R/A_{sph}$, where $A_{sph}=\pi D^2$, is shown in
figure~\ref{fig14}. 
Two regions of strongly negative and positive mean pressure
appear on the upper hemisphere (please recall the definition of
pressure in equation~\ref{equ-define-total-pressure}).
While the negative-valued region is slightly 
shifted downstream of the sphere, the positive region (i.e.\ with the
normal stress directed towards the center of the sphere) is located
somewhat upstream.
In case D50 this latter (high-pressure) region is almost confined
above $y=y_0$, while it is elongated in the spanwise
direction in case D120, reaching to smaller wall-normal distances on
the sides of the sphere. 
As it was previously noted in section \ref{sec1}, the mean flow penetrates deeper through the grooves between the roughness elements for increasing values of $\mut{Re_{bH}}$, 
causing the 
steady spanwise-vorticity structures to squeeze on the top of the spheres.
This explains the lateral spreading of the two regions in figures \ref{fig14}c,d characterised by peaks of the average \mut{pressure}.
By virtue of the distribution shown in figures \ref{fig14}c,d, it is possible to deduce that, for the square arrangement of spherical roughness elements, an effective position for two pressure probes in an analogous experiment should be in the 
center 
of the two regions presently highlighted.

For the simulation D120, a region of intense pressure 
fluctuations was detected on the upper upstream quarter of the sphere
surface, in correspondence to the region where 
the average pressure $\langle\overline{p}_{tot}\rangle_{B}$ is
positive. 
More specifically,
figure~\ref{fig24} shows the r.m.s.\ value of the pressure
fluctuations around the time-and-sphere-box average, 
$p_{tot}^{\prime\prime}$, at the sphere surface. 
The region of high pressure fluctuation intensity in case D120 is
found to coincide largely with the high mean-pressure region observed in
figure~\ref{fig14}$(c,d)$, while 
the fluctuations are of relatively small intensity over a
considerably large region around the top of the sphere. 
Although this evidence was not exhaustively investigated, it could be
inferred that in the fully-rough regime and for the present
arrangement of the spheres, most of the sphere-turbulence interaction
occurs in the upstream region at the top of the sphere, 
which might be a signature of the vortices shed from the
sphere located just upstream. 
%
\begin{figure}
\setlength{\unitlength}{0.353mm}
\begin{picture}(0,300)(0,0)
\put(5,148){
\put(0,0){\includegraphics[trim=0cm 0cm 0cm 0cm, clip, width=.438\textwidth]{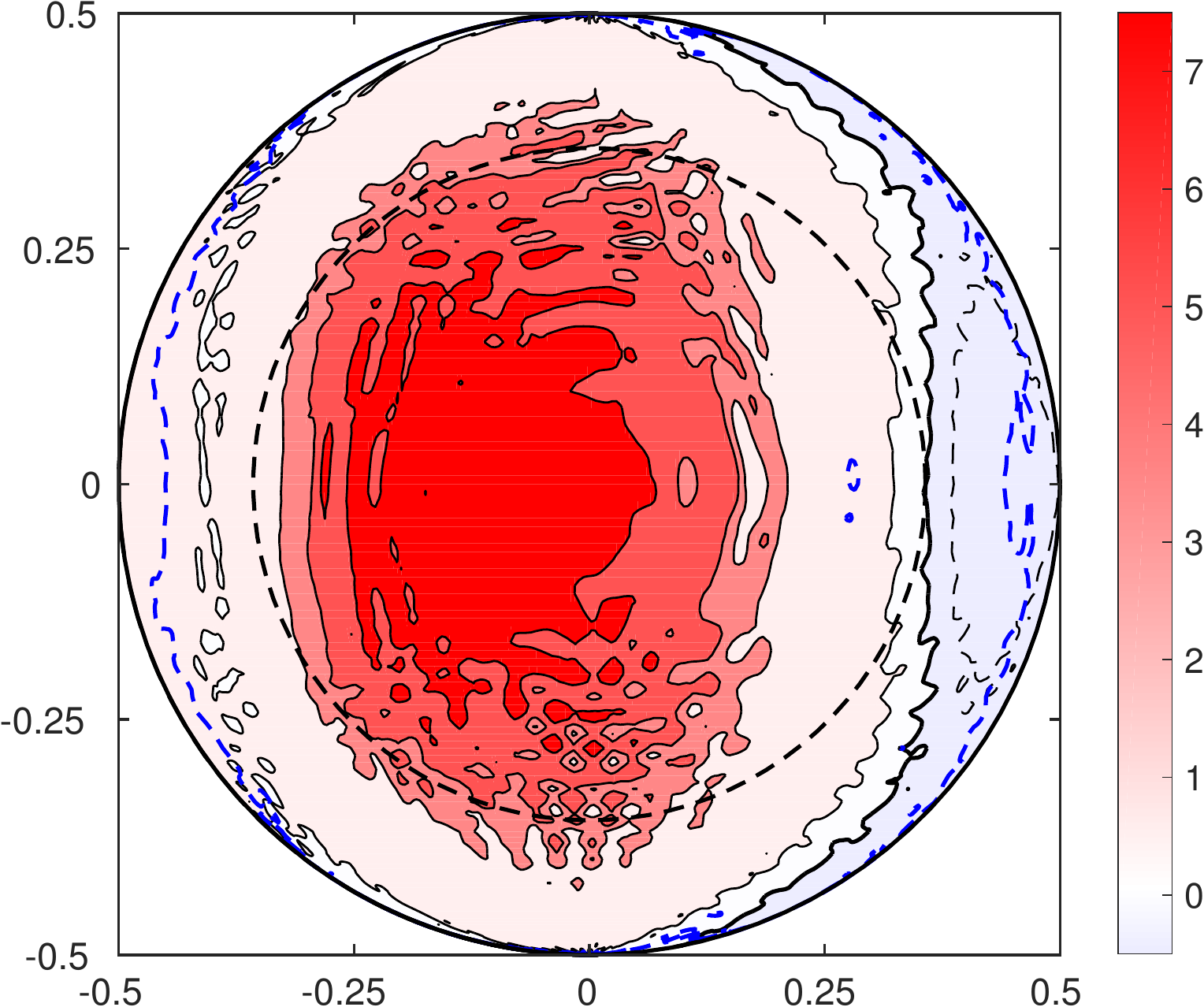}}
\put(-5,73){$\dfrac{\widetilde{z}}{D}$}
\put(-7,135){$a)$}
}
\put(200,148){
\put(0,0){\includegraphics[trim=0cm 0cm 0cm 0cm, clip, width=.43\textwidth]{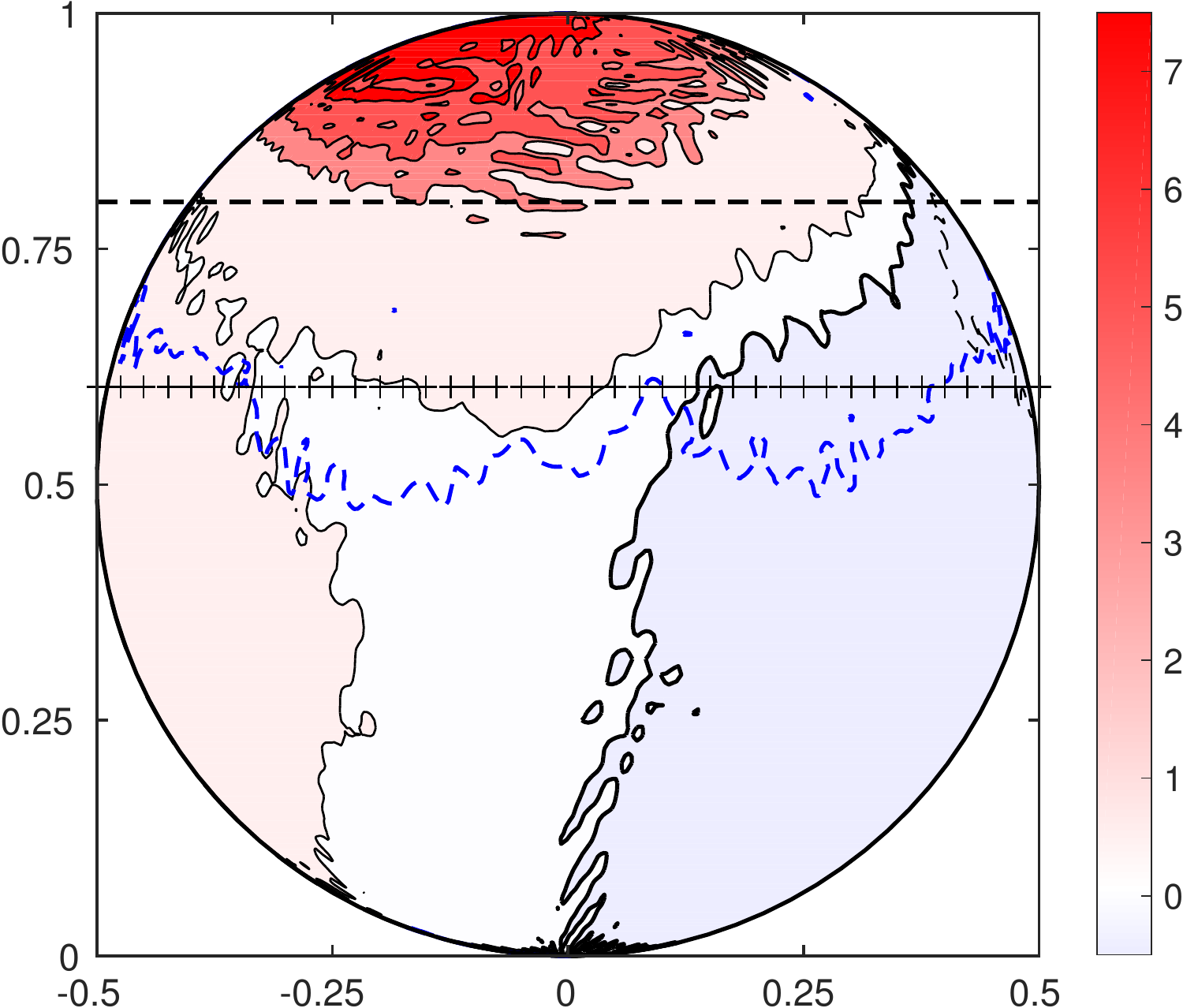}}
\put(-11,73){$\dfrac{y}{D}$}
\put(-8,119){\scriptsize $\dfrac{y_0}{D}$}
\put(3,117.5){\line(2,-1){10}}
\put(-23,103){\scriptsize \color{black!}$\left. \dfrac{y}{D}\right\vert_{\tau_{tot}=0}$}
\put(3,92.5){\color{black!}\line(2,-1){10}}
\put(-7,135){$b)$}
}
\put(5,3){
\put(0,0){\includegraphics[trim=0cm 0cm 0cm 0cm, clip, width=.438\textwidth]{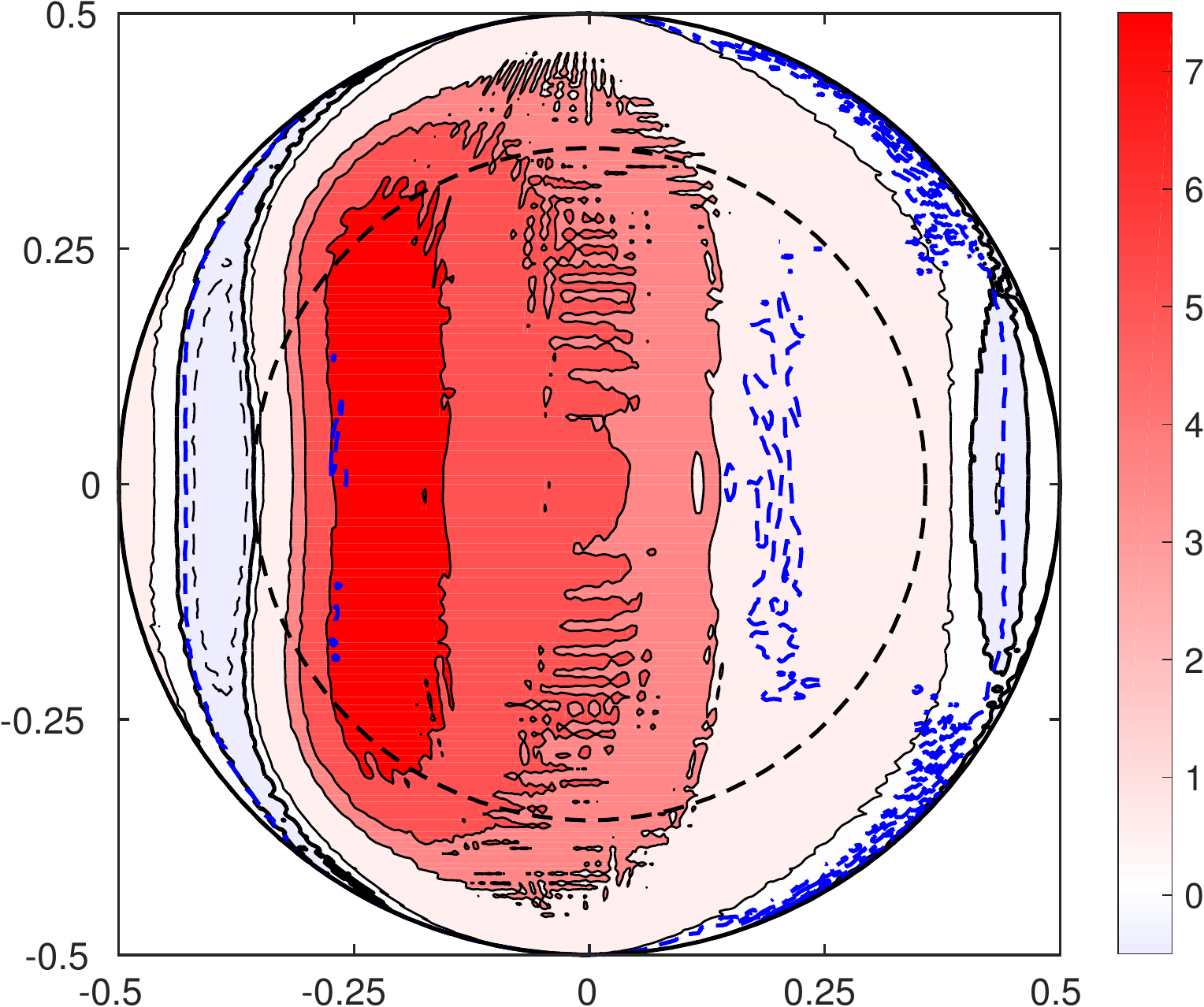}}
\put(72,-8){$\widetilde{x}/D$}
\put(-5,72){$\dfrac{\widetilde{z}}{D}$}
\put(-7,135){$c)$}
}
\put(200,3){
\put(0,0){\includegraphics[trim=0cm 0cm 0cm 0cm, clip, width=.43\textwidth]{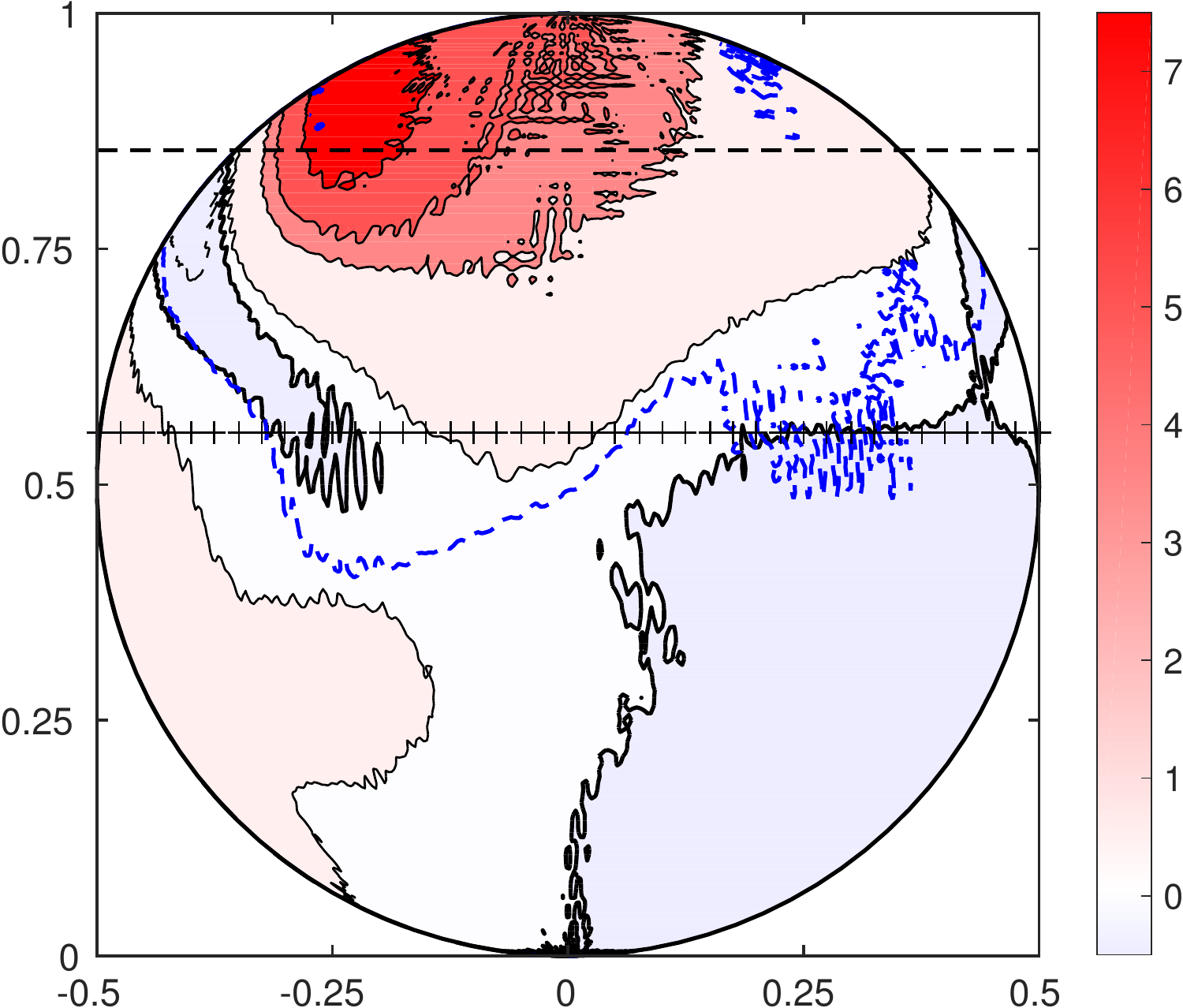}}
\put(70,-8){$\widetilde{x}/D$}
\put(-11,72){$\dfrac{y}{D}$}
\put(-8,123){\scriptsize $\dfrac{y_0}{D}$}
\put(3,125){\line(2,-1){10}}
\put(-23,96){\scriptsize \color{black!}$\left. \dfrac{y}{D}\right\vert_{\tau_{tot}=0}$}
\put(3,86){\color{black!}\line(2,-1){10}}
\put(-7,135){$d)$}
}
\end{picture}
\caption{Distribution of the streamwise component of $\widetilde{\boldsymbol{\tau}}_n$ normalised by $F_R/A_{sph}$. Panels $(a,c)$ show the top view while panels $(b,d)$ show the side view of the sphere. Black crosses and blue broken lines indicate the distance from the wall at which $\tau_{tot}=0$ and the values $\left\vert\boldsymbol{\tau}_n - (\boldsymbol{\tau}_n\cdot\boldsymbol{n})\boldsymbol{n}\right\vert A_{sph}/F_R=10^{-2}$ at the sphere surface, respectively. $(a,b)$ run D50; $(c,d)$ run D120. Thin contour lines at values $[-.5,\, 0.5,\, 3.5,\, 5.0,\, 7.5]$. The thick contour line indicates the value $0$.}
\label{fig11}
\end{figure}

  %
  Figure~\ref{fig11} shows the projection of the surface
  stress-vector $\widetilde{\boldsymbol{\tau}}_n$ upon the streamwise
  direction, i.e.\ the  map of the local surface-stress contribution
  to drag, $\widetilde{\boldsymbol{\tau}}_n\cdot\mathbf{e}_x$.
  It can be seen that in case D120 (figure~\ref{fig11}$c,d$) this
  quantity is largest in roughly the same region on the upper,
  upstream part of the sphere, where the mean surface-pressure 
  is large and positive (figure~\ref{fig14}$c,d$), 
  and where the surface-pressure fluctuations were found to be large
  (figure~\ref{fig24}).
  In the transitionally-rough case D50, on the contrary, the
  maximum contribution to drag is provided by the shear stress near the
  crest of the roughness elements (figure~\ref{fig11}$a,b$), in a zone
  less affected by high pressure (figure~\ref{fig14}$a,b$). 
  Therefore, although the values of the dimensionless drag force and
  the intensity of its fluctuations are similar in the transitionally-
  and in the fully-rough regimes, they clearly originate from
  different processes:
  the first is associated with the skin friction, while the latter is
  due to pressure. 
This picture can be 
interpreted as a manifestation of a ``form drag mechanism''
in case D120.  
It is also supported by the considerations previously formulated in
section \ref{sec1} about the shift of
$\left\langle\overline{u'''u'''}\right\rangle^+$ beneath the crest of
the spheres in the fully-rough regime. 

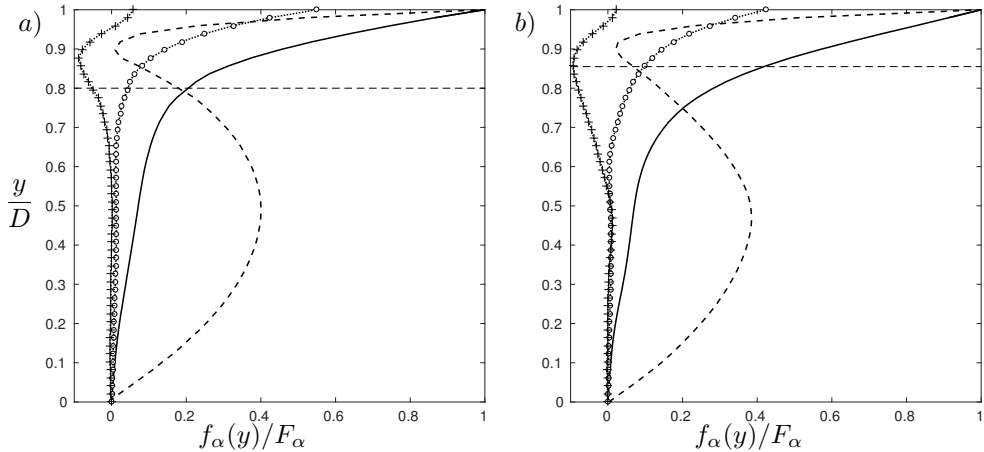
\begin{figure}
\setlength{\unitlength}{0.353mm}
\begin{picture}(0,164)(0,0)
\put(20,5){
\put(5,0){\includegraphics[trim=0cm 0cm 0cm 0cm, clip, width=.43\textwidth]{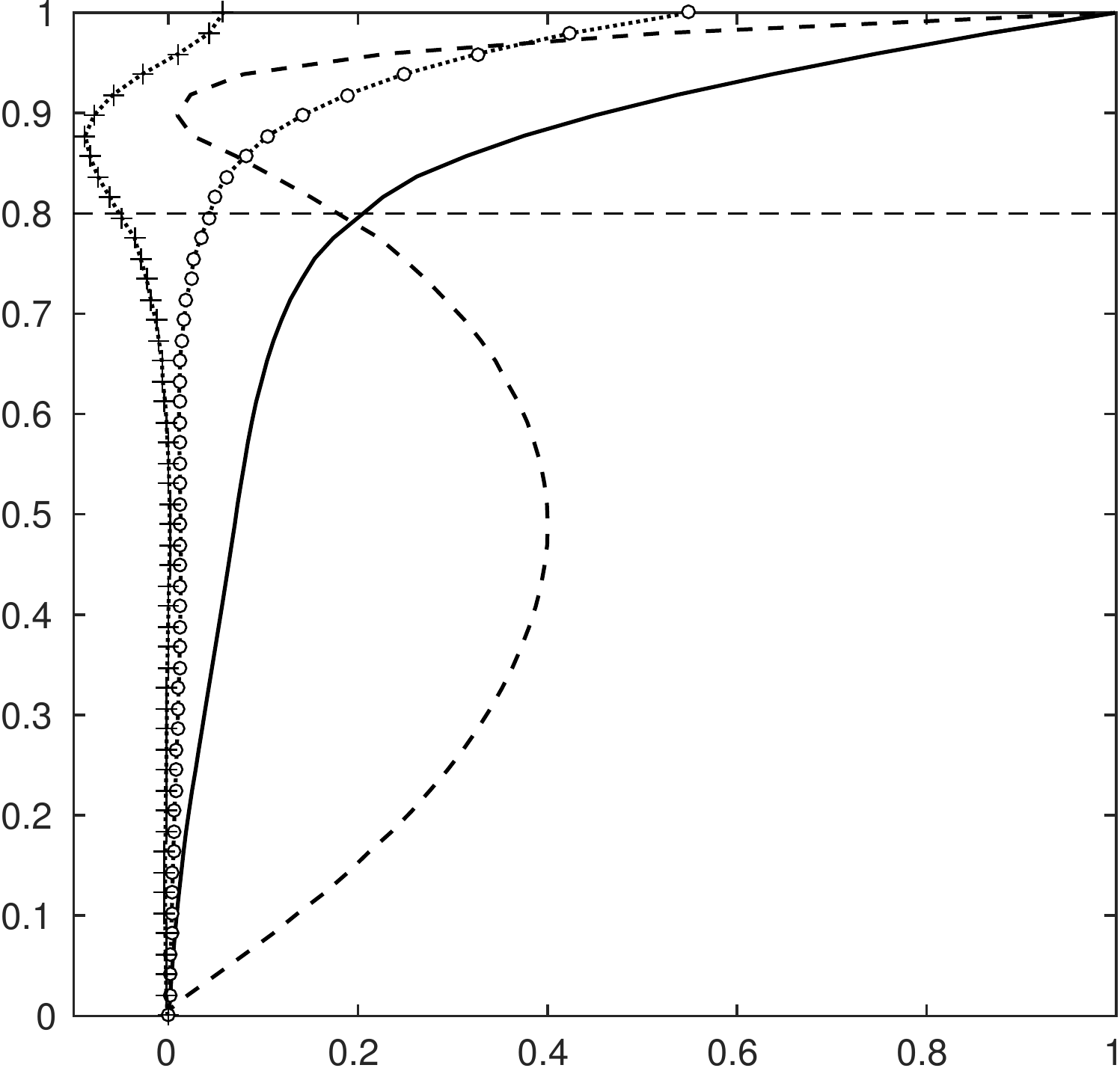}}
\put(190,0){\includegraphics[trim=0cm 0cm 0cm 0cm, clip, width=.43\textwidth]{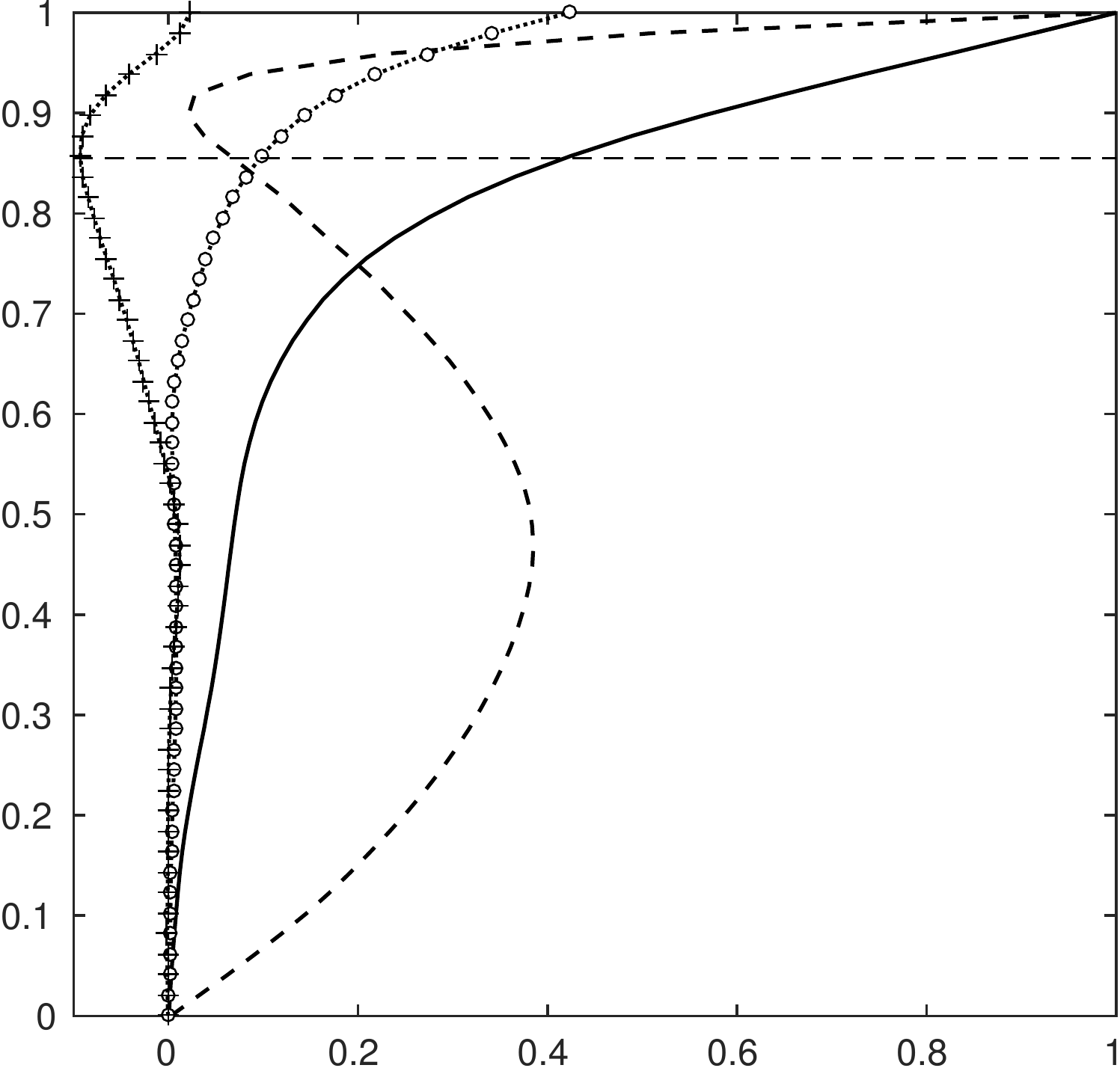}}
\put(-5,147){$a)$}
\put(180,147){$b)$}
\put(-10,80){$\dfrac{y}{D}$}
\put(62,-8){$f_\alpha (y)/F_\alpha$}
\put(247,-8){$f_\alpha (y)/F_\alpha$}
}
\end{picture}
\caption{Cumulative streamwise $[\boldsymbol{-}]$ and wall-normal $[$-
  - -$]$ components of \mut{$f_\alpha (y)$ normalised by $F_\alpha$,
    were $\alpha$ is replaced by $x$ and $y$, respectively}. The
  horizontal broken line is located at $y_0/D$.
  \mut{The lines
    $[\cdot\cdot${$\circ$}$\cdot\cdot]$ and
    $[\cdot\cdot${\footnotesize +}$\cdot\cdot]$ indicate the viscous
    contributions ${f}_\alpha^{(\nu)}(y)$ to the cumulative streamwise
    and wall-normal force, respectively.}
  $a)$ Run D50; $b)$ run D120.}
\label{fig15}
\end{figure}
  In order to further investigate the contributions of pressure and
  viscous stresses to the net force on the spheres, let us
  define a cumulative force $\boldsymbol{f}(y)$ as the 
  integral of the stress vector over the sphere surface up to height $y$, viz. 
  \begin{equation}
    \boldsymbol{f}(y) =
    \dfrac{D}{2}\displaystyle\int_0^{y}\int_{0}^{2\pi}
    \widetilde{\boldsymbol{\tau}}_n(\eta,\theta) \, d\theta d\eta
    \,,
    \label{eq8b}
  \end{equation}
  where $\eta$ denotes the wall-normal coordinate defined in the range
  $[0,D]$ and $\theta$ the azimuthal angle. Obviously
  $\boldsymbol{f}(D)=\boldsymbol{F}$, as defined in 
  (\ref{eq5a}). 
  The streamwise and wall-normal components of $\boldsymbol{f}(y)$, normalised by 
  the respective components of the total force $\boldsymbol{F}$, 
  are shown in figure~\ref{fig15}.
  For the purpose of comparison, the same figure also shows the
  viscous contributions $\boldsymbol{f}^{(\nu)}(y)$ to the cumulative
  force in the same scaling. 
  It can be observed that the cumulative viscous contributions to both
  drag and lift are negligible for small wall-distances. 
  At larger wall-distances 
  (for $y\gtrsim0.6D$ in case D50 and for $y\gtrsim0.7D$ in case D120)
  the viscous contribution to drag increases monotonically up to the
  final value listed in table~\ref{tab3}. 
  Contrarily, in both cases the cumulative viscous contribution to
  lift is negative for small wall-distances, and changes its sign only
  very close to the top of the spheres, most of the positive lift
  contribution being generated for $y\gtrsim0.85D$. 
  Note that the different distribution of the pressure with the
  distance from the wall in the present cases does not manifest
  itself in any significant difference in the normalised
  cumulative lift force profile, even though the vertical
  dimensionless lift $F_y/F_R$ is much larger in case D120. 

  \citet{chan2011} have introduced an analogy between the force and
  torque acting upon a square region of the wall in smooth-wall
  channel flow and the corresponding force/torque components acting
  upon a wall-mounted sphere (cf.\ the sketch in their figure~9).
  This ``smooth-wall analogy'' was formulated on the basis of the
  observation that in the transitionally-rough regime 
  essentially only the crests of the spheres were exposed to
  significant flow, and that viscous effects were dominating the
  surface stress.
  Hence, in their case the action of turbulent fluctuations on the
  upper part of the sphere surface could be modelled by analogy with that 
  occurring on a smooth-wall tile $\Gamma_s$ of side length $s$
  comparable to the sphere diameter $D$.
  In order to test to which extent this model still holds in the
  fully-rough regime, we use the data from the simulation of open-channel
  flow over a smooth-wall (at the same bulk Reynolds number as D120) 
  in order to compute the statistics of the 
  wall force and torque fluctuations as functions of $s$. 
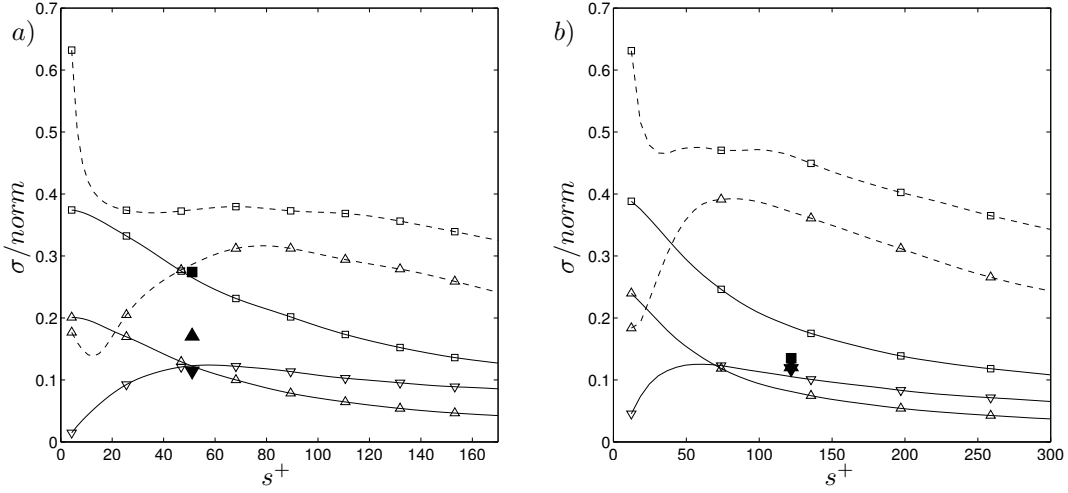
\begin{figure}
\setlength{\unitlength}{0.353mm}
\begin{picture}(0,180)(0,0)
\put(7,5){\includegraphics[trim=0cm 0cm 0cm 0cm, clip, height=.45\textwidth]{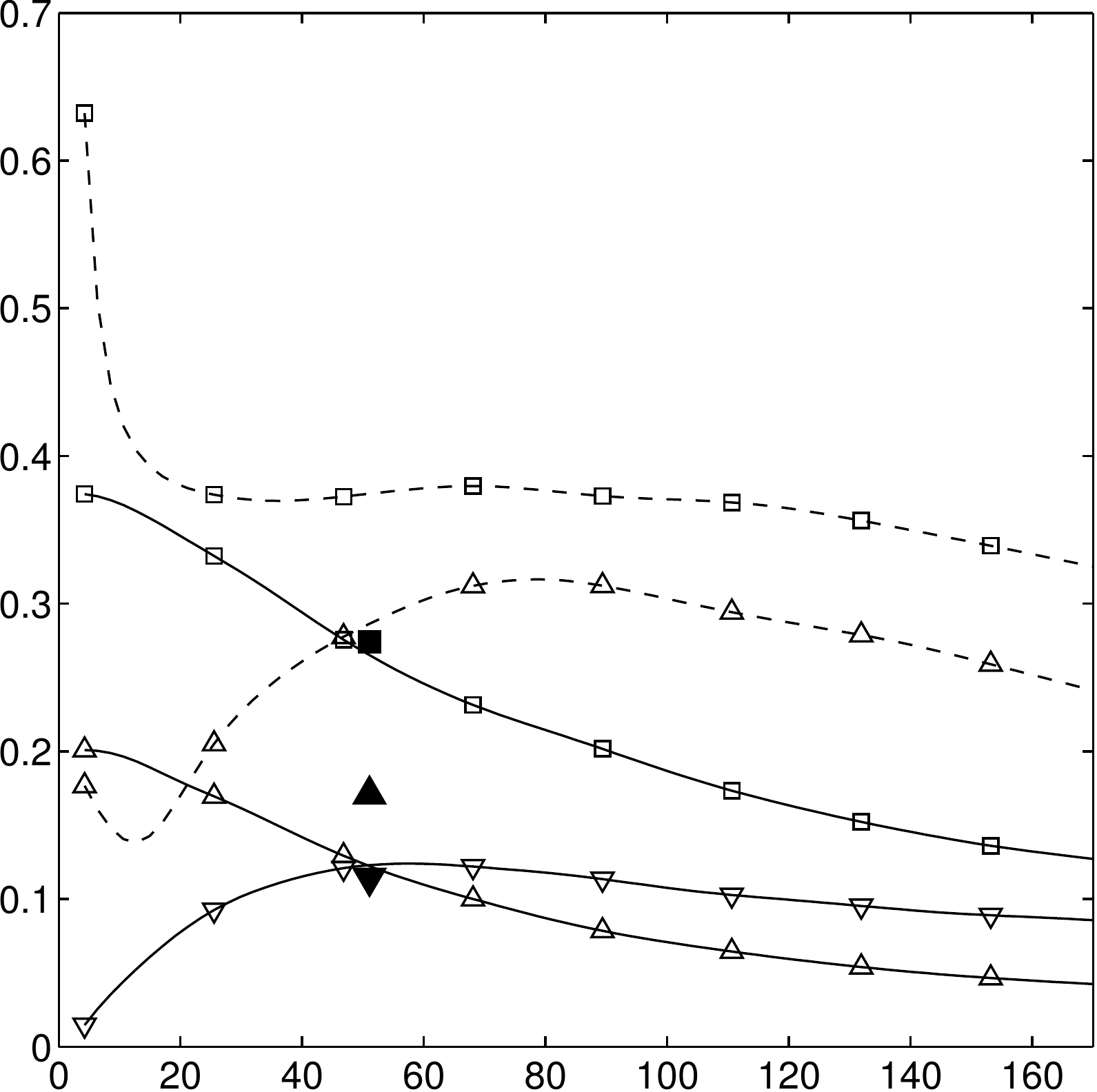}}
\put(213,5){\includegraphics[trim=0cm 0cm 0cm 0cm, clip, height=.45\textwidth]{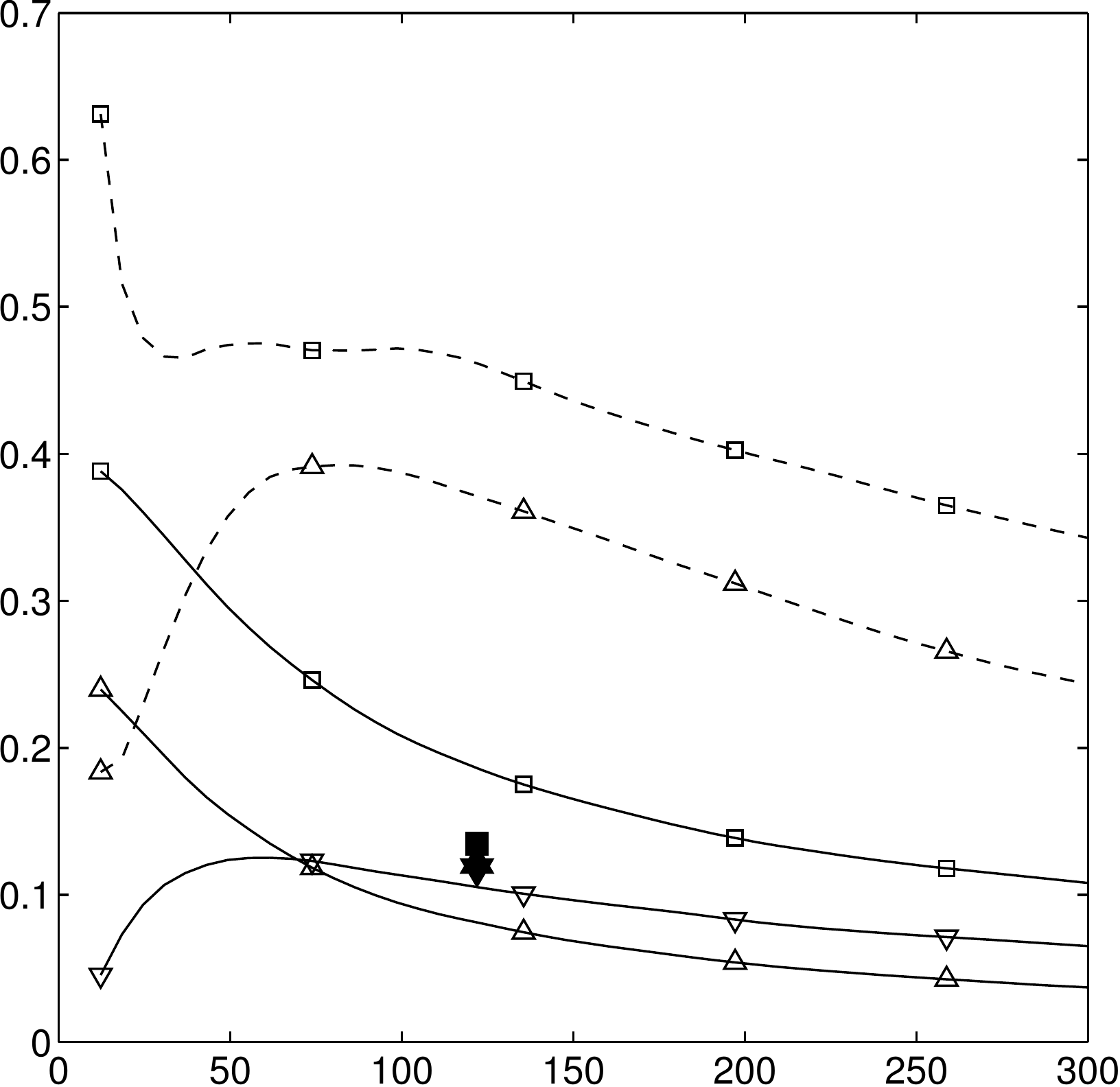}}
\put(-2,163){$a)$}
\put(200,163){$b)$}
\put(-5,75){\rotatebox{90}{$\sigma/norm$}}
\put(200,75){\rotatebox{90}{$\sigma/norm$}}
\put(91,-4){$s^+$}
\put(301,-4){$s^+$}
\end{picture}
\caption{Standard deviation of the torque acting on a smooth square
  tile of side $s^+$, with an ideal arm of length $s^+/2$, originated
  by the shear-stress, $\boldsymbol{\sigma}_{\mathcal{T}}$ (solid
  lines), or by the shear-stress and pressure,
  $\sigma_{\mathcal{T}^{(p)}_x}$ (broken lines), normalised by
  $\frac{1}{2}\varrho u_{\tau}^2 s^3$. The standard deviation of the
  torque acting on the sphere $\widetilde{\mathcal{S}}$, normalised by
  $T_R$, is also visualised with full symbols for the simulations
  $(a)$ D50 and $(b)$ D120. Symbols indicate the streamwise
  $[$---$\bigtriangleup$---$,\,\blacktriangle]$, the wall-normal
  $[$---$\bigtriangledown$---$,\,\blacktriangledown]$ and spanwise
  $[$---$\square$---$,\,\blacksquare]$ components of torque. Panel
  $(a)$: $\mut{Re_{bH}}\sim 2900$, run D50; panel $(b)$:
  $\mut{Re_{bH}}\sim 6900$, run D120.
} 
\label{fig18}
\end{figure}
In particular, the standard deviations 
of the torque components $\sigma_{T_x}/T_R$, $\sigma_{T_y}/T_R$ and 
$\sigma_{T_z}/T_R$ for the simulations D50 and D120 are compared with those of the torque:
\begin{equation}
\boldsymbol{\mathcal{T}} = \displaystyle\int^{s/2}_{-s/2}\int^{s/2}_{-s/2} \widetilde{\boldsymbol{r}}_s\times \left(\widetilde{\left\langle\overline{\boldsymbol{\tau}_\nu} \right\rangle}_B\vert_{y=0} \cdot\boldsymbol{e}_y\right)\, dx\,dz
\label{eq7a}
\end{equation}
and
\begin{equation}
\boldsymbol{\mathcal{T}}^{(p)} = \displaystyle\int^{s/2}_{-s/2}\int^{s/2}_{-s/2} \widetilde{\boldsymbol{r}}_s\times \left(\widetilde{\left\langle\overline{\boldsymbol{\tau}} \right\rangle}_B\vert_{y=0} \cdot\boldsymbol{e}_y\right)\, dx\,dz
\label{eq7b}
\end{equation}
obtained for the smooth-wall square tile $\widetilde{\Gamma}_{L_B}$ 
(i.e. of side $s=L_B$) about the ideal point $\widetilde{\boldsymbol{x}}_s=(0,-s/2,0)$ such that 
$\widetilde{\boldsymbol{r}}_s=\widetilde{\boldsymbol{x}}-\widetilde{\boldsymbol{x}}_s$ 
with $\widetilde{\boldsymbol{x}}\in \widetilde{\Gamma}_{L_B}$.
The symbol $\boldsymbol{e}_y$ in the equations (\ref{eq7a}-\ref{eq7b}) 
indicates the wall-normal unit vector of the canonical base while 
$\boldsymbol{\tau}$ and $\boldsymbol{\tau}_\nu$ denote the total stress 
tensor and the viscous stress tensor (i.e. without the pressure
contribution), 
respectively, cf.\ the definition in (\ref{equ-define-total-stress-tensor}). 
The comparison is visualised in figure~\ref{fig18}, where the full symbols 
indicate the values related to the simulations D50 and D120, respectively.
It is to be pointed out that the contribution of pressure fluctuations to 
$\boldsymbol{\mathcal{T}}$ was not accounted for (solid lines in figure~\ref{fig18}) 
in order to preserve the analogy, since pressure fluctuations do not contribute 
to torque on a sphere.
In fact, the trend for the standard deviations
$\sigma_{\mathcal{T}^{(p)}_x}(L^+_B)$ and
$\sigma_{\mathcal{T}^{(p)}_z}(L^+_B)$, obtained considering also the
contribution of pressure, normalised by $\frac{1}{2}\varrho u_\tau
s^3$, are definitely far from $T_x$ and $T_z$ obtained for the 
spheres (broken lines in figure~\ref{fig18}).
Indeed, this simple approach was found surprisingly satisfactory for 
flows in the transitionally-rough regime, but it can be seen that it 
fails for the simulation D120, because the mechanism of sphere-flow 
interaction in the fully-rough regime is not correctly interpreted by the model.
In particular, figure~\ref{fig18}b shows graphically that the standard 
deviations of the three components of torque, $\boldsymbol{\sigma}_T/T_R$, 
tend to collapse on the same value, as discussed above.
In other terms, the failure of the ``smooth-wall analogy'' in the fully-rough 
regime can be interpreted as an effect of the destruction of the buffer layer 
and the disappearance of significant viscous effects in the vicinity of the sphere crests.
\begin{figure}
\setlength{\unitlength}{0.353mm}
\begin{picture}(0,300)(0,0)
\put(5,147){
\put(0,4){\includegraphics[trim=0cm 0cm 0cm 0cm, clip, width=.44\textwidth]{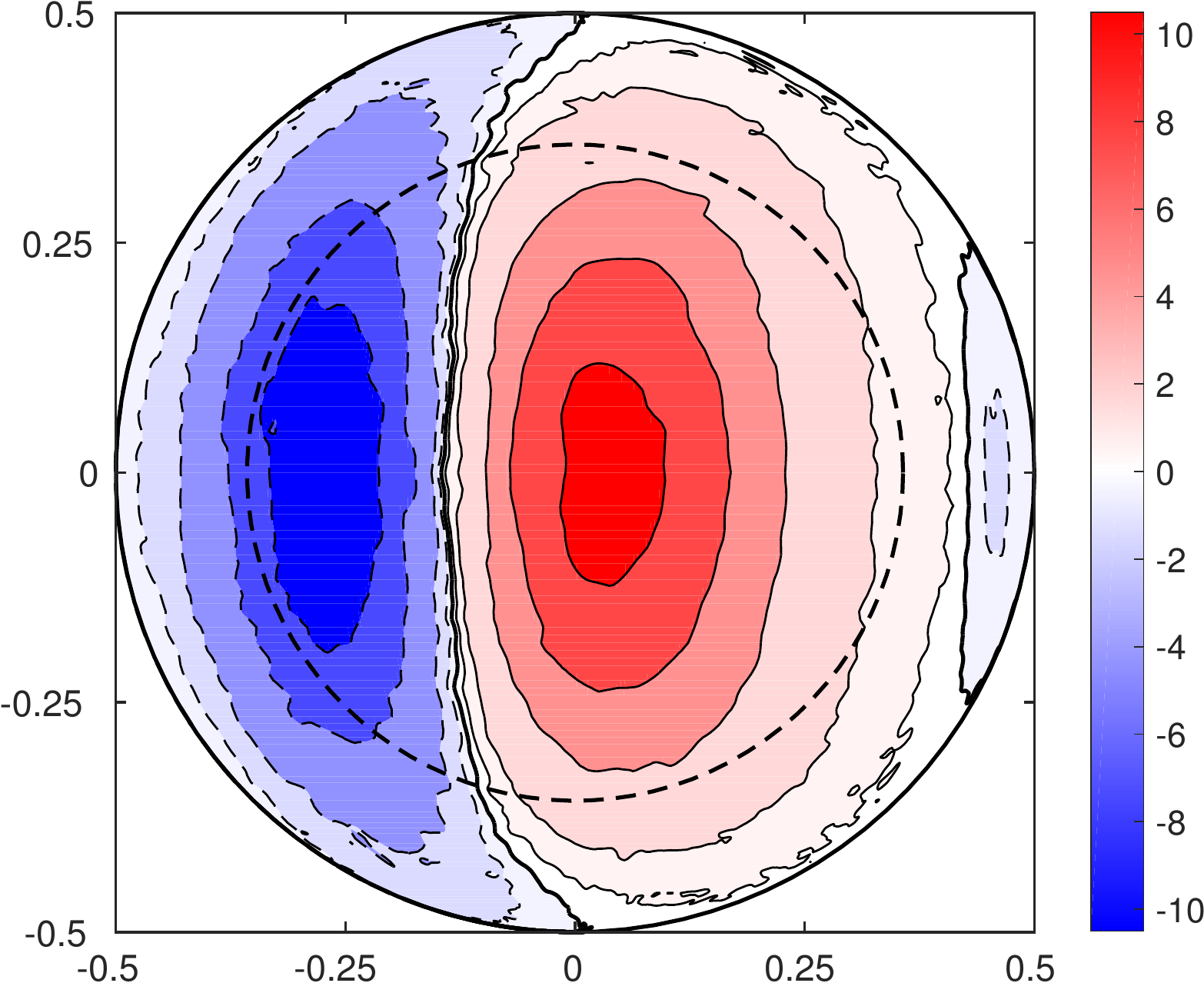}}
\put(-7,73){$\dfrac{\widetilde{z}}{D}$}
\put(-7,135){$a)$}
}
\put(200,147){
\put(0,4){\includegraphics[trim=0cm 0cm 0cm 0cm, clip, width=.43\textwidth]{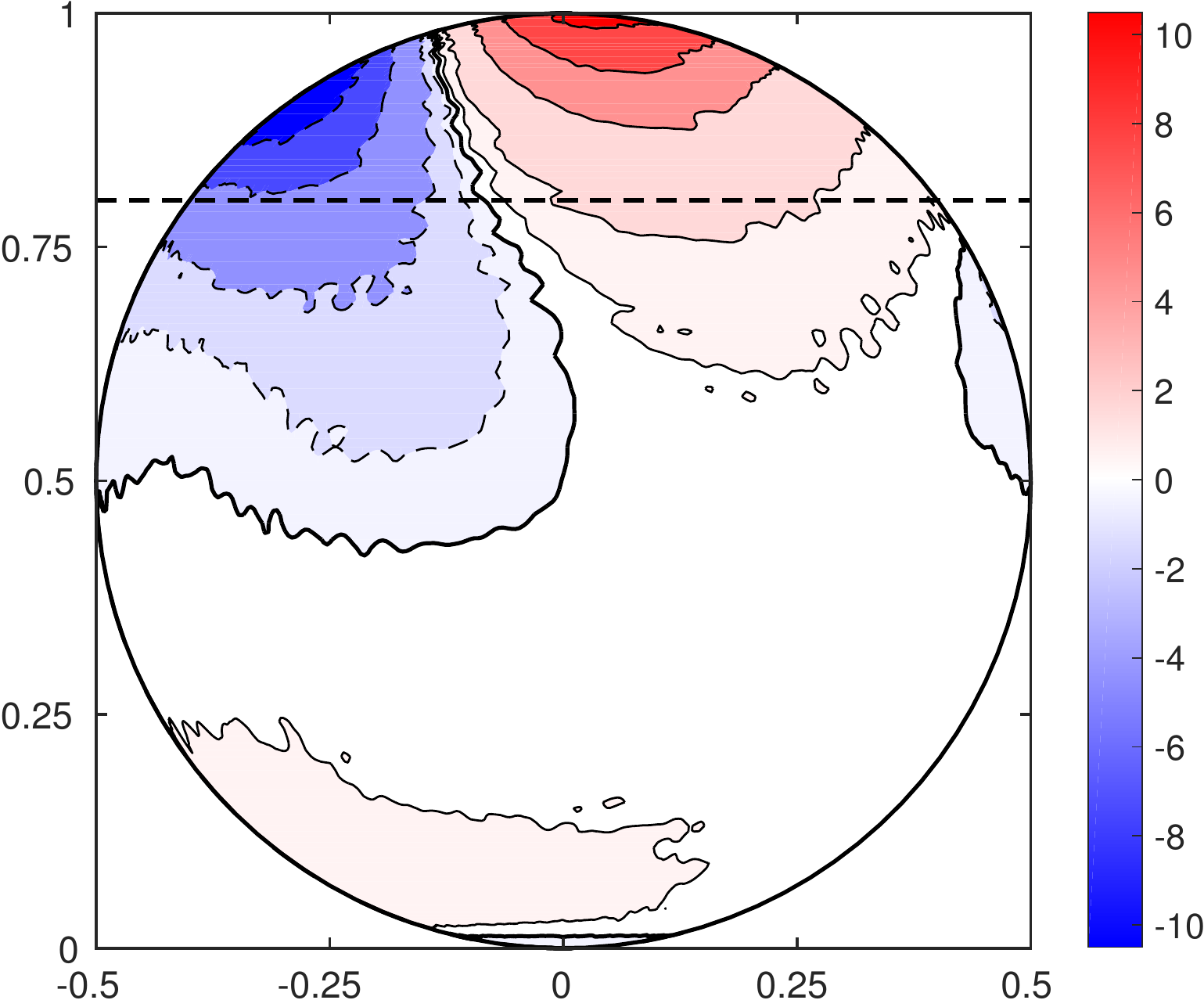}}
\put(-10,75){$\dfrac{y}{D}$}
\put(-8,116){\scriptsize $\dfrac{y_0}{D}$}
\put(3,117.5){\line(2,-1){10}}
\put(-7,135){$b)$}
}
\put(5,2){
\put(0,4){\includegraphics[trim=0cm 0cm 0cm 0cm, clip, width=.44\textwidth]{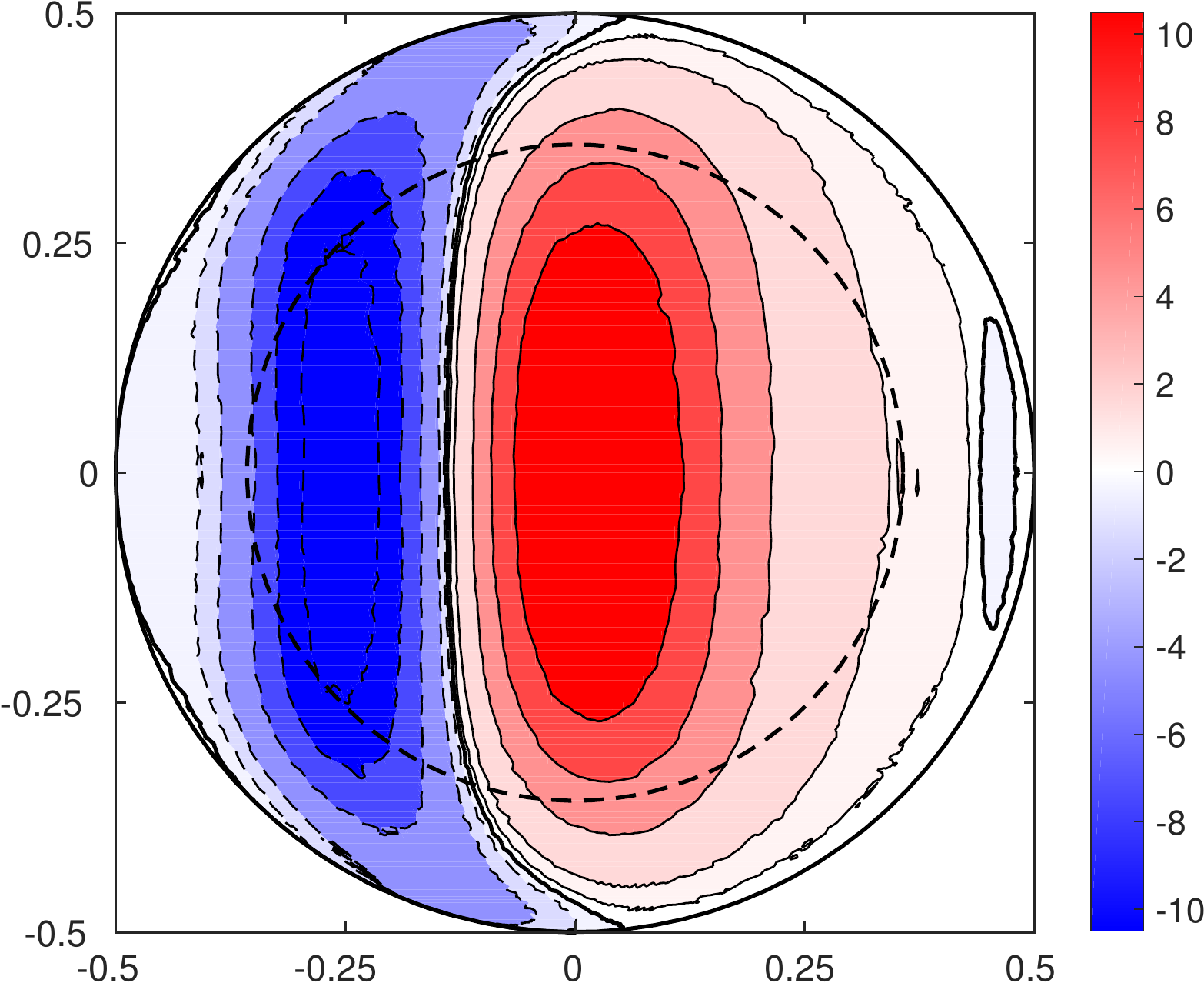}}
\put(70,-5){$\widetilde{x}/D$}
\put(-7,73){$\dfrac{\widetilde{z}}{D}$}
\put(-7,135){$c)$}
}
\put(200,2){
\put(0,4){\includegraphics[trim=0cm 0cm 0cm 0cm, clip, width=.43\textwidth]{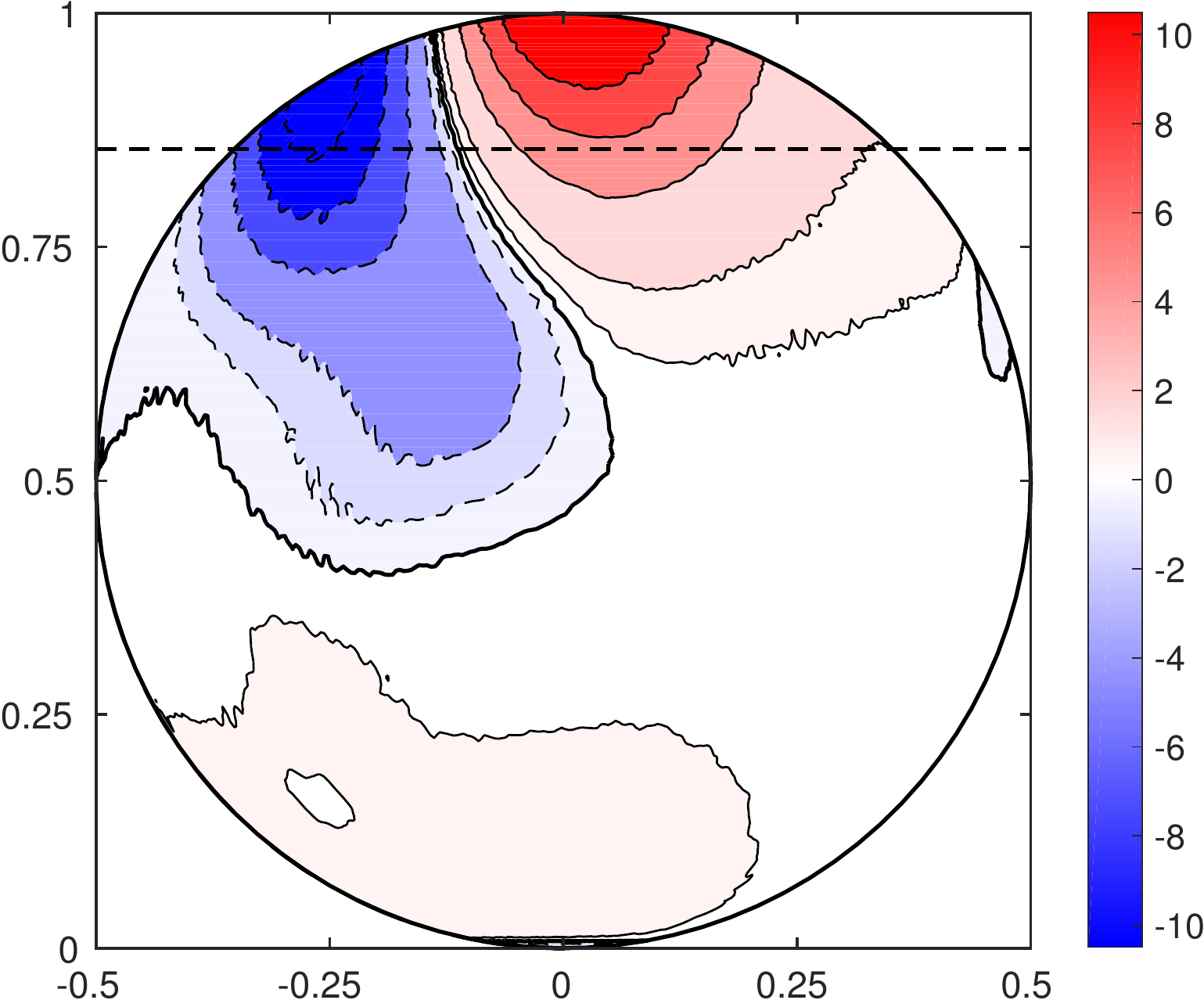}}
\put(70,-5){$\widetilde{x}/D$}
\put(-10,74){$\dfrac{y}{D}$}
\put(-8,123){\scriptsize $\dfrac{y_0}{D}$}
\put(3,124.5){\line(2,-1){10}}
\put(-7,135){$d)$}
}
\end{picture}
\caption{Distribution of the $y$-component of $\widetilde{\boldsymbol{\tau}}_n$ 
normalised by $F_R/A_{sph}$. Panels $(a,c)$ show the top view while panels $(b,d)$ 
show the side view of the sphere. $(a,b)$ run D50; $(c,d)$ run D120. 
Thin contour lines at values $\pm[0.5,\, 1.5,\, 4.5,\, 7.5,\, 10.5]$. 
The thick contour line indicates the value $0$.}
\label{fig12}
\end{figure}
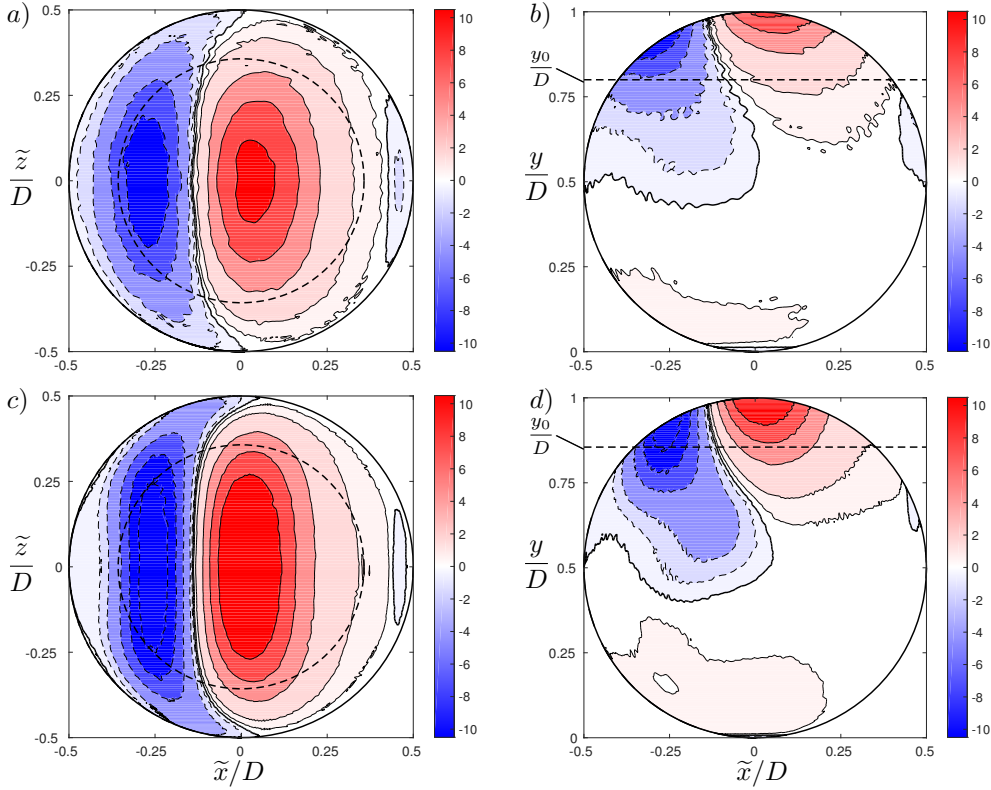

%
Figure~\ref{fig14} showed that most of the lift force can be attributed 
to the contributions of \mut{pressure} in the region near the top of the 
sphere for both the simulations D50 and D120.
In fact, the $y$-compontent of $\widetilde{\boldsymbol{\tau}}_n$ 
(this is the global wall-normal direction), which is visualised in figure~\ref{fig12}, 
exhibits essentially the same distribution as the pressure.  
\begin{figure}
\setlength{\unitlength}{0.353mm}
\begin{picture}(0,300)(0,0)
\put(5,147){
\put(0,3){\includegraphics[trim=0cm 0cm 0cm 0cm, clip, width=.44\textwidth]{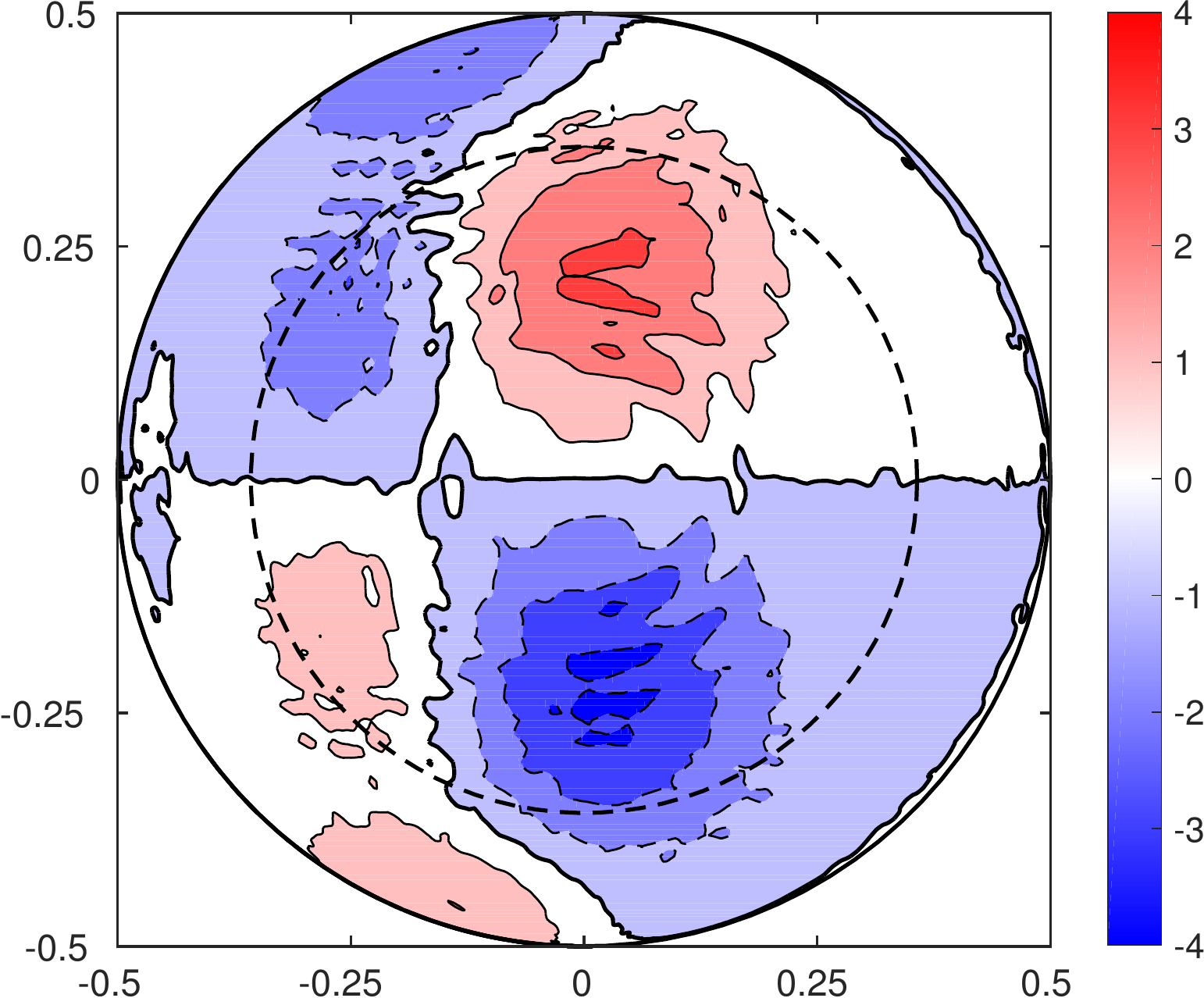}}
\put(-7,73){$\dfrac{\widetilde{z}}{D}$}
\put(-7,135){$a)$}
}
\put(200,147){
\put(0,3){\includegraphics[trim=0cm 0cm 0cm 0cm, clip, width=.43\textwidth]{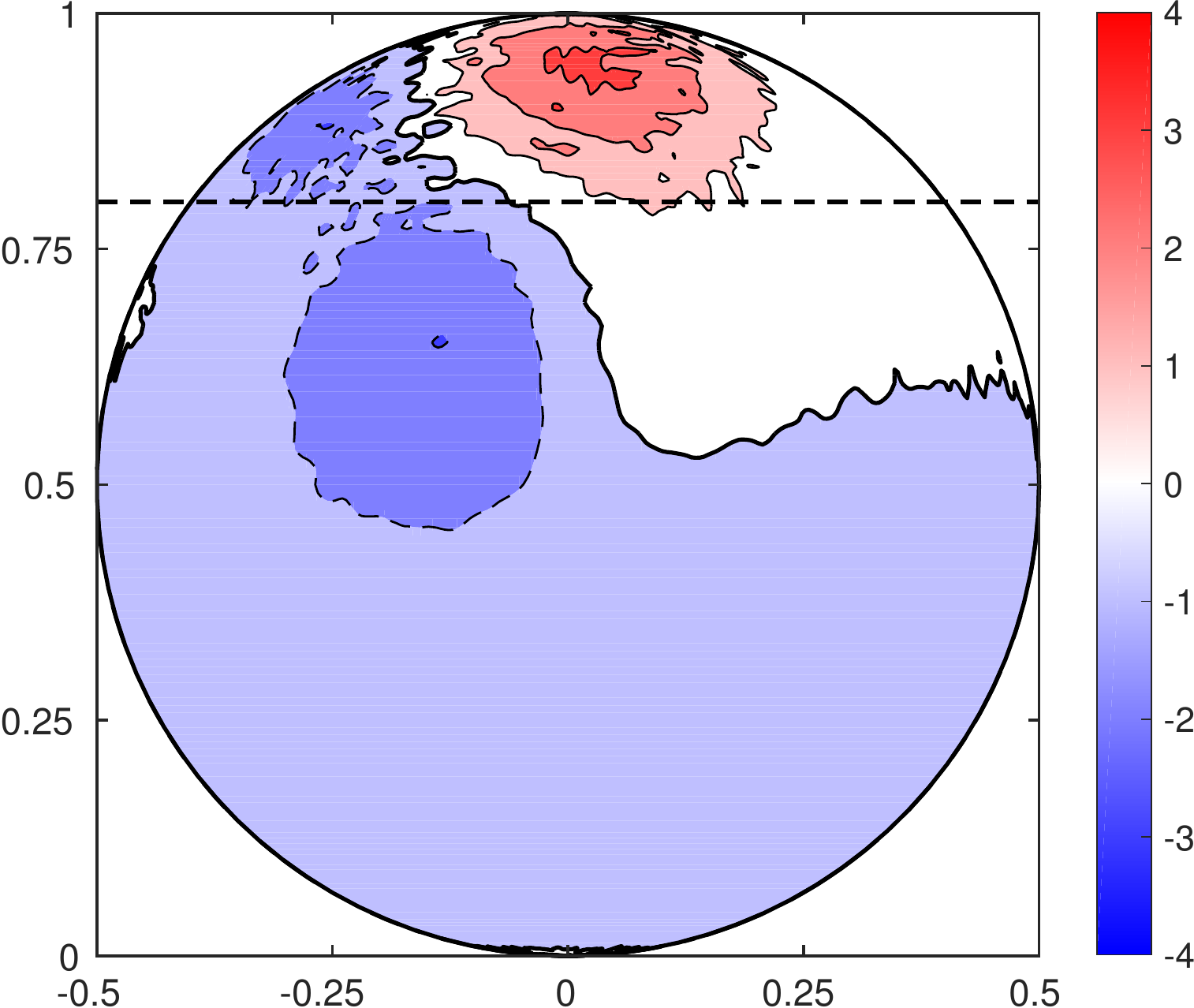}}
\put(-10,74){$\dfrac{y}{D}$}
\put(-8,116){\scriptsize $\dfrac{y_0}{D}$}
\put(3,118){\line(2,-1){10}}
\put(-7,135){$b)$}
}
\put(5,2){
\put(0,3){\includegraphics[trim=0cm 0cm 0cm 0cm, clip, width=.44\textwidth]{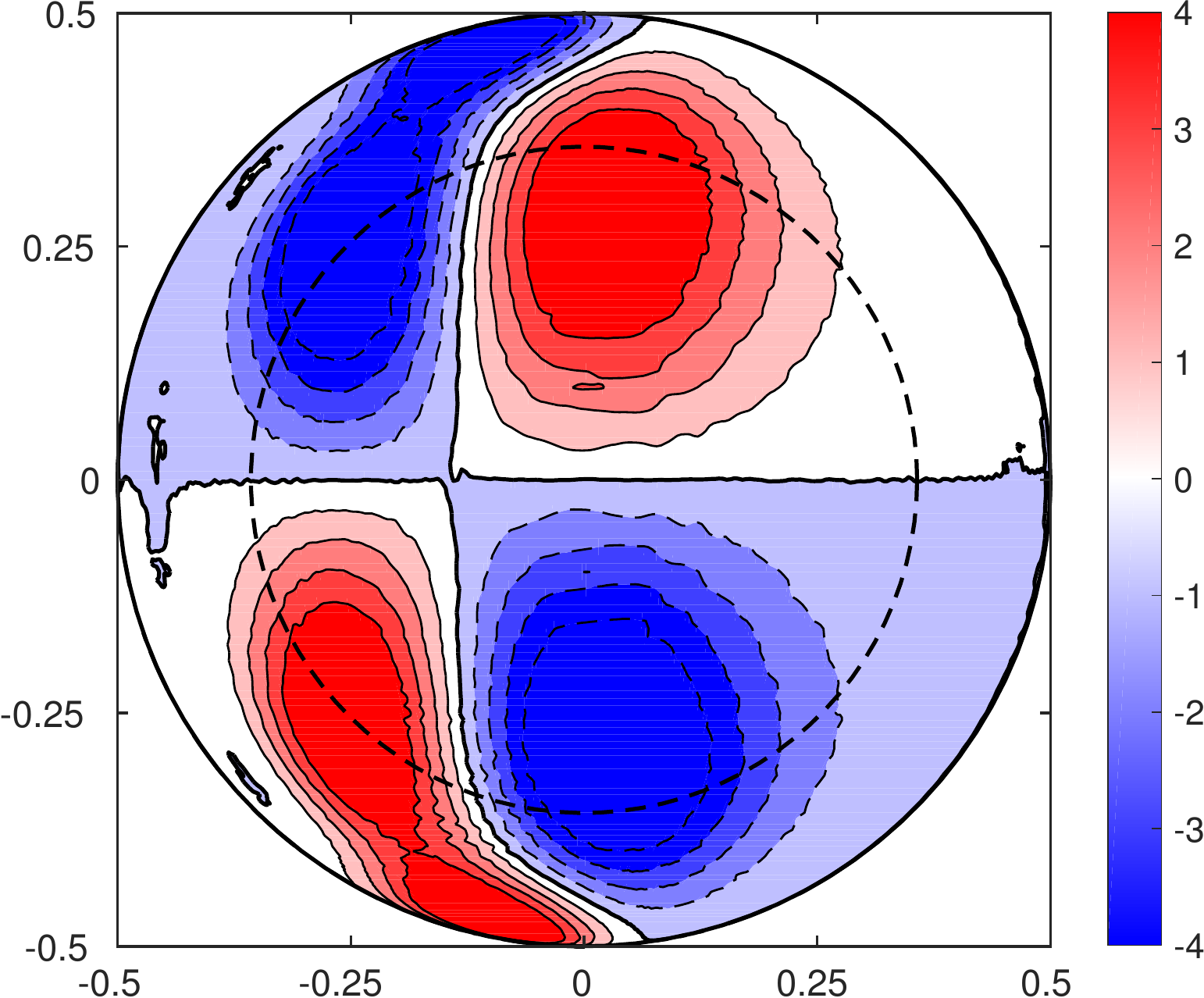}}
\put(70,-5){$\widetilde{x}/D$}
\put(-7,73){$\dfrac{\widetilde{z}}{D}$}
\put(-7,135){$c)$}
}
\put(200,2){
\put(0,3){\includegraphics[trim=0cm 0cm 0cm 0cm, clip, width=.43\textwidth]{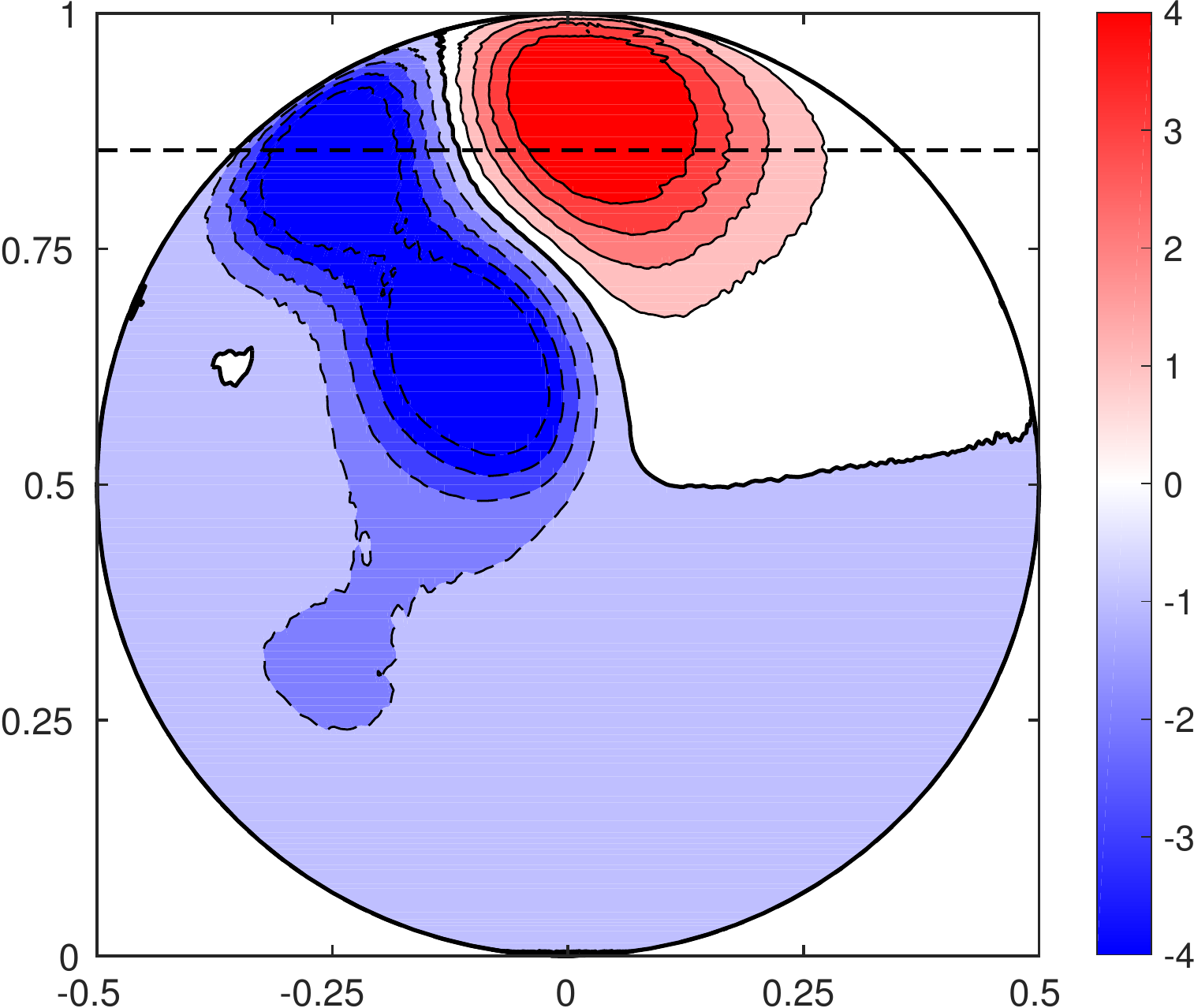}}
\put(70,-5){$\widetilde{x}/D$}
\put(-10,74){$\dfrac{y}{D}$}
\put(-8,123){\scriptsize $\dfrac{y_0}{D}$}
\put(3,125){\line(2,-1){10}}
\put(-7,135){$d)$}
}
\end{picture}
\caption{Distribution of the spanwise component of $\widetilde{\boldsymbol{\tau}}_n$ 
normalised by $F_R/A_{sph}$. Panels $(a,c)$ show the top view while panels $(b,d)$ 
show the side view of the sphere. $(a,b)$ run D50; $(c,d)$ run D120. 
Thin contour lines at values $\pm[1,\, 2,\, 3,\, 4]$. 
The thick contour line indicates the value $0$.}
\label{fig13}
\end{figure}
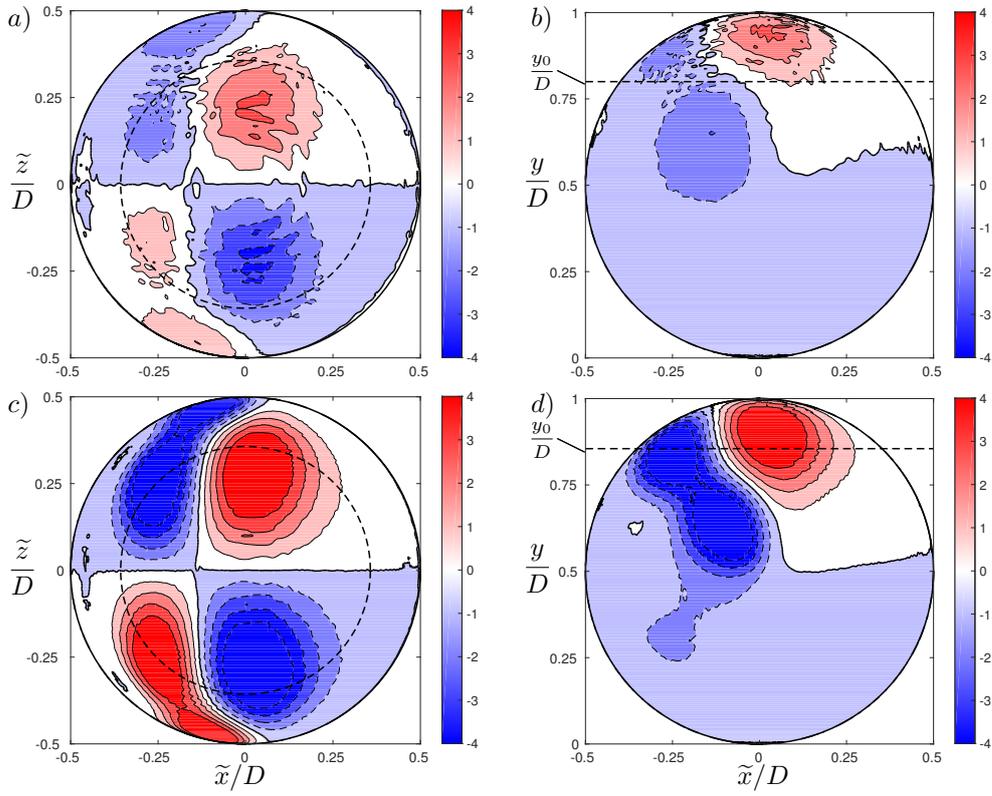

  Finally, let us return to the discussion of the effect which the
  structure of the average flow field in the vicinity of the spheres
  has upon the stress distribution on the spheres' surface.
  Please recall the strong wall-normal vorticity structures
  described in section \ref{sec1} and shown in figure~\ref{fig10}.
  Here we relate these features to the distribution of the spanwise 
  component of $\widetilde{\boldsymbol{\tau}}_n$ on the upper sides of
  the sphere, which is shown in figure~\ref{fig13}. 
  The latter figure confirms that these average vortical structures,
  which are more intense  at higher Reynolds numbers, and which tend
  to adhere closer to the sphere surface, cause stronger lateral force 
  contributions locally. 
  From these considerations it can also be understood that 
  the intensity of turbulent fluctuations of the lateral force
  increases from D50 to D120, resulting in the observed large values
  of  $\sigma_{F_z}/F_R$ and of $ \sigma_{T_y}/T_R $ (cf.\
  tables~\ref{tab3}-\ref{tab4}). 
  In particular, the increase of the value of $ \sigma_{T_y}/T_R $ in the 
  fully-rough regime along with the decrease of $ \sigma_{T_z}/T_R $ 
  provide further pieces of evidence of the breakdown of the
  ``smooth-wall analogy''. 
\subsection{What can be inferred from the force acting on the roughness elements?}\label{sec3}
In the following, a comparison of the results obtained in the present DNS 
with those from 
laboratory experiments is performed. 
As already pointed out, it is still not entirely clear how 
the details of the flow structure scale with 
the bulk Reynolds number, the particle Reynolds number and with the
relative submergence. 
However, \citet{amir2014}, who experimentally investigated the forces acting 
on wall-mounted spheres in hexagonal arrangement, identified quantities that 
are almost independent of some of these parameters in the fully-rough regime.
\begin{figure}
\setlength{\unitlength}{0.353mm}
\begin{picture}(0,160)(0,0)
\put(0,5){
\put(10,0){\includegraphics[trim=0cm 0cm 0cm 0cm, clip, width=.95\textwidth]{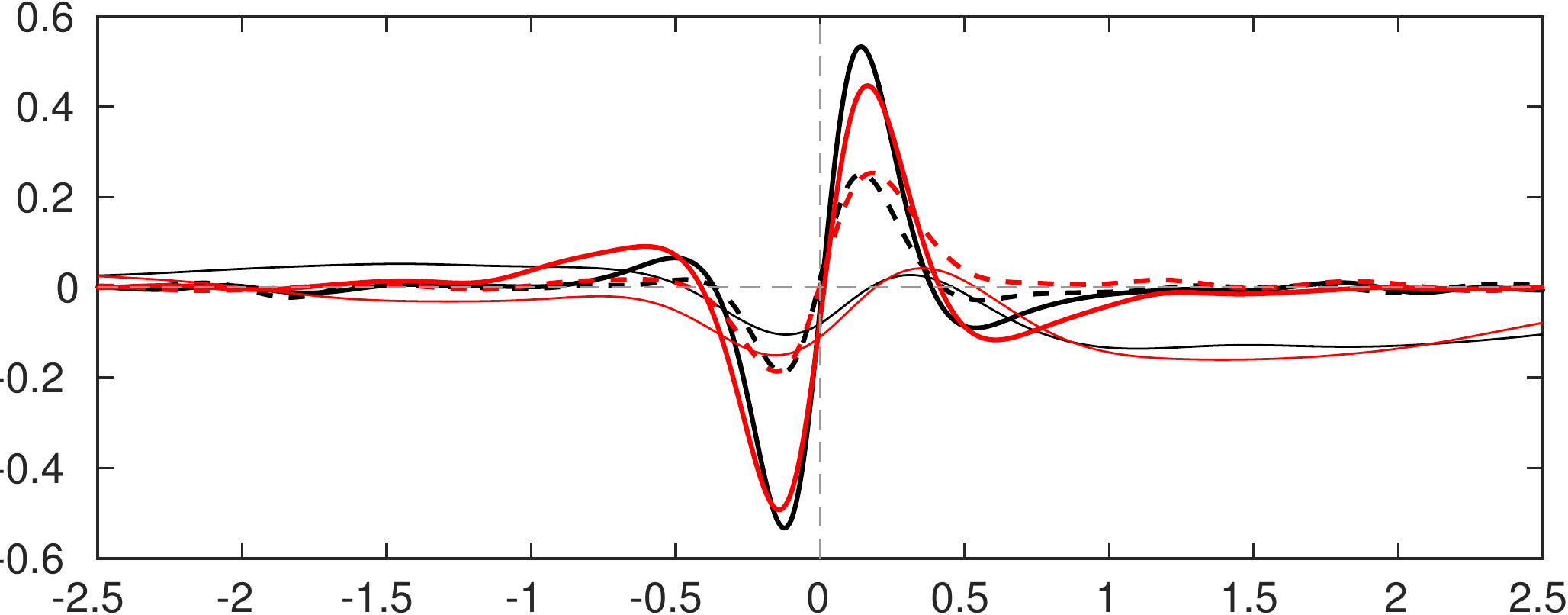}}
\put(182,-8){$\Delta_T U_{bH}/H$}
\put(0,50){\rotatebox{90}{$R_{F'_xF'_y}/(\sigma_{F_x}\sigma_{F_y})$}}
\put(206,81){$E$}
\put(210,74){\large $\star$}
\put(235,81){$F$}
\put(233,74){\large $\star$}
}
\end{picture}
\caption{
Cross-correlation $R_{F'_xF'_y}(0,\Delta r_z,\Delta_T)$ between drag and lift acting on the spheres as a function of time and for spanwise separation $\Delta r_z=0$ (thick solid lines) and $\Delta r_z=L_B$ (thick broken lines) for the runs D50 (black) and D120 (red). Thin lines indicate the function $R_{F'_xF'_y}$ referred to a square element of smooth wall (in absence of the spheres) of side-length $s^+=51$ (red) and $s^+=129$ (black) for the smooth-wall simulations at $\mut{Re_{bH}}\sim 2900$ and $\sim 6900$, respectively. For the sake of reference in the text, symbols E and F indicate the location of the zero-crossings for positive delay $\Delta_T$ in these two latter curves.
}
\label{fig23}
\end{figure}
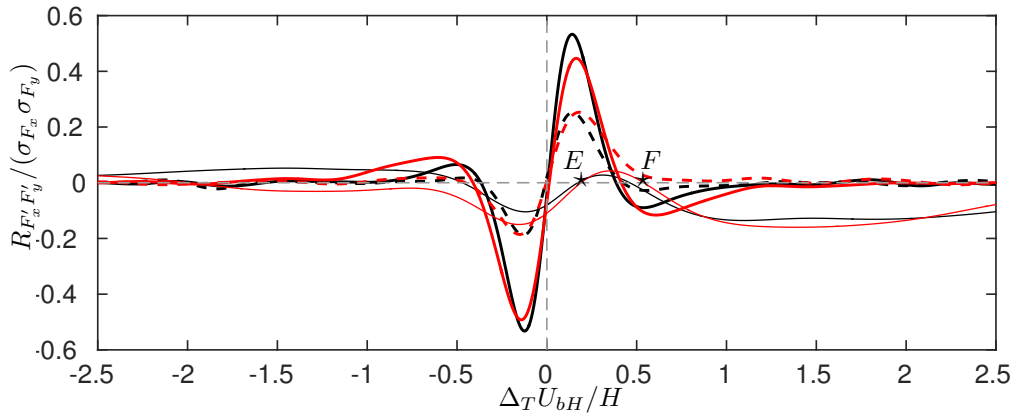
These authors estimated the temporal cross-correlation between 
neighbouring spheres on the basis of 
differential pressure measurements. 
In particular, they estimated the root mean square of the lift 
force fluctuations on the basis of the values of the difference 
between \mut{pressure} measured at the top and at the bottom of 
the spheres (as indicated with $A$ and $B$ in figure~\ref{fig24}), and
showed that it was almost directly proportional to $u_{\tau}^2$. By
normalizing the root mean square value of pressure fluctuations 
with $\varrho u_{\tau}^2$ they defined the coefficient $k_L$, which 
was independent of $\mut{U_{bh}}/u_\tau$ and also of $H/D$ for $H/D>5$, 
and found that it was equal to $3.4$.
In order to facilitate the comparison with the data of
\citet{amir2014}, we have computed the value of the coefficient
$k_L$ in an analogous manner (i.e.\ as the normalized r.m.s.\ pressure
difference between the two poles of the sphere at points $A$, $B$ in
figure~\ref{fig24}), and found that $k_L=1.1$ in case D120. 
It is reasonable that the present value is smaller than the value
obtained by  \citet{amir2014}, because the present spheres in a
square arrangement shelter each other more strongly than in the 
hexagonal arrangement investigated by those authors. 

A quantity which can be expected to be more robust with respect to
variations of the arrangement of the roughness elements is 
the convection velocity $U_c$ of larger vortex structures. 
\citet{amir2014} found a fair correlation of drag and lift between spheres 
separated by a distance of either $1$ or $2$ diameters according to the 
availability of instrumented neighbouring spheres (only a few spheres were 
instrumented with one-dimensional pressure-probes each).
Mimicking the technique described by \citet{amir2014} 
\citep[which is different from that used by][]{chan2013}, 
the convection velocity $U_{c}$ of force fluctuations which are induced by 
the interaction of roughness elements with turbulent vortex structures, was 
estimated on the basis of the lag of the peak of 
the temporal cross-correlation of 
drag force between neighbouring spheres aligned along the streamwise direction.
In particular, $U_{c}$ is found to measure $0.72\,U_{bH}$ and $0.64\,U_{bH}$ 
for the runs D50 and D120, respectively.
These values are in 
fair agreement with the value $0.66 U_{bH}$ obtained by
\citet{amir2014} for their experiments characterised by the 
relative submergence $H/D=5$. 
Since \citet{amir2014} showed that the speed $U_c$ was essentially independent 
of $\mut{U_{bh}}/u_\tau$ (see their figure 19b), the fact that the Reynolds 
number in the present DNS is substantially lower should not matter. 
Figure~\ref{fig23} shows the temporal cross-correlation of drag and lift force 
fluctuations, denoted by $R_{F'_xF'_y}(\Delta r_x=i\,L_B,\Delta r_z=j\,L_B,\Delta_T)$ 
with $i=0,\ldots,n_c-1$ and $j=0,\ldots,n_r-1$, for the runs D50 and D120 (thick lines) 
as well as for the respective simulations performed over a smooth wall (thin lines).
In the spirit of the smooth-wall analogy discussed in \S~\ref{sec2},
the lift force is associated with pressure fluctuations while drag
force fluctuations are associated with those of the viscous
shear-stress. 
Since \citet{amir2014} could equip each sphere with a one-direction pressure probe only, they could estimate $R_{F'_xF'_y}$ only from measurements made on distinct spheres, separated by a non-negative distance either in the streamwise or spanwise directions.
According to their hexagonal arrangement, the spanwise direction was more convenient.
Presently, $R_{F'_xF'_y}(0,L_B,\Delta_T)$ was computed for the simulations D50 and D120 (see broken lines in figure~\ref{fig23}).
Although the present value of $R_{F'_xF'_y}$ is approximately
two times larger than that observed by \citet{amir2014} (possibly due to the different geometrical configuration of the roughness), the time lags of the negative (or positive) peaks are found to be essentially the same, equal to $0.14\,U_{bH}/H$ and to $0.16\,U_{bH}/H$, respectively.
Moreover, the 
single-sphere cross-correlation 
function $R_{F'_xF'_y}(0,0,\Delta_T)$ was 
computed (see solid thick lines in figure~\ref{fig23}) which revealed the same time lag of the positive (negative) maximum 
($\Delta_T=0.16\,H/U_{bH}$ and $\Delta_T=-0.14\,H/U_{bH}$ for cases D50 and D120), and the presence of a secondary negative (positive) peak at $\Delta_T=0.5\,H/U_{bH}$ ($\Delta_T=-0.5\,H/U_{bH}$).
The trend of $R_{F'_xF'_y}$ indicates that a positive (negative) 
fluctuation of lift is most probably preceded by a positive (negative) 
fluctuation of drag and followed by a negative (positive) fluctuation of drag, 
but also that lift fluctuations are practically uncorrelated with any drag 
fluctuation at the same instant.
Similar statistics were observed also by \citet{hofland2005}, \citet{dwivedi2010} 
as well as by \citet{amir2014}, while the cross-correlation function for case D50 
was also shown in figure~8 of \citet{chan2013}. 
%

%
%

In case D120 the average time interval between positive and negative fluctuations of lift, equal to $\Delta_T=0.33\,H/U_{bH}$, was obtained from the occurrence of the global minimum of the auto-correlation function of lift, $R_{F_yF_y}(0,0,\Delta_T)$ (not shown here). 
  In \citet{chan2013} the value of $\Delta_T=0.42 H/U_{bH}$ is measured for their case D50. 
%
Consistently, nearly after the same time interval ($\Delta_T=0.37\,H/U_{bH}$), the cross-correlation $R_{F'_xF'_y}$ is found to vanish in both cases D50 and D120. 
Furthermore, the symmetry of $R_{F'_xF'_y}$ with respect to the point $\Delta_T=0$ suggests that forthcoming and past fluctuations have similar spatial and temporal scales as well as the same intensity. 
  The fact that highly auto-correlated lift fluctuations and cross-correlated drag-lift fluctuations are synchronised suggests that they are associated with the same turbulent event.
  Thus, a 
%
linear relationship between length and time scales of turbulent fluctuations through the convection velocity $U_c$, i.e. the validity of Taylor's frozen turbulence hypothesis, can be assumed \citep[e.g.][]{chan2013}.
Thereby, it is possible to estimate the length scale, 
(henceforth denoted as $\ell$) 
as the product of the time interval $\Delta_T=0.3\,H/U_{bH}$ between drag (or lift) fluctuations of opposite sign and the convection velocity $U_c$ previously estimated.
For the runs D50 and D120, $\ell$ was equal to $0.22\,H$ ($1.24\,D$) and $0.19\,H$ ($1.03\,D$), respectively.
Furthermore, in the spanwise direction the two-point correlation of lift and drag was found negligible at the separation distance $2\,D$ in both the cases D50 and D120 (not shown here). 
Finally, since $\ell$ depends on $U_c$, which is in turn affected by the value of $H/D$, the drag-lift cross-correlation function is possibly affected by $H/D$. %

Therefore, although the evolution of the coherent turbulent structures 
propagating over the roughness elements has not been presently studied, 
the curves of figure~\ref{fig23} suggest that the fluctuations of drag 
and wall-normal lift forces are predominantly caused by the interaction 
between the roughness elements and vortices of specific size.
In particular, the size of these vortices is comparable to the length scale 
$\ell$ which we have estimated in the aforementioned way, and, therefore, 
to the size of the spheres. %
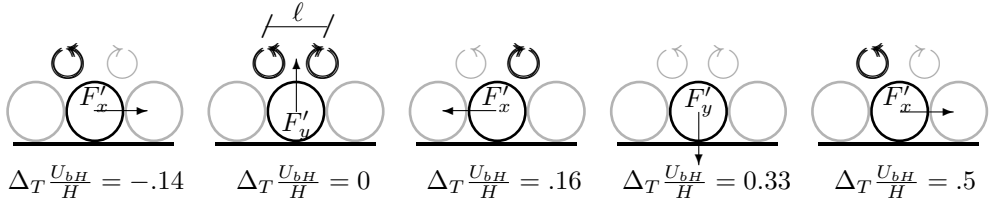
\begin{figure}
\setlength{\unitlength}{0.353mm}
\begin{picture}(0,75)(0,0)
\put(10,5){
\put(10,0){
\put(6,40){\LARGE $\pmb{\circlearrowleft}$}
\put(26,40){\LARGE\color{gray!70} $\circlearrowright$}
\put(10,20){\Huge $\bigcirc$}
\put(-12,20){\color{gray!60}\Huge $\bigcirc$}
\put(32,20){\color{gray!60}\Huge $\bigcirc$}
\put(22,26.5){\vector(1,0){20}}
\linethickness{.4mm}
\put(-8,14){\line(1,0){60}}
\thinlines
\put(-10,-3){$\Delta_T\frac{U_{bH}}{H}=-.14$}
\put(16.5,28.5){$F'_x$}
}
\put(85,0){
\put(6,40){\LARGE $\pmb\circlearrowleft$}
\put(26,40){\LARGE $\pmb\circlearrowright$}
\put(10,20){\Huge $\bigcirc$}
\put(-12,20){\color{gray!60}\Huge $\bigcirc$}
\put(32,20){\color{gray!60}\Huge $\bigcirc$}
\put(22.5,26){\vector(0,1){20}}
\linethickness{.4mm}
\put(-8,14){\line(1,0){60}}
\thinlines
\put(0,-3){$\Delta_T\frac{U_{bH}}{H}=0$}
\put(17,19){$F'_y$}
\put(11,58){\line(1,0){23}}
\put(9,56){\small$\boldsymbol{/}$}
\put(32,56){\small$\boldsymbol{/}$}
\put(21,60){$\ell$}
}
\put(160,0){
\put(6,40){\LARGE\color{gray!70} $\circlearrowleft$}
\put(26,40){\LARGE $\pmb\circlearrowright$}
\put(10,20){\Huge $\bigcirc$}
\put(-12,20){\color{gray!60}\Huge $\bigcirc$}
\put(32,20){\color{gray!60}\Huge $\bigcirc$}
\put(22,26.5){\vector(-1,0){20}}
\linethickness{.4mm}
\put(-8,14){\line(1,0){60}}
\thinlines
\put(-4,-3){$\Delta_T\frac{U_{bH}}{H}=.16$}
\put(16.5,28.5){$F'_x$}
}
\put(235,0){
\put(6,40){\LARGE\color{gray!70} $\circlearrowleft$}
\put(25,40){\LARGE\color{gray!70} $\circlearrowright$}
\put(10,20){\Huge $\bigcirc$}
\put(-12,20){\color{gray!60}\Huge $\bigcirc$}
\put(32,20){\color{gray!60}\Huge $\bigcirc$}
\put(22.5,26){\vector(0,-1){20}}
\linethickness{.4mm}
\put(-8,14){\line(1,0){60}}
\thinlines
\put(-6,-3){$\Delta_T\frac{U_{bH}}{H}=0.33$}
\put(17,28){$F'_y$}
}
\put(310,0){

\put(6,40){\LARGE $\pmb\circlearrowleft$}
\put(25,40){\LARGE\color{gray!70} $\circlearrowright$}
\put(10,20){\Huge $\bigcirc$}
\put(-12,20){\color{gray!60}\Huge $\bigcirc$}
\put(32,20){\color{gray!60}\Huge $\bigcirc$}
\put(22.5,26){\vector(1,0){20}}
\linethickness{.4mm}
\put(-8,14){\line(1,0){60}}
\thinlines
\put(-2,-3){$\Delta_T\frac{U_{bH}}{H}=.5$}
\put(16,28.5){$F'_x$}
}
}
\end{picture}
\caption{
Conceptual scheme of the sequence of positive-to-negative drag and lift 
fluctuations ideally induced by disturbances propagating with the speed $U_c$. %
The values of time intervals were obtained for the simulation D120.
}
\label{fig23a}
\end{figure}

With the purpose to help the
interpretation of figure~\ref{fig23}, a conceptual model of the response 
of the spheres to the action of these vortices is sketched in figure~\ref{fig23a}
\mut{
  where, in a schematic side view, spheres are represented by circles,
  vortices by circular arrows and hydrodynamic forces by vectors
  applied to the sphere center. %
}
In this model, turbulent coherent structures induce pulsations of the 
intensity of the vortices characterizing the average flow field. %
These pulsations propagate downstream with speed $U_c$ and 
produce on the spheres the sequence of fluctuations of drag and wall-normal 
lift selected by the cross-correlation curves of figure~\ref{fig23}. %
\mut{
  Vortices involved in the intensity pulsation at different propagation phases are 
  highlighted in black in the sequences of figure~\ref{fig23a}.
}
Hence, the period of the pulsation can be defined as 
the time interval between two consecutive positive (or negative) 
peaks of the cross-correlation function in figure~\ref{fig23}, which measures $0.64 H/U_{bH}$. %
Let us note that disturbances which do not result in highly cross-correlated 
fluctuations of the net drag and lift forces acting on the spheres are not 
considered by the model. %
The conceptual model of interaction between turbulence structures 
and roughness elements sketched in figure~\ref{fig23a} is consistent with 
the statistical evidences related to the hydrodynamic forces and with 
the topology of the average vorticity field described above. 

\mut{
  However, so far our argument only relates to pulsations of the mean flow in the vicinity
  of the spheres, not to the actual coherent structures which cause the pulsations. In order
  to go further, a detailed analysis of the correlation between the
  particle forces and the flow field should be carried out, as has been
  done by \citet{chan2013} for the transitionally-rough case. 
  In the present case, one should specifically identify those turbulent
  events which have a maximum correlation with the occurrence of a
  strong positive lift force fluctuation, and which are the same events
  that have a maximum
  correlation with a strongly positive (negative) particle drag
  fluctuation at a fixed negative (positive) delay time of 
  $\Delta_T\sim 0.15 H/U_{bH}$. This additional analysis, which is beyond the scope of
  the present work, would yield information on the flow structure
  responsible (in an average sense) for the said cross-correlation.
}

\vspace{1ex}
Finally, the cross-correlation function of drag and lift, $R_{F'_xF'_y}$, was
also calculated for the smooth-wall simulations at $\mut{Re_{bH}}=2900$ and
$6900$ 
  in order to estimate length and time scales of force fluctuations
  acting on a square tile (with side length $L_B$) of a smooth channel
  wall at a corresponding Reynolds number. 
%
The thin solid lines of figure~\ref{fig23} show the cross-correlation for 
the smooth-wall cases.
Both the curves for the smooth-wall simulations at $\mut{Re_{bH}}=2900$ 
and $6900$ show a negative correlation at the time interval $\Delta_T=0$, 
which means that a lift fluctuation is most probably associated with a drag 
fluctuation of opposite sign.
Indeed, these combinations bring to mind the action of sweep-like and
ejection-like events which 
dominate the Reynolds stress statistics in a turbulent wall-bounded
flow over a smooth wall. 
%
As opposed to 
the rough-wall cases, the cross-correlation of lift
and drag for the smooth-wall simulations do not 
exhibit the symmetry with respect to the origin of the axes, suggesting that
forthcoming and past turbulent events are in general characterised by 
different length and time scales.
However, from the diagrams of figure~\ref{fig23}, it is possible to 
distinguish large turbulent structures, characterised by length scale 
much larger than $L_B$, from ``smaller'' turbulent structures characterised 
by the time scale $\Delta^{(F)}_T-\Delta^{(E)}_T$ equal to $0.30$ and $0.36$ 
for the simulations at $\mut{Re_{bH}}=2900$ and $6900$, respectively, 
where $\Delta^{(E)}_T$ and $\Delta^{(F)}_T$ denote the time intervals at 
the points $E$ and $F$, marked by a star in figure~\ref{fig23}, when the 
cross-correlation function vanishes. 
Then, by using the same procedure described for the rough bottom simulations, 
the convective velocity of turbulent structures $U_c$ was estimated for the 
smooth-wall cases equal to $0.71\,U_{bH}$ and $0.69\,U_{bH}$, respectively. 
Hence, the length scale of the ``smaller'' turbulent structures was found 
equal to $0.20\,H$ and to $0.25\,H$ for the simulations at $\mut{Re_{bH}}=2900$ 
and $6900$, respectively, which are nearly equal to $L_B$.
Indeed, \citet{chan2011} noted that the average operator over a tile of size 
$L_B$ filtered out turbulent structures much smaller or much larger than $L_B$.
Such a filter effect was also actually produced by the sphere surface in the 
rough bottom cases.
However, in the limits of the cross-correlation analysis, the present
results show that the mechanism of flow-roughness interaction both in
the transitionally- and fully-rough regimes appears significantly
different from 
  the interaction between the turbulent flow and a smooth wall.  
%
%
\iffourteen
\begin{figure}
\setlength{\unitlength}{0.353mm}
\begin{picture}(0,300)(0,0)
\put(5,145){
\put(0,0){\includegraphics[trim=0cm 1cm .5cm 1cm, clip, width=.43\textwidth]{\figdir F50_spantorque_distrib_contour_top_JFM}}
\put(-7,74){$\widetilde{z}^+$}
\put(-7,135){$a)$}
}
\put(200,145){
\put(0,0){\includegraphics[trim=0cm 1cm .5cm 1cm, clip, width=.43\textwidth]{\figdir F50_spantorque_distrib_contour_side_JFM}}
\put(-7,74){$y^+$}
\put(-8,122){\scriptsize $y^+_0$}
\put(3,121.5){\line(2,-1){10}}
\put(-7,135){$b)$}
}
\put(5,0){
\put(0,0){\includegraphics[trim=0cm 1cm .5cm 1cm, clip, width=.43\textwidth]{\figdir F120_spantorque_distrib_contour_top_JFM}}
\put(73,-4){$\widetilde{x}^+$}
\put(-7,74){$\widetilde{z}^+$}
\put(-7,135){$c)$}
}
\put(200,0){
\put(0,0){\includegraphics[trim=0cm 1cm .5cm 1cm, clip, width=.43\textwidth]{\figdir F120_spantorque_distrib_contour_side_JFM}}
\put(73,-4){$\widetilde{x}^+$}
\put(-7,74){$y^+$}
\put(-8,122){\scriptsize $y^+_0$}
\put(3,121.5){\line(2,-1){10}}
\put(-7,135){$d)$}
}
\end{picture}
\label{fig14}
\caption{Distribution of the spanwise component of torque on the surface of 
roughness elements normalised by $T_R/A_{sph}$. Panels $(a,c)$ show the top 
view while panels $(b,d)$ show the side view of the sphere. $(a,b)$ run D50; 
$(c,d)$ run D120.}
\end{figure}
\fi
\section{Conclusions}\label{sec4}
Direct numerical simulation of open-channel flow has been performed
over a bed of spheres in square arrangement for values of the bulk and
roughness Reynolds numbers of approximately $6900$ and $120$,
respectively, for which the fully-rough flow regime was attained. 
The objective of the present work is to quantify the differences in
the flow structure which arise when passing from the
transitionally-rough flow regime to the fully-rough one, providing us with an
enhanced picture of the flow-roughness interaction.
Therefore, one of the DNSs performed by \citet{chan2011} in the
transitionally-rough regime ($\mut{Re_{bH}}=2900$ and $D^+=50$),
characterised by the same shallowness and arrangement of the spheres
as the present DNS, is extensively referred to. 
Furthermore, smooth-wall data from simulations at the same
bulk Reynolds numbers and computational domain size are used for the
purpose of comparison. 
The two DNSs, in the transitionally- and fully-rough regimes, were
compared by using different statistical tools. 
In particular, three average operators were employed, namely the
simple time-average, the periodic-sphere-box-average and the wall-normal
plane-average (the latter two combined with the time-average), 
each extracting different statistical features of the flow. 

When increasing the Reynolds number it is observed that the flow
penetrates deeper into the crevices between the roughness elements,
causing a secondary flow in the vicinity of the spheres which 
consists of two pairs of recirculation cells.
A downward shift of the mean flow profile is observed, 
which results in the disappearance of the viscous sublayer
and of the buffer layer, while viscous effects were confined to the
region below the crests of the spheres where velocity fluctuations
were found to be strongly correlated.
In fact, velocity fluctuations around the
periodic-sphere-box-averaged flow field (i.e.\ those which exclude
the average flow field at the roughness element scale) were
found to be of nearly negligible intensity below the crests, whereas
those around the plane-average were large. 
This fact indicates that vortex structures in the interstitial flow
are weakly turbulent, but mostly associated with the average shear-layer
originating from the sphere surface. 
Although, the damping of turbulent fluctuations in the interstitial
fluid region does not significantly affect the flow structure in the
transitionally-rough regime, because normal \mut{stresses} attain the
maximum intensity over the crest of the roughness elements, it
definitely does in the fully-rough regime, since the peak of normal \mut{stresses} 
would have been placed below the crest of the spheres. %
This is
a clear indicator of the destruction of the buffer layer and, consequently, that 
the flow regime is fully-rough %
\pbt{%
  (only a small relative maximum of the root mean square of the streamwise 
  velocity is present above the crest of the spheres). %
}%

Moreover, the flow in the \mut{log-law} region was investigated.
%
%
It was found that, for the present 
configuration, the von
K\'arm\'an constant deviates \mut{somewhat} from the value that is 
attained in absence of the roughness, attaining the value $0.381$ 
in the fully-rough regime.
\mut{
  Note that this value is still within the bounds
  ($\kappa=0.39\pm0.02$) determined by \citet{marusic2013} from a
  data-set including various boundary layer and pipe flow experiments
  as well as measurements in the atmospheric surface layer, covering a
  considerable range of Reynolds numbers under nominally smooth and,
  in the latter case, transitionally-rough conditions ($k_s^+\approx21$).
}
By comparing the velocity profile obtained from simulations over a smooth and a 
fully-rough wall it was possible to quantify the conductivity of the
present 
configuration with respect to the ideal case of maximum conductivity.
The integration constant $C_I^+$ appearing in the law of the wall \eqref{eq1} tends to zero 
with increasing particle Reynolds number while the 
conductivity $C_{II}^+$ (which includes the effects associated with the geometrical 
configuration of the domain) approaches $11.6$.
Presently, the value of $C_{II}^+$ was found to be equal to $10.7$, 
indicating that, for the present values of $H/D$ and for a 
well-packed square arrangement of spheres, 
the roughness geometry is highly conductive. 

The vorticity field, averaged in time and over periodic boxes around
the spheres, has been discussed in detail. 
It was found that on average the vortical structures in the spheres'
wakes reach significantly further downstream at the larger particle
Reynolds number, clearly reconnecting with the surface of the
downstream neighbor sphere.
Furthermore, in case D120 a second set of four
counter-rotating vortices appears in the inter-particle grooves due
to the fact that the primary mean flow velocities are enhanced in
that region. 

Low- and high-speed streaks were observed over the crest of the
spheres similarly to those forming over a smooth wall.
These structures are characterised by a spanwise spacing 
increasing linearly with the distance from the spheres.
In particular, the spacing between the streaks at the crest level
($y^+=D^+$) is almost the same as that 
observed over a smooth wall at the same Reynolds number when the same
distance from the virtual wall is chosen (i.e.\
$y^+_{smooth}=(D^+-y_0^+)_{rough}$). 
In general, the relationships between 
the minima locations of the spanwise two-point correlations of the
different velocity components  
$\lambda_{ux}^+$,
$\lambda_{uy}^+$ and $\lambda_{uz}^+$ shown by \citet{KMM1987} for  
channel-flow over a smooth wall were found to hold also over a fully-rough wall, but
their individual magnitudes are modulated by the
presence of the roughness. 

  The action of the flow on individual roughness elements
  has been investigated in detail, including an investigation of the
  spatial distribution of stress statistics on the surface of the
  spheres. 
  The coefficient of the average drag force acting on the spheres remains
  almost constant when the bulk Reynolds number is increased from
  $2900$ to $6900$ (for a fixed value of the relative submergence),  
  which is in line with the Darcy-Weisbach equation. 
  At the same time the average lift coefficient increases by
  approximately $30\%$ in this interval, 
  mostly due to the contribution of pressure acting on the upper
  upstream part of the sphere surface.
  Concerning the distribution of the streamwise component of the
  stress vector at the sphere surface,
  we have observed that in the fully-rough regime
  the most intense contribution shifts towards
  smaller wall-distances (downwards) and towards the upstream-facing
  part of the upper side of the spheres, 
  as compared to the transitionally-rough case. 
  This same region approximately coincides with the region where the
  average pressure on the sphere's surface attains relatively large 
  positive values (note the definition of pressure in
  \ref{equ-define-total-pressure}). 
  Compared to the transitionally-rough reference case, the
  contribution of the skin-friction to the net drag force
  acting on the roughness elements is reduced by nearly $15\%$ in the
  fully-rough regime, while the pressure contribution becomes dominant.  
  
  An analogy with the force/torque acting upon an element of a
  smooth-wall channel flow, which was successfully used by
  \citet{chan2011} to explain e.g.\ the statistical moments of the
  torque acting on the roughness elements, 
  is found to break down in the fully-rough regime. 
  The breakdown of the analogy with a smooth wall, which is
  not unexpected, can be attributed to the roughness-induced
  destruction of the buffer layer and to the reduction of the
  importance of viscous effects in the vicinity of the sphere crests. 

The results obtained through DNS in the fully-rough regime were
compared with those of experiments by \citet{amir2014} in an
open-channel flow over a bed of immobile spheres in a hexagonal
arrangement at larger particle Reynolds numbers. 
The convective velocity of turbulent structures, $U_c$, was estimated
from the time correlation of the hydrodynamic forces acting on
neighbouring spheres and showed values in fair agreement with those
obtained by \citet{amir2014}. 
This supports the observation that the convective velocity depends
only on the relative submergence 
while it is independent of the Reynolds number and, on
the basis of the present results, it also appears to be largely
independent of the precise arrangement of the roughness elements. 
Indeed, this was, to the knowledge of the authors, the first time that
results obtained by DNS were used to supplement experimental
observations in a fully-rough open-channel flow. 

Finally, we have investigated the cross-correlation between
fluctuations of drag and wall-normal lift forces, which are known to
be highly correlated for non-zero separation times.
Our data reveals a periodic sequence of events with an average
period of approximately $0.64H/U_{bH}$. 
Such a cyclic process was explained 
as a manifestation of turbulence below the crest of the spheres,  
consisting in pulsations of the intensity of the spanwise-oriented 
vortices adhering to the sphere surface which
were identified in the sphere-box/time-averaged flow field.

Although, DNS in the fully-rough regime at 
even larger Reynolds numbers than presently considered 
are already possible, we belive that it would be more relevant for the purpose of
studying the flow resistance in natural rivers to try to remove the
condition of regularity of the arrangement of the roughness elements. 
Roughness elements of similar size and shape in a ``random'' 
arrangement will presumably have a different interaction with 
the flow with respect to the present cases.
Moreover, the shelter effect of neighbouring roughness elements on each other
could be investigated and quantified 
when considering a random bed. 

\vspace*{1ex}
This study has been funded by the Deutsche Forschungsgemeinschaft
(project UH~242/4-2).  
We acknowledge the generous support from 
the Leibniz-Rechenzentrum (LRZ, Munich) for the
extensive computational resources on SuperMUC 
(project \textit{pr87yo})
and from the Steinbuch Centre for Computing (SCC, Karlsruhe), in
particular for generous allotment of storage space. 
The authors wish to thank Clemens Chan-Braun, Manuel 
Grac\'ia-Villalba and Aman Kidanemariam for the helpful discussions 
throughout the elaboration of the present work. 
Thanks is also due to Clemens Chan-Braun for initially setting up the 
simulation D120. 
\appendix
\section{Definition of average operators}\label{apxA}
Let $q(\boldsymbol{X},T)$ be a quantity defined in the present
computational domain as a function of the random variables
$\boldsymbol{X}$ and $T$, and let us consider the following events 
\begin{eqnarray}\label{eqA1-1}
E_1 &=& \lbrace\boldsymbol{X}-\boldsymbol{x}^{(i)}_c \in
\mathcal{B}\rbrace = \lbrace\boldsymbol{X} \in
\mathcal{B}+\boldsymbol{x}^{(i)}_c\rbrace\,,\\\label{eqA1-2}
E_2 &=& \lbrace\phi(\boldsymbol{X})=1\rbrace \,,\\
E_3 &=& \lbrace (X,y,Z)=(x,y,z)\rbrace\,,
\label{eqA1}
\end{eqnarray}
where $\boldsymbol{x}^{(i)}_c$ are the coordinates of the center of
the $i-$th sphere, $\mathcal{B}$ is the rectangular domain defined in
section \ref{sec0}, and $\phi$ denotes 
an indicator function which measures unity at points occupied by fluid
and vanishes otherwise. 
Note that in the statement (\ref{eqA1-1})  the position of the (fixed)
sphere 
$\boldsymbol{x}^{(i)}_c$ is not a random variable, and, due to the
perfectly square arrangement, turbulence
statistics are independent of the position of the $i-$th
sphere, and the probabilty $\mathbb{P}(E_1)$ is constant and equal to
$1/N_s$. 
Moreover, $q$ is assumed to be an ergodic process such that its
statistics converge for a ``sufficiently long'' 
sampling time and the dependence on time 
can be removed by applying the temporal average operator 
\begin{equation}
\overline{q}(\boldsymbol{X})=\dfrac{1}{\mathcal{T}}\displaystyle\int_{\mathcal{T}}q\,dt
\label{eqA2}
\end{equation}
where $\mathcal{T}$ denotes the time interval during which $q$ is sampled. For the sake of clarity, hereafter the dependence on time is implied if an overline is not present.\\
Then, two space average operators are defined on the basis of the 
events (\ref{eqA1-1}-\ref{eqA1}): 
\begin{itemize}
\item The \textit{sphere-box-average} $\widetilde{\left\langle q\right\rangle}_B(\widetilde{\boldsymbol{x}})$ is the expected value of $\phi q$ when the conditioned event $E_1\vert E_2$ occurs
\begin{eqnarray}
\widetilde{\left\langle q\right\rangle}_B(\widetilde{\boldsymbol{x}}) & = &
\displaystyle\sum\limits_{i=1}^{N_s} \phi(\boldsymbol{x}) q(\boldsymbol{x}-\boldsymbol{x}^{(i)}_c) \mathbb{P}\left(E_1\vert E_2\right) \nonumber\\
& = & \displaystyle\sum\limits_{i=1}^{N_s} \phi(\boldsymbol{x}) q(\boldsymbol{x}-\boldsymbol{x}^{(i)}_c) \mathbb{P}(E_2\vert E_1)\mathbb{P}(E_1)/\mathbb{P}(E_2)
\label{eqA3}
\end{eqnarray}
where $\widetilde{\boldsymbol{x}}\in \mathcal{B}$ and $\phi$ is
\begin{equation}
\phi \left(\boldsymbol{X}\right) = 
\left\lbrace
\begin{array}{l l}
  1 &\mathrm{if} \quad \left\vert\boldsymbol{X}-\boldsymbol{x}^{(j)}_c\right\vert > D/2 \quad \mathrm{for}\: \mathrm{every}\quad j=1,\ldots,N_s \\
  0 &\mathrm{otherwise}\:\: .
\end{array}
\right.
\label{eqA4}
\end{equation}
Because of the streamwise and spanwise periodicity of the sphere arrangement and of the boundary conditions, we note that the function $\phi\left(\boldsymbol{X}\right)$ 
is unaffected by the translation of coordinates 
$\phi(\boldsymbol{X}-\boldsymbol{x}^{(i)}_c)$ because $\phi$ is the periodic extension of the function $\widetilde{\phi}$ defined in the random variable $\widetilde{\boldsymbol{X}}\in \mathcal{B}$. Similarly, $\widetilde{\left\langle q\right\rangle}_B$ can be extended periodically in the streamwise and spanwise directions such that the resulting sphere-box-averaged flow field $\left\langle q\right\rangle_B(\boldsymbol{x})$ equals $\widetilde{\left\langle q\right\rangle}_B(\boldsymbol{x}-\boldsymbol{x}^{(i)}_c)$ for every $i=1,\ldots,N_s$. 
Thus, since $\mathbb{P}(\phi=1)=\mathbb{P}(\widetilde{\phi}=1)$ and
$\mathbb{P}(E_2\vert E_1)=\mathbb{P}(E_2)$, the sphere-box-average of
$q$ can be equally 
expressed in the two following forms 
\begin{eqnarray}
\left\langle q\right\rangle_B(\boldsymbol{x}) & = &
\dfrac{1}{N_s} \displaystyle\sum\limits_{i=1}^{N_s} \phi(\boldsymbol{x}) q(\boldsymbol{x}-\boldsymbol{x}^{(i)}_c)\\
\mathrm{or}\nonumber\\
\widetilde{\left\langle q\right\rangle}_B(\widetilde{\boldsymbol{x}}) & = &
\dfrac{1}{N_s} \displaystyle\sum\limits_{i=1}^{N_s} \widetilde{\phi}(\widetilde{\boldsymbol{x}}) q(\boldsymbol{x}-\boldsymbol{x}^{(i)}_c)\:\: .
\label{eqA5}
\end{eqnarray}
\item The \textit{plane-average} $\left\langle q\right\rangle(y)$ is defined as the expected value of $\phi q$ when the conditioned event $E_3\vert E_1$ occurs, i.e. the mean value of $\phi q$ over wall-parallel planes excluding the points where $\phi$ vanishes. Similarly the streamwise- and spanwise-averages can be defined under the further conditions $Z=z$ and $X=x$ and indicated by $\left\langle q\right\rangle_x(y,z)$ and $\left\langle q\right\rangle_z(x,y)$, respectively.
\end{itemize}
\bibliographystyle{apalike}
\bibliography{\refdir Refbib_channel}

\end{document}